\documentclass[12pt,a4paper,JHEP3]{article}

\pdfoutput=1
\usepackage{jheppub}
\usepackage{slashed,cancel,lscape,caption,array,graphicx,subcaption}
\usepackage[utf8]{inputenc}
\usepackage{amsmath}
\usepackage{nicefrac}  
\usepackage{mathtools}
\usepackage{multirow}
\usepackage{booktabs} 
\usepackage{chngcntr}
\usepackage{nicefrac}

\usepackage{tikz}
\usetikzlibrary{decorations.pathmorphing}
\usetikzlibrary{decorations.markings}
\usetikzlibrary{positioning, shapes, snakes, arrows}

\tikzset{
	graviton/.style={line width=.8pt, -latex,decorate, decoration={snake, segment length=4pt,amplitude=1.8pt, pre length=.15cm, post length=.25cm}},
	worldline/.style={gray, line width=1pt},
	worldlineBold/.style={black, line width=.6pt},
	zUndirected/.style={line width=1pt},
	zParticle/.style={line width=1pt,postaction={decorate},decoration={markings,mark=at position .6 with {\arrow[#1]{latex}}}},
	zParticleF/.style={line width=1pt,postaction={decorate}},
	cscalar/.style={line width=1pt,postaction={decorate},decoration={markings,mark=at position .6 with {\arrow[#1]{latex}}}},
	cscalar2/.style={line width=1pt,postaction={decorate},decoration={markings,mark=at position .8 with {\arrow[#1]{latex}}}},
	photon/.style={line width =.8pt, decorate, decoration={snake, segment length=4pt, amplitude=1.8pt,  pre length=.1cm, post length=.1cm}},
	photonRed/.style={red, line width =.8pt, decorate, decoration={snake, segment length=4pt, amplitude=1.8pt,  pre length=.1cm, post length=.1cm}},
	cross/.style={cross out, line width =.8pt, draw=black, minimum size=2*(#1-\pgflinewidth), inner sep=0pt, outer sep=0pt},
cross/.default={4pt}
}

\DeclareFontFamily{OT1}{pzc}{} 
\DeclareFontShape{OT1}{pzc}{m}{it}{<-> s * [1.350] pzcmi7t}{}
\DeclareMathAlphabet{\mathpzc}{OT1}{pzc}{m}{it}

\def\cO{\mathcal{O}}

\def\eps{\epsilon}

\def\d{\mathrm{d}}
\def\D{\mathrm{D}}

\def\dd{\delta\!\!\!{}^-\!}

\def\d{\mathrm{d}}
\def\eps{\epsilon}

\def\braket#1{\langle #1 \rangle}

\newcommand*\Bell{\ensuremath{\boldsymbol\ell}}

\def\nn{\nonumber}

\def\ie{i.e. }

\def\eqn#1{eq.~\eqref{#1}}
\def\eqns#1#2{eqs.~\eqref{#1} and~\eqref{#2}}

\def\Fig#1{Fig.~{\ref{#1}}}

\def\sec#1{section~{\ref{#1}}}

\def\App#1{Appendix~{\ref{#1}}}
\def\rcite#1{ref.~\cite{#1}}
\def\rcites#1{refs.~\cite{#1}}

\newcommand{\vev}[1]{\langle #1\rangle}

\newcommand{\xdot}{{\dot x}}

\newcommand{\be}{\begin{equation}}
\newcommand{\ee}{\end{equation}}
\newcommand{\ba}{\begin{align}}
\newcommand{\ea}{\end{align}}
\ifx\genfrac\sdflkaj\else\fi
\newcommand{\sfrac}[2]{{\textstyle\frac{#1}{#2}}}

\begin{document}

\begin{flushright}
\begingroup\footnotesize\ttfamily
	HU-EP-22/24-RTG
\endgroup
\end{flushright}

\vspace{15mm}

\begin{center}
{\LARGE\bfseries 
	All Things Retarded: Radiation-Reaction in Worldline Quantum Field Theory
\par}

\vspace{15mm}

\begingroup\scshape\large 
	Gustav Uhre Jakobsen,${}^{1,2,3}$ Gustav~Mogull,${}^{1,2,3}$
	Jan~Plefka${}^{1,3}$ and Benjamin Sauer${}^{1}$  
\endgroup
\vspace{5mm}
					
\textit{${}^{1}$Institut f\"ur Physik und IRIS Adlershof, Humboldt-Universit\"at zu Berlin, \phantom{${}^2$}\\
  Zum Gro{\ss}en Windkanal 2, D-12489 Berlin, Germany} \\[0.25cm]
\textit{${}^{2}$Max-Planck-Institut f\"ur Gravitationsphysik
(Albert-Einstein-Institut)\\
Am M\"uhlenberg 1, D-14476 Potsdam, Germany } \\[0.25cm]
\textit{${}^{3}$Kavli Institute for Theoretical Physics, University of California,\\
Santa Barbara, CA 93106, USA } \\[0.25cm]

\bigskip
  
\texttt{\small\{gustav.uhre.jakobsen@physik.hu-berlin.de, gustav.mogull@aei.mpg.de, 
jan.plefka@hu-berlin.de, benjamin.sauer@hu-berlin.de\}}

\vspace{10mm}

\textbf{Abstract}\vspace{5mm}\par
\begin{minipage}{14.7cm}
We exhibit an initial-value formulation of the worldline quantum field theory (WQFT) approach to the classical 
two-body problem in general relativity. We show that the Schwinger-Keldysh (in-in) formalism leads to purely
retarded propagators in the evaluation of observables in the WQFT. Integration technology for retarded master integrals
is introduced at third post-Minkowskian (3PM) order. As an application we compute the complete radiation-reacted impulse and 
radiated four momentum
for the scattering of two non-spinning neutron stars including tidal effects at 3PM order, as well as the leading (2PM) 
far-field gravitational waveform. 
\end{minipage}\par

\end{center}
\setcounter{page}{0}
\thispagestyle{empty}
\newpage
 
\tableofcontents

\section{Introduction}

The gravitational two-body problem lies at the origin of physics as a scientific discipline 
itself \cite{Newton:1687eqk}. The Newtonian case is well known to be integrable and supports bound or unbound orbits depending on the conserved energy of the 
system. In general relativity (GR) the same is true in principle, yet the problem is more complicated due to the loss of energy and (angular) momentum through gravitational radiation. In bound systems the continuous emission of gravitational radiation leads to
an ever-shrinking inspiral followed by a merger of the two massive objects ---
black holes, neutron stars or stars. With the gravitational waves emitted by such violent encounters 
now routinely being detected by
the LIGO-Virgo-KAGRA  observatories \cite{LIGOScientific:2021djp},
a new era of gravitational wave (GW) astronomy has begun.
This calls for high precision predictions of the emitted waveforms 
in binary encounters depending on the source's parameters.

Due to the non-linearity of Einstein's
equations and radiative effects exact results are out of reach,
and one is forced to use approximate perturbative or numerical approaches.
In bound systems the inspiral phase typically runs over many cycles,
and so perturbative methods are preferable:
one often uses a \emph{post-Newtonian} (PN) expansion in Newton's constant
$G$ and relative velocity $v$ (with $\frac{Gm}{r} \sim v^{2}\ll 1$),
which works well for near-circular orbits
\cite{Blanchet:2013haa,Schafer:2018kuf,Futamase:2007zz,Pati:2000vt}
\cite{Goldberger:2004jt,Goldberger:2006bd,Goldberger:2009qd,Kol:2007bc,Galley:2009px,Foffa:2019hrb,Blumlein:2020pyo,Bini:2020wpo,Bini:2020hmy}.
On the other hand, for unbound systems ---
i.e.~gravitational scattering events, in which the two massive bodies fly past each other ---
the \emph{post-Minkowskian} (PM) weak-field expansion in powers of Newton's constant $G$ is more natural
\cite{Westpfahl:1985tsl,Bel:1981be,Ledvinka:2008tk}.
While the non-periodic gravitational Bremsstrahlung produced by such
scattering events appears to be out of reach of current-generation GW detectors,
it may be detectable in the third generation of GW detectors~\cite{Kocsis:2006hq,Mukherjee:2020hnm,Zevin:2018kzq,Gamba:2021gap,Khalil:2022ylj}.
The close relationship between bound and unbound two-body systems also allows one to recycle
scattering observables into useful input for the bound problem~\cite{Kalin:2019rwq,Kalin:2019inp,Cho:2021arx}.
This is especially useful for highly elliptic bound orbits where the relative velocity
$v$ is large at the point of closest approach~\cite{Khalil:2022ylj}.

In an effective field theory (EFT) setting,
well-separated black holes (BHs) or neutron stars (NSs) can be idealized
as two massive point particles moving on trajectories $x^{\mu}_{i}(\tau)$ coupled to gravity.
Finite-size and tidal effects are included by coupling an infinite series of higher-dimensional operators to the 
particle's worldline theory with free Wilson coefficients (Love numbers) $c^{(i)}_{E^{2}/B^{2}}$.
Using the Polyakov formulation with einbein $e$ the one-dimensional worldline theory reads
 \be
 S_{\text{pm}}^{(i)}=- m_{i}\!\int\!\d\tau_{i} \bigg[ \frac{1}{2e}g_{\mu\nu}(x_{i}) {\dot x}_{i}^{\mu}(\tau_{i}) {\dot x}_{i}^{\nu}(\tau_{i})
+\frac{e}{2} 
+\frac{c_{E^2}^{(i)}}{e^{3}} E_{\mu \nu}^{(i)} E^{(i) \mu \nu}
+ \frac{c_{B^2}^{(i)}}{e^{3}} B_{\mu \nu}^{(i)} B^{(i) \mu \nu}
 +\ldots \bigg],\nn
 \ee
where we have included the first layer of tidal effects which are quadratic
in the Riemann tensor and quartic in $\dot x_{i}^{\mu}$ ---
see \eqn{eq:action} for a definition.
The complete gravitational field theory consists of these two point-mass worldline actions $S_{\text{pm}}^{(i)}$ ($i=1,2$)
coupled to the four-dimensional bulk Einstein-Hilbert action $S_{\text{EH}}$.

Expanding the metric around a flat Minkowski vacuum,
$g_{\mu\nu}(x)=\eta_{\mu\nu}+\kappa\,h_{\mu\nu}(x)$ with $\kappa=\sqrt{32\pi G}$,
one solves the resulting equations of motion for the trajectories $x^{\mu}_{i}(\tau_{i})$
and the graviton waveform $h_{\mu\nu}(x)$ perturbatively in $G$.
In a scattering scenario the initial boundary conditions are straight-line
trajectories of the two bodies, $x^{\mu}_{i}=b_{i}^{\mu}+ v^{\mu}_{i}\tau_{i}$,
with no incoming radiation $h_{\mu\nu}=0$.
Provided the impact parameter is large compared with the intrinsic size of the massive bodies,
i.e.~$|\mathbf{b}|=|\mathbf{b_{1}}-\mathbf{b_{2}}|\gg Gm_{i}$,
the weak-field expansion and point-particle approximations are valid.
Key observables in the gravitational scattering problem are: (i) the waveform $h_{\mu\nu}=\frac{f(t,\vec{x})}{r} +
\ldots$  in the far zone $r\gg|\mathbf{b}|$, (ii) the impulse or momentum deflection 
$\Delta p_{i}^{\mu}:= m_{i}[\xdot^{\mu}_{i}(\tau=+\infty)-\xdot^{\mu}_{i}(\tau=-\infty)]$ of the $i$'th body (which also contains the scattering angle $\theta$),
and for spinning bodies (iii) the ``spin kick'' $\Delta a^{\mu}_{i}$, 
i.e.~the change of the $i$'th body's spin vector.
These quantities also encode the total energy, momentum and angular momentum 
dissipated via gravitational radiation (Bremsstrahlung). A special feature of gravity is that the 
field $h_{\mu\nu}$ does not relax to zero after the scattering event;
instead, it takes on a new finite value --- the memory effect.
This traditional GR approach of solving the equations of motion perturbatively has been used to establish the
leading-order (LO) gravitational waveform \cite{1975ApJ...200..245T,1977ApJ...215..624C,Kovacs:1977uw,Kovacs:1978eu}, 
the deflection angle up to 2PM \cite{Bel:1981be,Westpfahl:1985tsl,Ledvinka:2008tk}
and the explicit trajectory up to 2PM order \cite{Gralla:2021qaf} ---
see also recent work on the electromagnetic analogue at 3PM order \cite{Saketh:2021sri}
.

Quantum field theory (QFT) techniques employing a classical limit of the perturbatively evaluated
path integral have also proven themselves highly effective.
Initially introduced for the computation of an effective worldline action in the PN expansion
\cite{Goldberger:2004jt,Goldberger:2006bd,Goldberger:2009qd,Kol:2007bc,Galley:2009px,Foffa:2019hrb,Blumlein:2020pyo,Bini:2020wpo,Bini:2020hmy} --- see \cite{Porto:2016pyg,Levi:2018nxp} for reviews --- 
these have now been successfully extended to the scattering problem in a PM expansion \cite{Kalin:2020mvi}.
Conservative observables including spin effects have been computed at 2PM order~\cite{Cho:2021mqw},
at 3PM order including tidal effects \cite{Kalin:2020fhe,Kalin:2020lmz} and at 4PM order \cite{Dlapa:2021npj,Dlapa:2021vgp}.
In this worldline EFT approach graviton fluctuations $h_{\mu\nu}$ are integrated out of the path integral
of the above action, $S_{\rm EH} + \sum_{i=1}^{2} S_{\text{pm}}^{(i)}$,
yielding an effective action $\Gamma[x_{i}]$ describing only the the massive bodies' positions.
The equations of motion for $\Gamma[x_{i}]$ are then perturbatively solved in $G$,
yielding the desired deflection observables.
Radiative observables at 3PM order have also been derived from the LO waveform
\cite{Mougiakakos:2021ckm,Riva:2021vnj,Mougiakakos:2022sic,Riva:2022fru}.
However, a systematic inclusion of radiation-reaction effects at higher PM orders requires the introduction
of a Schwinger-Keldysh ``in-in'' path-integral formalism~\cite{Schwinger:1960qe,Keldysh:1964ud},
as the classical problem to be solved is an initial-value problem. 

In a series of papers involving three of the present authors we have developed
an efficient extension to the worldline EFT approach known as \emph{worldline quantum field theory} (WQFT)
\cite{Mogull:2020sak,Jakobsen:2021smu,Jakobsen:2021lvp,Jakobsen:2021zvh,Jakobsen:2022fcj}.
WQFT implements the philosophy of the modern scattering amplitudes program
\cite{Dixon:1996wi,Elvang:2013cua,Henn:2014yza,Bern:2019prr,Travaglini:2022uwo}
by directly computing \emph{on-shell observables}: translating this to the worldline EFT calls for integrating
out fluctuations of the worldline fields $x^{\mu}_{i}$. All classical observables,
including the impulse, spin kick and waveform
are derived from \emph{tree-level one-point functions} in WQFT. Indeed, the one-point functions directly
yield the desired solutions of the classical equations of motion of the system once the correct boundary conditions are implemented. 
In essence, the WQFT formalism provides an
efficient diagrammatic framework for solving the classical equations of motion of the spinning gravitational two-body system
perturbatively.
Moreover, in WQFT the spin degrees of freedom are captured by introducing anti-commuting worldline fields that display an extended
supersymmetry \cite{Jakobsen:2021zvh}. Using this state-of-the-art results for the waveform with \cite{Jakobsen:2021lvp} and without spin \cite{Jakobsen:2021smu} as well as the 3PM impulse and spin kick up to quadratic order in spin \cite{Jakobsen:2022fcj}
have been established. Further works have addressed  light bending \cite{Bastianelli:2021nbs}, the double copy \cite{Shi:2021qsb}
and applications to the eikonal phase~\cite{Wang:2022ntx}.

In this paper we describe how the WQFT systematically incorporates radiation-reaction
effects by implementing the Schwinger-Keldysh ``in-in'' path-integral formalism.
The outcome is to be expected, and was already implied in our previous works:
when computing tree-level one-point functions using the WQFT in-in formalism
one uses exactly the same Feynman rules as the WQFT in-out formalism,
but with \emph{retarded} propagators pointing towards the outgoing line.
This follows from a simple diagrammatic argument explained in \sec{sec:inin}.
We therefore extend modern Feynman loop integration technology to handle retarded propagators,
initiating this program in \sec{sec:retInt} for the specific case of a 3PM integral family.
In \sec{sec:tidal} we also apply the complete WQFT formalism to compute
non-spinning radiative observables (impulse, scattering angle, waveform) at 3PM 
\emph{including tidal effects}.
By contrast, in the traditional worldline EFT approach one goes through the
intermediate stage of an effective action and a laborious separation
into advanced and retarded propagators in a doubled field approach is required.

A complementary and rapidly developing alternative approach
to the classical gravitational scattering problem is based on scattering amplitudes
\cite{Neill:2013wsa,Luna:2017dtq,Bjerrum-Bohr:2013bxa,Bjerrum-Bohr:2018xdl,Bern:2019nnu,Bern:2019crd,Cheung:2020gyp,Bjerrum-Bohr:2021din,DiVecchia:2021bdo}. 
Here, massive scalar fields are the avatars of spinless BHs or NSs,
and one obtains scattering data from $2\to 2$ amplitudes taken in the classical $\hbar\to0$ limit.
The innovations of the scattering amplitude program for constructing tree- and loop-level
amplitudes in perturbatively quantized GR have
recently enabled quick progress to higher PM orders.
The conservative effective potential has been established at 3PM~\cite{Bern:2019nnu,Bern:2019crd,Cheung:2020gyp} 
and recently at 4PM order \cite{Bern:2021dqo,Bern:2021yeh} in the non-spinning case.
The scattering of higher-spin fields may also be used to emulate the compact objects' spin
\cite{Guevara:2018wpp,Guevara:2019fsj,Aoude:2021oqj,Chen:2021kxt,Bern:2022kto,Aoude:2022trd,FebresCordero:2022jts}. 
However, here the inclusion of radiation-reaction effects here poses a challenge
\cite{Damour:2020tta, DiVecchia:2020ymx,Herrmann:2021tct,Bjerrum-Bohr:2021din,DiVecchia:2021ndb},
and so far the complete impulse $\Delta p_i^\mu$ has only been produced in the non-spinning case.
Radiative corrections to the 3PM scattering angle $\theta$ have, nevertheless,
been produced to arbitrary orders in spin~\cite{Alessio:2022kwv}.
The KMOC formalism addresses this problem by introducing an in-in style formalism in this context
\cite{Kosower:2018adc,Maybee:2019jus,Cristofoli:2021vyo,Cristofoli:2021jas},
but matching to the worldline EFT Wilson coefficients remains subtle.
Recently, amplitude techniques were also succesfully applied to the radiation-reaction
problem (tail effect) in the PN approach \cite{Edison:2022cdu}.

Another recently developed approach uses heavy-particle effective theories (HEFTs)
to describe the massive bodies \cite{Damgaard:2019lfh,Brandhuber:2021eyq}.
State-of-the-art results here include the spinless 3PM scattering angle \cite{Brandhuber:2021eyq},
spinning results with tidal effects
\cite{Aoude:2020ygw} and at 2PM with arbitrary spin orders of one particle \cite{Aoude:2022thd}. 

\bigskip

\noindent \emph{Note added:} While finalizing  this paper we became
aware of the work \cite{Kalin:2022hph} which overlaps with some of our results.

\section{In-in formalism for WQFT}
\label{sec:inin}

To simplify the discussion we consider as a toy model a
free massless scalar field theory coupled to an external physical source $Q(x)$
\be
\label{Sq}
S[\phi;Q] = \frac{1}{2} \int\!\d^{4}x\, \partial_{\mu}\phi\partial^{\mu}\phi 
+ S_{\text{int}}[\phi;Q]\,.
\ee
The WQFT analogues of $\phi$ are the fluctuating fields on the worldlines
$z_i^{\mu}(\tau)$ and the graviton $h_{\mu\nu}(x)$ in the bulk, while the source $Q$ emerges from the
background configurations $b_i^{\mu} +v_i^{\mu}\tau$ on the worldlines that
couple non-linearly to $z_i^{\mu}$ and $h_{\mu\nu}$.
However, let us for now consider an arbitrary interaction term $S_{\text{int}}[Q,\phi]$
that depends on an external source $Q(x)$ in an unspecified manner.
We are interested in finding solutions to the classical equations of motion
\be
\partial^2 \phi_{\text{class}}(x) = \left.\frac{\delta S_{\text{int}}[Q,\phi]}{\delta\phi(x)}\right|_{\phi=\phi_{\text{class}}}
\ee
in a perturbative expansion of $S_{\text{int}}[Q,\phi]=\sum_{n} g^{n}S_{\text{int}}^{[n]}[Q,\phi]$ with a coupling $g$. The QFT path integral provides
an efficient diagrammatic way to do this. Coupling the theory to a fiducial source $J(x)$
we may compute the generating functional $W[J]$ in a perturbative loop expansion via
\be
e^{\frac{i}{\hbar} W[J]} =
\int\D\phi\, \exp \Bigl \{\frac{i}{\hbar}\Big(S[\phi;Q] + \int\!\d^{4}x J(x) \phi(x)\Big)\Bigr \} \, .
\ee
The one-point function follows as 
\be
\label{inoutexpval}
\vev{\hat\phi(x)} = \left.\frac{\delta W[J]}{\delta J(x)}\right|_{J=0}\, ,
\ee
while the effective action $\Gamma[\phi]$ is related to $W[J]$ via a Legendre
transform
\be
\Gamma[\phi] = \frac{i}{\hbar}\int\!\d^{4}x\,J(x)\phi(x) - W[J]\, .
\ee
It is a central QFT result that the equations of motion of the (full quantum)
effective action $\Gamma[\phi]$ are solved by the one-point function of \eqref{inoutexpval}:
\be
\left.\frac{\delta \Gamma[\phi]}{\delta \phi(x)}\right|_{\phi(x)= \vev{\hat\phi(x)}}=0 \, ,
\ee
in the absence of the fiducial sources $J(x)$. Importantly, the tree-level
approximation to the effective action \emph{is} the classical action, the loop contributions
being suppressed by powers of $\hbar$:
\be
\Gamma(\phi) = 
\frac{1}{2} \int\!\d^{4}x\, \partial_{\mu}\phi\partial^{\mu}\phi 
+ S_{\text{int}}[\phi;Q]
 + \cO(\hbar)\, .
\ee
Hence, a constructive way to establish the classical solution $\phi_{\text{class}}(x)$ is
to evaluate the one-point function $\vev{\phi(x)}$ at tree level:
\be
\phi_{\text{class}}(x) = \vev{\hat\phi(x)}\Bigr |_{\text{tree-level}}\,.
\ee
This simple QFT relation is the basis for our use of QFT to efficiently solve 
classical physics problems. Note, that the tree-level one-point function may 
still have a non-trivial perturbative expansion in the couplings $g$ of 
$S_{\text{int}}[\phi;Q]$ in the presence of physical sources $Q$ ---
as is the case in WQFT.
However, the one-point function $\langle \hat \phi (x) \rangle$ in the standard ``in-out''
path integral formalism is the solution to a boundary value problem;
not the initial-value problem  that is required in a classical setting.
This is because the in-out one-point function $\langle \hat \phi (x) \rangle$
--- as opposed to its notation ---
is not a true expectation value of the field operator $\hat \phi(x)$. 
Rather, it is the matrix element of $\hat \phi(x)$ between the in-vacuum
and the out-vacuum state to which the in-vacuum has evolved,
and these two states need not be identical.
Realization of the true expectation value of the field operator
in the in-vacuum state as a one-point function is achieved using the Schwinger-Keldysh ``in-in''
path integral formalism~\cite{Schwinger:1960qe,Keldysh:1964ud}, which we shall now review ---
following closely the pedagogical expositions found in refs.~\cite{Jordan:1986ug,Weinberg:2005vy,Galley:2009px}.

\subsection{One-point functions in the in-out formalism} 

Let us first review a few important details regarding the in-out formalism.
In the interaction picture the state of the system is time dependent. Let $|\Psi\rangle$
be the state of the system in the infinite past ($T=-\infty$),
when the interaction and Heisenberg picture are taken to coincide.
The time evolution of the state vector in the interaction picture
is governed by the interaction Hamiltonian $\hat{H}_{\rm int}$, 
$|\Psi(t)\rangle = \hat U(t,-\infty)|\Psi\rangle$,
with the time-evolution operator
\be
\label{UPIdef}
\hat U(T',T) = \mathcal{T} \exp \Bigr [ \frac{i}{\hbar} \int_{T}^{T'}\!\!\d t\! \int \!\d^{3}x 
\hat H_{\text{int}}[Q(t,\mathbf{x}), \hat \phi_{I}(t,\mathbf{x})] \Bigr ]\,.
\ee
$\mathcal{T}$ denotes time ordering and
$\hat \phi_{I}(t,\mathbf{x})$ is the field operator in the interaction picture (see e.g.~\rcite{Schwartz:2014sze} chapter 7).
It is related to the field in the Heisenberg picture $\hat \phi_{H}(t,\mathbf{x})$ via
\be
\hat \phi_{I}(t,\mathbf{x})= \hat U(t,-\infty) \hat \phi_{H}(t,\mathbf{x})  \hat U(-\infty,t)\, .
\ee
We now add an auxiliary external linear source $J(x)$ to the system
\be
\hat U_{J}(T',T) = \mathcal{T} \exp \Bigr [ \frac{i}{\hbar} \int_{T}^{T'}\!\!\d t\! \int \!\d^{3}x \left (
\hat H_{\text{int}}[Q,\hat\phi_{I}] + J(x) \hat \phi_{I}(x) \right )\Bigr ] \,.
\ee
The vacuum transition amplitude in the presence of sources has the in-out path-integral representation
\be
\label{Wdef}
\langle 0| \hat U_{J}(\infty,-\infty) | 0\rangle = 
\int\D\phi\, \exp \Bigl \{\frac{i}{\hbar}\Big(S[\phi] + \int\!\d^{4}x J(x) \phi(x)\Big)\Bigr \} =: e^{\frac{i}{\hbar} W[J]}\, ,
\ee
Here, $|0\rangle$ represents the vacuum state in the infinite past when the interactions are assumed to turn off.
The in-out one-point function of the Heisenberg field is
\begin{align}
\begin{aligned}
\langle  \hat \phi_{H} (t,\mathbf{x}) \rangle_{\text{in-out}}  
:= \frac{\delta W}{\delta J(t,\mathbf{x})}\Bigr |_{J=0}
&= \langle 0| \hat U(\infty,t) \hat \phi_{I}(t, \mathbf{x})
\hat U(t,-\infty) | 0\rangle \\
& =  \langle 0| \hat U(\infty,-\infty) \hat \phi_{H}(t, \mathbf{x})| 0\rangle\, .
\label{inoutres}
\end{aligned}
\end{align}
This simple example illustrates that we have \emph{not} evaluated the vacuum expectation value of the
Heisenberg field operator, but rather the matrix element between the in-vacuum $|0\rangle$ 
and the state  $\hat U(-\infty,\infty)|0\rangle$, i.e.~the out-vacuum.
Hence, the tree-level one-point function $\langle  \hat \phi_{H} (t,\mathbf{x}) \rangle_{\text{in-out}}  $ cannot be identified 
with the classical field configuration $\phi_{\text{class}}(x)$, which should be given by a vacuum expectation value
of the field operator in the Heisenberg picture, $\langle 0 |  \hat \phi_{H} (t,\mathbf{x}) |0\rangle$.

This is easily verified by specializing the interacting piece of the action \eqref{Sq}
to a linear term $S_{\text{int}}[Q,\phi]= \int\d^{4}x\,Q(x) \phi(x)$.
Then the path integral of \eqn{inoutres} may be performed exactly, yielding
\be
\langle  \hat \phi_{H} (x) \rangle_{\text{in-out}} = \int\!\d^{4}y \, D_{F}(x-y)\, Q(y)
\ee
with the Feynman propagator (or Green's function) $D_{F}(x)$.
While this solves the classical equations of motion
$\partial^{2}\phi(x)= Q(x)$  emerging from \eqref{Sq} (as it should), 
it is not the causal solution one usually seeks for in classical physics:
for this, we instead need a \emph{retarded} propagator $D_{\rm ret}(x-y)$.

\subsection{One-point functions in the in-in formalism} 

The Schwinger-Keldysh in-in formalism is designed to yield a one-point function that \emph{is} a true expectation
value of the field in the Heisenberg picture, i.e.~we demand that
\begin{align}
\label{truevev}
\langle  \hat \phi_{H} (t,\mathbf{x}) \rangle_{\text{in-in}}  
:= \langle 0 | \hat \phi_{H} (t,\mathbf{x}) |0\rangle
= \langle 0| \hat U(-\infty,t) \hat \phi_{I}(t, \mathbf{x})
\hat U(t,-\infty) | 0\rangle \, .
\end{align}
To achieve this, we introduce a generating functional $W[J_{1},J_{2}]$ depending on two auxiliary external 
linear sources that generalizes \eqn{Wdef}:
\be
\label{WJJ}
e^{\frac{i}{\hbar} W[J_{1},J_{2}]}= \langle 0| \hat U_{J_{2}}(-\infty,\infty)\, \hat U_{J_{1}}(\infty,-\infty)  | 0\rangle \, .
\ee
Note that we maintain only one external source field $Q$.
The in-vacuum is evolved to the infinite future in the presence of sources $J_{1}$,
and $Q$ is then evolved backwards in time to the infinite past in the presence of $J_{2}$ and $Q$.
One may then express the expectation value \eqn{truevev} in two equivalent ways:
\begin{align}
\begin{aligned}
\frac{\delta W[J_{1},J_{2}]}{\delta J_{1}(t,\mathbf{x})}\Bigr |_{J_{i}=0}
&=\langle 0| \hat U(-\infty,\infty) \hat U(\infty,t) \hat \phi_{I}(t, \mathbf{x})
\hat U(t,-\infty) | 0\rangle \\
=
\frac{\delta W[J_{1},J_{2}]}{\delta J_{2}(t,\mathbf{x})}\Bigr |_{J_{i}=0} &=
\langle 0| \hat U(-\infty,t) \hat \phi_{I}(t, \mathbf{x})
\hat U(t,\infty) \hat U(\infty,-\infty) | 0\rangle \\
&=\langle 0| \hat U(-\infty,t) \hat \phi_{I}(t, \mathbf{x})
\hat U(t,-\infty)  | 0\rangle
=\langle  \hat \phi_{H} (t,\mathbf{x}) \rangle_{\text{in-in}} \, .
\end{aligned}
\end{align}

The generating functional $W[J_{1},J_{2}]$ has a path integral 
representation upon doubling the fields. It is established by inserting the unit operator $\mathbf{1}=\sum_{|\psi\rangle} |\psi\rangle \, \langle \psi|$
into \eqn{WJJ} using \eqn{Wdef}:
\begin{align}
e^{\frac{i}{\hbar} W[J_{1},J_{2}]} &= \sum_{|\psi\rangle} \langle 0| \hat U_{J_{2}}(-\infty,\infty)\,  |\psi\rangle \, \langle \psi|\hat U_{J_{1}}(\infty,-\infty)  | 0\rangle  \\
&=\int\!\D[\phi_{1},\phi_{2}] \exp \Bigl \{\frac{i}{\hbar} \Bigl [
S[\phi_{1}] - S[\phi_{2}] + \int\!\d^{4}x \Bigl( J_{1}(x) \phi_{1}(x)- J_{2}(x) \phi_{2}(x) \Bigr) \Bigr]\nonumber
\Bigr \} \,.
\end{align}
The $\phi_{1}$ field propagates forward in time, the $\phi_{2}$ field backwards.
Importantly, in the path integral above the two fields are linked via the boundary condition at future infinity
$
\phi_{1}(t=+\infty,\mathbf{x})=\phi_{2}(t=+\infty,\mathbf{x})
$,
while at past infinity both fields vanish $
\phi_{1}(t=-\infty,\mathbf{x})=\phi_{2}(t=-\infty,\mathbf{x})=0
$. This is a consequence of the sum over all states $|\psi\rangle$ in the first line above.

In order to set up the in-in perturbation theory we need to establish the propagator structure
in the free theory. We encounter
a $2\times 2$ propagator matrix $D^{AB}(x,y)$ related to the doubled fields. It is 
most easily derived from the free-field generating functional $W_{0}[J_{1},J_{2}]$ in the operator
representation \eqref{WJJ}~\cite{Jordan:1986ug,Weinberg:2005vy,Galley:2009px} ---
see \App{appA} for a derivation ($A,B=1,2$):
\begin{align}
\label{12matrix}
\langle \phi_{A}(x)\, \phi_{B}(y) \rangle
= \left ( \begin{matrix} \vev{0|\mathcal{T} \phi(x) \phi(y)|0} & 
\vev{0|\phi(y) \phi(x)|0} \\ \vev{0|\phi(x) \phi(y)|0}  & \vev{0|\mathcal{T}^{\ast} \phi(x) \phi(y)|0} \end{matrix} \right )
=
\left ( \begin{matrix} D_{F}(x,y) & 
D_{-}(x,y)\\ D_{+}(x,y)  & D_{D}(x,y) \end{matrix} \right )\!,
\end{align}
with the Feynman $D_{F}(x,y)$ and Dyson (or anti-time-ordered) $D_{D}(x,y)$ Green's 
function appearing on the diagonal. The off-diagonal entries are known as the
Wightman Green's functions, $D_{+}(x,y) = \vev{0|\phi(x) \phi(y)|0}=D_{-}(y,x)$.
Note that here $|0\rangle$ is the Fock vacuum of the \emph{free} theory which is stationary
under time evolution.

We find it convenient to adopt the Keldysh basis \cite{Keldysh:1964ud}
by introducing the sum and difference of the two fields and sources:
\begin{align}
\begin{aligned}
\phi_{-}&= \phi_{1}- \phi_{2}\, , &\qquad\qquad
\phi_{+}&= \sfrac 12 (\phi_{1}+ \phi_{2})\, ,\\
J_{-}&= J_{1}- J_{2}\, , &\qquad\qquad
J_{+}&= \sfrac 12 (J_{1}+ J_{2})\, .
\end{aligned}
\end{align}
The propagator matrix in the Keldysh basis then becomes ($a,b=+,-$) \cite{Galley:2009px}
\be
\label{Keldyshprop}
\langle \phi_{a}(x)\, \phi_{b}(y) \rangle
=  \left ( \begin{matrix} \sfrac 1 2 D_{H}(x,y) & D_{\text{ret}}(x,y) \\ - D_{\text{adv}}(x,y) & 
0 \end{matrix} \right )\,,
\ee
with the advanced $D_{\rm adv}(x,y)$ and retarded $D_{\rm ret}(x,y)$ Greens's functions as well as the symmetric Hadamard 
function $D_{H}(x,y)=\vev{0|\{ \phi(x), \phi(y)\}|0}$.
In the Keldysh basis the generating functional of the interacting theory takes the form
\begin{align}
\label{ininPI}
&e^{\frac{i}{\hbar} W[J_+,J_{-}]}\\
&=
\int \D[\phi_{+},\phi_{-}] \exp \Bigl \{\frac{i}{\hbar} \Bigl [
S[\phi_{+}+\sfrac1 2 \phi_{-}] - S[\phi_{+}-\sfrac1 2 \phi_{-}]
+ \int\!\d^{4}x \Bigl( J_{+}\phi_{-}+ J_{-}\phi_{+} \Bigr) \Bigr]
\Bigr \}  \, .\nn
\end{align}
The true vacuum expectation value of the Heisenberg field operator
may now be computed 
from the one-point function of $\phi_{+}$ at vanishing sources $J_{\pm}$ by
\begin{align}
\label{vevinin}
\langle  \hat \phi_{H} (t,\mathbf{x}) \rangle_{\text{in-in}} &= 
\langle  \hat \phi_{H\, +} (t,\mathbf{x}) \rangle_{\text{in-in}}\Bigr |_{J_{\pm}=0}
=\frac{\delta W[J_+,J_{-}]}{\delta J_{-}}\Bigr |_{J_{\pm}=0}\\
&=
\int \D[\phi_{+},\phi_{-}]  \phi_{+}(t,\mathbf{x})\, \exp \Bigl \{\frac{i}{\hbar} \Bigl [
S[\phi_{+}+\sfrac1 2 \phi_{-}] - S[\phi_{+}-\sfrac1 2 \phi_{-}]  \Bigr]
\Bigr \}  \, , \nn
\end{align}
using $\vev{\phi_{1}}_\text{in-in}|_{J_{i}=0}=\vev{\phi_{2}}_\text{in-in}|_{J_{i}=0}$ in the Schwinger basis.
Importantly, the in-in effective action $\Gamma[\vev{\phi_{\pm}}]$ is obtained as the Legendre transform of the
generating functional
\be
\Gamma[\vev{\phi_{+}},\vev{\phi_{-}}] = W[J_+,J_{-}] - \int\!\d^{4}x\,\left(J_{-} \vev{\phi_{+}}+ J_{+} \vev{\phi_{-}}\right)\, .
\ee
Finally, at tree level the in-in effective action gives rise to the classical equations of motion that are solved by the expectation
value \eqref{vevinin}:
\be
\label{EffsolvesEOM}
0= \frac{\delta \Gamma}{\delta \vev{\phi_{-}}} \Bigr |_{\vev{\phi_{-}}=0, \, \vev{\phi_{+}}=\phi_{\text{class}},\,  J_{\pm}=0}\, .
\ee
Note that at tree level the $\vev{\phi_{+}\phi_{+}}$ component $D_{H}(x,y)$  of the
Keldysh propagator matrix \eqref{Keldyshprop} does not contribute ---
in momentum space it is  $\tilde D_{H}(k)= \dd(k^{2})$ and only has on-shell
support --- as we will show this in the next section.
Hence, the tree-level or 
classical physics result can only depend on advanced and retarded propagators.
This is the key relation to exploit for our purposes:
applying the in-in formalism to the computation of the one-point functions
$\vev{z_i^{\mu}}$ and $\vev{h_{\mu\nu}}$ of WQFT yields a PM perturbative, diagrammatic procedure to establish the solutions
$z_{\text{class}}(\tau)$ and $h^{\mu\nu}_{\text{class}}(t,\mathbf{x})$
to the equations of motion of the classical two-body scattering problem.

\subsection{In-in one-point functions in background field theory}

Evaluating the in-in path integral  can be quite laborious due to the need to
establish ``doubled'' Feynman rules from \eqn{ininPI} --- now adopting the Keldysh basis.
These involve vertices dressed with plus- and minus-labeled legs and novel symmetry factors
(the tensorial structure of the vertices remains inert with respect to the standard in-out Feynman rules).
As for the propagators, in momentum space the retarded and advanced propagators are
\begin{align}\label{eq:ininprop}
\begin{aligned}
    \widetilde{D}_{\text{ret}}(k)=
	\begin{tikzpicture}[baseline={(current bounding box.center)}]
	\coordinate (x) at (-.7,0);
	\coordinate (y) at (0.5,0);
	\draw [zParticle] (x) -- (y) ;
	\draw [fill] (x) circle (.06) node [above] {};
	\draw [fill] (y) circle (.06) node [above] {};
		\draw [fill] (x) circle (.06) node [below] {$-$};
	\draw [fill] (y) circle (.06) node [below] {$+$};
	\end{tikzpicture}&=\frac{-i}{(k^{0}+i0)^{2}-\mathbf{k}^2}\,, \\
	 \widetilde{D}_{\text{adv}}(k)=
	\begin{tikzpicture}[baseline={(current bounding box.center)}]
	\coordinate (x) at (-.7,0);
	\coordinate (y) at (0.5,0);
	\draw [zParticle] (y) -- (x) ;
	\draw [fill] (x) circle (.06) node [above] {};
	\draw [fill] (y) circle (.06) node [above] {};
		\draw [fill] (x) circle (.06) node [below] {$+$};
	\draw [fill] (y) circle (.06) node [below] {$-$};
	\end{tikzpicture}&=\frac{-i}{(k^{0}-i0)^{2}-\mathbf{k}^2} \, ,
\end{aligned}
\end{align}
with the direction of the arrow above indicating the direction of causality flow ---
$i0$ denotes a small \emph{positive} imaginary part.
However, for the special case of tree-level one-point functions in a background field theory
the in-in diagrammatics are very simple:
here we demonstrate that one needs only draw tree-level in-out graphs,
but with retarded propagators replacing Feynman propagators.

Sticking to the scalar field example,
the interaction vertices emerge from the in-in action~\eqref{ininPI}.
In perturbation theory the interacting part of the Lagrangian is polynomial in the fields.
Hence, expanding the Lagrangian~\eqref{ininPI} gives rise to a series
of vertices dressed with plus- and minus-labeled legs.
Expanding $S_{\text{in-in}}[\phi_{-},\phi_{+};Q]=S[\phi_{+}+\sfrac1 2 \phi_{-}] - S[\phi_{+}-\sfrac1 2 \phi_{-}]$
in the minus-labeled fields $\phi_{-}$ it is obvious that only vertices with an \emph{odd} number of minus legs will arise.
At linear order in minus fields the vertex structure is particularly simple:
\begin{align}
\label{ininscalarsimp}
S_{\text{in-in,int}}[\phi_{-},\phi_{+};Q]&=
\phi_{-} \left ( \frac{\delta S_{\text{int}}[\phi;Q]}{\delta \phi}\right )_{\phi\to\phi_{+}} 
+
\cO(\phi_{-}^{3})\, ,
\end{align}
and importantly $S_{\text{int}}[\phi;Q]$ is the original interacting piece of the (in-out) action~\eqref{Sq}.
Hence, for the single-minus leg vertices we find precisely the same Feynman rules (including symmetry factors) 
as in the in-out formalism, with the distinction
that these are extended by dressing each leg successively by a minus label and all others by a plus label.

Importantly, starting with the one-point vertices connected to the background Q the emitted graviton field
always carries a minus label:
\be
 \begin{tikzpicture}[baseline={(current bounding box.center)}]
    \coordinate (x) at (0,0);
    \node (k) at (0,-1.3) {$\phi_{-}$};
     \draw [zParticle] (x) -- (k);
    \draw (0,0) node[cross,red] {};
     \end{tikzpicture}
  \ee
where the cross symbolizes the background field $Q$.
On the contrary, the outgoing leg relevant for the one-point functions $\vev{\phi_{H}}$ is strictly
plus labeled, c.f.~\eqref{vevinin}. This leads to the following causality flow for 
a tree-level one-point function in the $Q$ background: 
\be
\vev{\phi(k)} =
 \begin{tikzpicture}[baseline={(current bounding box.center)}]
    \tikzstyle{blob}=[
    circle,
    minimum size =.5cm,
    draw=black,
    thick,
    fill=lightgray,
    text=black
]
  \coordinate (inA) at (-1,1);
  \coordinate (outA) at (6,1);
  \coordinate (inB) at (-1,-1);
  \coordinate (outB) at (6,-1);
  \coordinate (xA) at (0,1);
  \coordinate (yA) at (1,1);  
  \coordinate (zA) at (2,1);
  \coordinate (xB) at (0,-1);
  \coordinate (yB) at (1,-1);
  \coordinate (zB) at (2,-1);
  \coordinate (l1) at (2.5,0.75);
  \coordinate (l2) at (2.5,0.45);
  \coordinate (l3) at (2.5,0.15); 
  \coordinate (ll3) at (2.5,-0.15);
  \coordinate (ll2) at (2.5,-0.45);
  \coordinate (ll1) at (2.5,-0.75);
 \draw (xA) node[cross,red] {};
 \draw (yA) node[cross,red] {};
 \draw (zA) node[cross,red] {};
 \draw (xB) node[cross,red] {};
 \draw (yB) node[cross,red] {};
 \draw (zB) node[cross,red] {};
 \draw [dotted] (0,0.7) -- (0,-0.7);
   \draw [zParticle] (xA) -- (l3);
  \draw [zParticle] (yA) -- (l2);
  \draw [zParticle] (zA) -- (l1);
  \draw [zParticle] (xB) -- (ll3);
  \draw [zParticle] (yB) -- (ll2);
  \draw [zParticle] (zB) -- (ll1);
  \draw [zParticle] (4.5,0) -- (6,0) node [midway,below] {$k$} node [very near end, above] {$+$} node[very near start,above] {$-$};
  \fill [draw=black,
    thick,
    fill=lightgray] (2.5,-1) rectangle (4.5,1);
      \draw [fill] (4.5,0) circle (.08);
  \draw [fill] (l1) circle (.08);
  \draw [fill] (l2) circle (.08);
  \draw [fill] (l3) circle (.08);
  \draw [fill] (ll1) circle (.08);
  \draw [fill] (ll2) circle (.08);
  \draw [fill] (ll3) circle (.08);
  \end{tikzpicture}\,.
\ee
The grey shaded rectangle subsumes all tree-level interactions containing two-valent, three-valent and
higher-point vertices that connect the $n$ ingoing retarded propagators emerging from the background $Q$ to the
single outgoing leg.

The crucial insight is that at tree level only vertices with a 
\emph{single minus leg} can contribute to these one-point functions.
Any tree-level graph will have the
topological structure of a rooted tree, i.e.~take a form such as
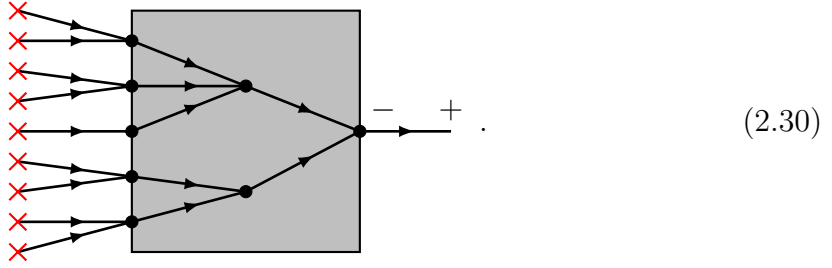
\be
\begin{tikzpicture}[baseline={(current bounding box.center)}]
\coordinate (L01) at (-1,1.2);
\coordinate (L02) at (-1,0.8);
\coordinate (L03) at (-1,0.4);
\coordinate (L04) at (-1,-0.0);
\coordinate (L05) at (-1,-0.4);
\coordinate (L06) at (-1,-0.8);
\coordinate (L07) at (-1,-1.2);
\coordinate (L08) at (-1,-1.6);
\coordinate (L09) at (-1,-2.0);
\coordinate (L11) at (0.5,0.8);
\coordinate (L12) at (0.5,0.2);
\coordinate (L125) at (0.5,-0.4);
\coordinate (L13) at (0.5,-1.0);
\coordinate (L14) at (0.5,-1.6);
\coordinate (L21) at (2,0.2);
\coordinate (L22) at (2,-1.2);
\coordinate (L31) at (3.5,-0.4);
\coordinate (L41) at (4.7,-0.4);
 \fill [draw=black,
    thick,
    fill=lightgray] (0.5,-2) rectangle (3.5,1.2);
\draw [zParticle] (L01)--(L11);
\draw [zParticle] (L02)--(L11);
\draw [zParticle] (L03)--(L12);
\draw [zParticle] (L04)--(L12);
\draw [zParticle] (L05)--(L125);
\draw [zParticle] (L06)--(L13);
\draw [zParticle] (L07)--(L13);
\draw [zParticle] (L08)--(L14);
\draw [zParticle] (L09)--(L14);
\draw [zParticle] (L125)--(L21) ;
\draw [zParticle] (L11)--(L21)  ;
\draw [zParticle] (L12)--(L21) ;
\draw [zParticle] (L13)--(L22) ;
\draw [zParticle] (L14)--(L22)  ;
\draw [zParticle] (L21)--(L31)  ;
\draw [zParticle] (L22)--(L31)  ;
\draw [zParticle] (L31)--(L41)  ;
\draw [fill] (L11) circle (.08);
\draw [fill] (L12) circle (.08);
\draw [fill] (L13) circle (.08);
\draw [fill] (L14) circle (.08);
\draw [fill] (L21) circle (.08);
\draw [fill] (L22) circle (.08);
\draw [fill] (L31) circle (.08);
\draw [fill] (L125) circle (.08);
\draw [fill] (L01) node[cross,red] {};
\draw [fill] (L02) node[cross,red] {};
\draw [fill] (L03) node[cross,red] {};
\draw [fill] (L04) node[cross,red] {};
\draw [fill] (L05) node[cross,red] {};
\draw [fill] (L06) node[cross,red] {};
\draw [fill] (L07) node[cross,red] {};
\draw [fill] (L08) node[cross,red] {};
\draw [fill] (L09) node[cross,red] {};
\draw (L31) node [above right] {$-$};
\draw (L41) node [above] {$+$};
\end{tikzpicture}\,.
\ee
From this structure it is immediately clear that inserting a vertex with three (or more) minus-labeled
(outgoing) legs in the shaded box inevitably leads to a loop-level graph,
as we just have a single outgoing leg. 
So we learn that \emph{at tree level only single-minus vertices may contribute}
to the one-point functions. 
Similarly, the $\vev{\phi_{+}\phi_{+}}$ Hadamard propagator cannot make an appearance, as conecting to
plus labeled (ingoing) legs of a vertex inevitably yields a loop diagram as every vertex has at least one minus labeled leg.
The consequence is that \emph{exclusively retarded propagators} appear in the 
computation if one assigns a momentum flow according to causality from $Q$ sources to
the outgoing operator line.
Therefore, in practical computations of one-point functions in a background field theory
one may effectively forget about the in-in formalism altogether.
One simply applies the usual (in-out) Feynman rules and uses retarded propagators everywhere,
with the direction of causality always pointing towards the outgoing line.

In hindsight, this fact is not surprising.
As we showed in \eqn{EffsolvesEOM},
it is a well-known fact that at tree level the path integral is dominated by
solutions to the classical equations of motion of the theory (the saddle-point approximation).
The sum of rooted tree diagrams is then simply a visual interpretation of a perturbative
expansion of the classical solution in powers of the coupling constant.
From a purely classical perspective,
using retarded propagators is then necessary to ensure fixing of boundary conditions at past infinity.

\subsection{In-in formalism for WQFT: observables}
\label{sec:ininObservables}

We now implement the in-in formalism for the WQFT.
In the non-spinning case there are two key observables to be computed which we typically evaluate in momentum space:
The impulse and the waveform. In the spinning case this is augmented by the spin-kick.
For the worldline coordinate we expand $x^{\mu}_{i}(\tau_{i})= b_{i}^{\mu} + v_{i}^{\mu}\tau_{i} + z^{\mu}_{i}(\tau_{i})$ with $i=1,2$ denoting
the two compact objects. In the in-in path integral we are led to double the fluctuation field $z^{\mu}_{i}\to \{ z^{\mu}_{i+},
z^{\mu}_{i-}\}$ but we do not double the background $b_{i}^{\mu} + v_{i}^{\mu}\tau_{i}$. 
This is justified, as in the end of the calculation the expectation value of the minus fields are set to zero, cp.~\eqref{EffsolvesEOM}.
We proceed analogously in the
spin case \cite{Jakobsen:2021lvp,Jakobsen:2021zvh}, where we have the background field expansion
$\psi^{\mu}_{i}= \Psi^{\mu}_{i} +\psi^{\prime\mu}_{i}$,
and double according to $\psi^{\prime\mu}_{i}\to \{ \psi^{\prime\mu}_{i+},\psi^{\prime\mu}_{i-}  \} $
in the Keldysh basis. Finally, the graviton field is doubled according to $h^{\mu\nu}\to\{ h^{\mu\nu}_{+}, h^{\mu\nu}_{-}\}$.

Let us now establish the Feynman rules for the in-in WQFT. 
As we argued in the last section if one is interested in tree-level one-point functions (as we are) we only need
the retarded propagators. We then have for the graviton
\begin{align}\label{eq:gravProp}
	\begin{tikzpicture}[baseline={(current bounding box.center)}]
	\coordinate (x) at (-.7,0);
	\coordinate (y) at (0.5,0);
	\draw [->,graviton] (x) -- (y) node [midway, below] {$k$} node [midway, above] {$\rightarrow$};
	\draw [fill] (x) circle (.06) node [above] {$\mu\nu$};
	\draw [fill] (y) circle (.06) node [above] {$\rho\sigma$};
		\draw [fill] (x) circle (.06) node [below] {$-$};
	\draw [fill] (y) circle (.06) node [below] {$+$};
	\end{tikzpicture}=i\frac{P_{\mu\nu;\rho\sigma}}{(k^{0}+i0)^{2}-\mathbf{k}^2}\,,
\end{align}
with $P_{\mu\nu;\rho\sigma}:=\eta_{\mu(\rho}\eta_{\sigma)\nu}-
\sfrac1{D-2}\eta_{\mu\nu}\eta_{\rho\sigma}$.
The retarded worldline  propagators for the fluctuations $z_i^\mu(\omega)$
and anti-commuting vectors $\psi_i^{\prime\mu}(\omega)$ are respectively
\begin{align}\label{eq:Propagators}
	\begin{tikzpicture}[baseline={(current bounding box.center)}]
	\coordinate (in) at (-0.6,0);
	\coordinate (out) at (1.4,0);
	\coordinate (x) at (-.2,0);
	\coordinate (y) at (1.0,0);
	\draw [zUndirected] (x) -- (y) node [midway, below] {$\omega$} node [midway, above] {$\rightarrow$};
	\draw [dotted] (in) -- (x);
	\draw [dotted] (y) -- (out);
	\draw [fill] (x) circle (.06) node [above] {$\mu$};
	\draw [fill] (y) circle (.06) node [above] {$\nu$};
	\draw [fill] (x) circle (.06) node [below] {$-$};
	\draw [fill] (y) circle (.06) node [below] {$+$};
	\end{tikzpicture}=-i\frac{\eta^{\mu\nu}}{m_i(\omega+i0)^2}\,, &&
		\begin{tikzpicture}[baseline={(current bounding box.center)}]
		\coordinate (in) at (-0.6,0);
		\coordinate (out) at (1.4,0);
		\coordinate (x) at (-.2,0);
		\coordinate (y) at (1.0,0);
		\draw [zParticle] (x) -- (y) node [midway, below] {$\omega$} node [midway, above] {$\rightarrow$};
		\draw [dotted] (in) -- (x);
		\draw [dotted] (y) -- (out);
		\draw [fill] (x) circle (.06) node [above] {$\mu$};
		\draw [fill] (y) circle (.06) node [above] {$\nu$};
		\draw [fill] (x) circle (.06) node [below] {$-$};
		\draw [fill] (y) circle (.06) node [below] {$+$};
		\end{tikzpicture}=-i\frac{\eta^{\mu\nu}}{m_i(\omega+i0)}\, .
\end{align}
Note that now the direction of the arrow above the propagators indicates the causality flow.
The retarded propagators were already used in \cite{Mogull:2020sak,Jakobsen:2021smu,Jakobsen:2021lvp,Jakobsen:2021zvh,Jakobsen:2022fcj}.

For the in-in WQFT vertices at linear order in minus fields the vertex structure is particularly simple ---
generalizing the scalar field discussion of \eqn{ininscalarsimp}:
\begin{align}
S^{\text{WQFT}}_{\text{in-in, int}}\Bigr |_{\text{lin $-$}}&=
h_{-}^{\mu\nu}\left ( \frac{\delta S^{\text{WQFT}}_{\text{int}}[h,z,\psi^\prime]}{\delta h^{\mu\nu}}\right )_{+}
+ \sum_{i=1}^{2}
z_{i-}^{\mu}\left ( \frac{\delta S^{\text{WQFT}}_{\text{int}}[h,z,\psi^\prime]}{\delta z^{\mu}_{i}}\right )_{+}\nn\\ &\quad
+ \sum_{i=1}^{2}
\psi^{\prime\mu}_{i-}\left ( \frac{\delta S^{\text{WQFT}}_{\text{int}}[h,z,\psi^\prime]}{\delta \psi^{\prime\mu}_{i}}\right )_{+}\, .
\end{align}
We find precisely the same Feynman rules (including symmetry factors) as in the in-out formalism, with the distinction
that these are extended by dressing each leg successively by a minus label and all others by a plus label.
Importantly, starting with the one-point functions connected to the background trajectories the connecting graviton field
always carries a minus label:
\be
 \begin{tikzpicture}[baseline={(current bounding box.center)}]
    \coordinate (in) at (-1,0);
    \coordinate (out) at (1,0);
    \coordinate (x) at (0,0);
    \node (k) at (0,-1.3) {$h_{\mu\nu}(k)$};
    \draw [dotted] (in) -- (x);
    \draw [dotted] (x) -- (out);
    \draw [graviton] (x) -- (k);
    \draw [fill] (x) circle (.08) node [above] {$-$};
  \end{tikzpicture}\,,
  \ee
while the tensorial structure remains as before~\cite{Mogull:2020sak,Jakobsen:2021smu,Jakobsen:2021lvp,Jakobsen:2021zvh,Jakobsen:2022fcj}. 
Again, the outgoing leg relevant for the one-point functions
$\vev{h_{+}^{\mu\nu}(k)}$, $\vev{z_{i+}^{\mu}(\omega)}$ or $\vev{\psi_{i+}^{\prime\mu}(\omega)}$ is strictly
plus by virtue of \eqref{vevinin}.

This leads to the following causality structure for the WQFT in the in-in formalism:
for the graviton emission 
\be
\vev{h^{\mu\nu}(k)} =
 \begin{tikzpicture}[baseline={(current bounding box.center)}]
    \tikzstyle{blob}=[
    rectangle,
    minimum size =.5cm,
    draw=black,
    thick,
    fill=lightgray,
    text=black
]
  \coordinate (inA) at (-1,1);
  \coordinate (outA) at (6,1);
  \coordinate (inB) at (-1,-1);
  \coordinate (outB) at (6,-1);
  \coordinate (xA) at (0,1);
  \coordinate (yA) at (1,1);  
  \coordinate (zA) at (2,1);
  \coordinate (xB) at (0,-1);
  \coordinate (yB) at (1,-1);
  \coordinate (zB) at (2,-1);
  \coordinate (l1) at (2.5,0.75);
  \coordinate (l2) at (2.5,0.45);
  \coordinate (l3) at (2.5,0.15); 
  \coordinate (ll3) at (2.5,-0.15);
  \coordinate (ll2) at (2.5,-0.45);
  \coordinate (ll1) at (2.5,-0.75);
  \draw [dotted] (inA) -- (outA);
  \draw [dotted] (inB) -- (outB);
  \draw [fill] (xA) circle (.08) node [above] {$-$};
  \draw [fill] (yA) circle (.08) node [above] {$-$};
  \draw [fill] (zA) circle (.08) node [above] {$-$};
  \draw [fill] (xB) circle (.08) node [below] {$-$};
  \draw [fill] (yB) circle (.08) node [below] {$-$};
  \draw [fill] (zB) circle (.08) node [below] {$-$};
  \draw [graviton] (xA) -- (l3);
  \draw [graviton] (yA) -- (l2);
  \draw [graviton] (zA) -- (l1);
  \draw [graviton] (xB) -- (ll3);
  \draw [graviton] (yB) -- (ll2);
  \draw [graviton] (zB) -- (ll1);
  \draw [graviton] (4.5,0) -- (6,0) node [midway,below] {$k$} node [very near end, above] {$+$} node[very near start,above] {$-$} node [right] {${}_{\mu\nu}$};
  \fill [draw=black,
    thick,
    fill=lightgray] (2.5,-1) rectangle (4.5,1);
      \draw [fill] (4.5,0) circle (.08);
  \draw [fill] (l1) circle (.08)  node [right] {$+$};
  \draw [fill] (l2) circle (.08) node [right] {$+$} ;
  \draw [fill] (l3) circle (.08) node [right] {$+$};
  \draw [fill] (ll1) circle (.08) node [right] {$+$};
  \draw [fill] (ll2) circle (.08) node [right] {$+$};
  \draw [fill] (ll3) circle (.08) node [right] {$+$};
  \end{tikzpicture}
\ee
as well as for the worldline one-point functions
\begin{align}\label{eq:ininZ1}
\vev{z_1^{\mu}(\omega)} &=
 \begin{tikzpicture}[baseline={(current bounding box.center)}]
    \tikzstyle{blob}=[
    rectangle,
    minimum size =.5cm,
    draw=black,
    thick,
    fill=lightgray,
    text=black
]
  \coordinate (inA) at (-1,1);
  \coordinate (outA) at (6,1);
  \coordinate (inB) at (-1,-1);
  \coordinate (outB) at (6,-1);
  \coordinate (xA) at (0,1);
  \coordinate (yA) at (1,1);  
  \coordinate (zA) at (2,1);
  \coordinate (xB) at (0,-1);
  \coordinate (yB) at (1,-1);
  \coordinate (zB) at (2,-1);
  \coordinate (l1) at (2.5,0.75);
  \coordinate (l2) at (2.5,0.45);
  \coordinate (l3) at (2.5,0.15); 
  \coordinate (ll3) at (2.5,-0.15);
  \coordinate (ll2) at (2.5,-0.45);
  \coordinate (ll1) at (2.5,-0.75);
  \draw [dotted] (inA) -- (outA);
  \draw [dotted] (inB) -- (outB);
  \draw [fill] (xA) circle (.08) node [above] {$-$};
  \draw [fill] (yA) circle (.08) node [above] {$-$};
  \draw [fill] (zA) circle (.08) node [above] {$-$};
  \draw [fill] (xB) circle (.08) node [below] {$-$};
  \draw [fill] (yB) circle (.08) node [below] {$-$};
  \draw [fill] (zB) circle (.08) node [below] {$-$};
  \draw [graviton] (xA) -- (l3);
  \draw [graviton] (yA) -- (l2);
  \draw [graviton] (zA) -- (l1);
  \draw [graviton] (xB) -- (ll3);
  \draw [graviton] (yB) -- (ll2);
  \draw [graviton] (zB) -- (ll1);
  \draw [zUndirected] (4.5,1) -- (6,1) node [midway,below] {$\omega$} node [very near end, above] {$+$} node[very near start,above] {$-$} node [right] {${}_{\mu}$} node [midway,above] {$\rightarrow$};
  \fill [draw=black,
    thick,
    fill=lightgray] (2.5,-1) rectangle (4.5,1);
      \draw [fill] (4.5,1) circle (.08);
   \draw [fill] (l1) circle (.08)  node [right] {$+$};
  \draw [fill] (l2) circle (.08) node [right] {$+$} ;
  \draw [fill] (l3) circle (.08) node [right] {$+$};
  \draw [fill] (ll1) circle (.08) node [right] {$+$};
  \draw [fill] (ll2) circle (.08) node [right] {$+$};
  \draw [fill] (ll3) circle (.08) node [right] {$+$};
  \end{tikzpicture}\\
  \vev{\psi^{\prime\mu}_{1}(\omega)} &=
 \begin{tikzpicture}[baseline={(current bounding box.center)}]
    \tikzstyle{blob}=[
    rectangle,
    minimum size =.5cm,
    draw=black,
    thick,
    fill=lightgray,
    text=black
]
  \coordinate (inA) at (-1,1);
  \coordinate (outA) at (6,1);
  \coordinate (inB) at (-1,-1);
  \coordinate (outB) at (6,-1);
  \coordinate (xA) at (0,1);
  \coordinate (yA) at (1,1);  
  \coordinate (zA) at (2,1);
  \coordinate (xB) at (0,-1);
  \coordinate (yB) at (1,-1);
  \coordinate (zB) at (2,-1);
  \coordinate (l1) at (2.5,0.75);
  \coordinate (l2) at (2.5,0.45);
  \coordinate (l3) at (2.5,0.15); 
  \coordinate (ll3) at (2.5,-0.15);
  \coordinate (ll2) at (2.5,-0.45);
  \coordinate (ll1) at (2.5,-0.75);
  \draw [dotted] (inA) -- (outA);
  \draw [dotted] (inB) -- (outB);
  \draw [fill] (xA) circle (.08) node [above] {$-$};
  \draw [fill] (yA) circle (.08) node [above] {$-$};
  \draw [fill] (zA) circle (.08) node [above] {$-$};
  \draw [fill] (xB) circle (.08) node [below] {$-$};
  \draw [fill] (yB) circle (.08) node [below] {$-$};
  \draw [fill] (zB) circle (.08) node [below] {$-$};
   \draw [graviton] (xA) -- (l3);
  \draw [graviton] (yA) -- (l2);
  \draw [graviton] (zA) -- (l1);
  \draw [graviton] (xB) -- (ll3);
  \draw [graviton] (yB) -- (ll2);
  \draw [graviton] (zB) -- (ll1);
  \draw [zParticle] (4.5,1) -- (6,1) node [midway,below] {$\omega$} node [very near end, above] {$+$} node[very near start,above] {$-$} node [right] {${}_{a}$} node [midway,above] {$\rightarrow$};
  \fill [draw=black,
    thick,
    fill=lightgray] (2.5,-1) rectangle (4.5,1);
  \draw [fill] (l1) circle (.08)  node [right] {$+$};
  \draw [fill] (l2) circle (.08) node [right] {$+$} ;
  \draw [fill] (l3) circle (.08) node [right] {$+$};
  \draw [fill] (ll1) circle (.08) node [right] {$+$};
  \draw [fill] (ll2) circle (.08) node [right] {$+$};
  \draw [fill] (ll3) circle (.08) node [right] {$+$};
  \end{tikzpicture}
\end{align}
As we learnt in the previous section, inside the grey shaded rectangle we should only insert single-minus-leg vertices.
Ironically we may then effectively forget about the in-in formalism  altogether:
simply apply the standard (in-out) Feynman rules,
assign momenta according to the causality flow (from  one-point worldline vertices to the outgoing line)
and use retarded propagators everywhere.  This is what was done (and its correctness
implicitly assumed) in our previous works \cite{Mogull:2020sak,Jakobsen:2021smu,Jakobsen:2021lvp,Jakobsen:2021zvh,Jakobsen:2022fcj}.
Here we have given a derivation of this property.

\subsection{In-in formalism for WQFT: free energy}

The eikonal phase in the scattering amplitudes approach to classical gravitational scattering
\cite{DiVecchia:2020ymx,DiVecchia:2021ndb,DiVecchia:2021bdo,Heissenberg:2021tzo} 
plays the role of a generating functional for the observables (impulse and spin kick).
The eikonal phase generalizes in WQFT to the free energy 
$\chi(b^{\mu},v_{i}^{\mu},\Psi_{i}^{\mu})$, which is a function of the background parameters.
In \rcites{Mogull:2020sak,Jakobsen:2021zvh} it was shown to yield the impulse and
spin kick  by differentiation with respect to $b^\mu$ and $\Psi_i^\mu$ respectively.
However, in the presence of radiation-reaction we need to generalize
this construction to the in-in formalism.
We therefore now modify the in-in prescription for the background 
configurations:
we introduce a plus- and minus-labeled background, i.e.~we expand $x^{\mu}_{i\pm}(\tau_i)=b^{\mu}_{i\pm}
+ v^{\mu}_{i\pm}\tau_{i} + z^{\mu}_{i\pm}(\tau_{i})$ and $\psi^{\mu}_{i\pm}(\tau_i)= \Psi^{\mu}_{i\pm}+\psi^{\prime\mu}_{i\pm}(\tau_{i})$. Importantly, the in-in free energy or zero-point function $\chi_{\text{in-in}}$ then needs to be evaluated only 
to \emph{linear} order in the minus background fields $b_{i-}, v_{i-}$ and $\Psi_{i-}$. This is so as the observables are derived from the in-in free energy via
\be
\Delta p_{\mu, i}= -\frac{\partial \chi_{\text{in-in}}}{\partial b^{\mu}_{i-}}\Bigr |_{b_{-}=v_{-}=\Psi_{-}=0}\, ,
\qquad
im_i\Delta \psi_{\mu, i}= \frac{\partial \chi_{\text{in-in}}}{\partial \bar\Psi^{\mu}_{i-}}\Bigr |_{b_{-}=v_{-}=\Psi_{-}=0}\, ,
\ee
identifying the plus-labeled background parameters with the physical ones:
$b^\mu_{i+}=b^\mu_{i}$, $v^\mu_{i+}=v^\mu_{i}$ and $\Psi^\mu_{i+}=\Psi^\mu_{i}$.

In summary we find the
following prescription: write down all zero-point diagrams with the in-out Feynman rules and successively assign to every vertex
on the worldlines a minus-labeled background. This single ``hot'' worldline vertex in a diagram then acts as a ``sink'' for the causality flow, meaning that all propagators
directed towards it are taken to be retarded. This prescription manifestly guarantees the differential relations above.
Yet, effectively this prescription is no more efficient than computing the observables directly as discussed above, as the so-defined
in-in free energy takes the form
\be
\chi_{\text{in-in}} = \sum_{i=1}^{2} -b_{i-}^{\mu}\, \Delta p_{\mu, i} 
+im_i\bar\Psi_{i-}^{\mu}\, \Delta\psi_{\mu, i} +
\ldots\,,
\ee
where the dots refer to terms quadratic in the minus background or linear in $v_{i-}$. Thus, there is no gain in efficiency
in evaluating the in-in free energy to the direct computation of the observables: deflection and spin kick.

\section{Retarded integrals}\label{sec:retInt}

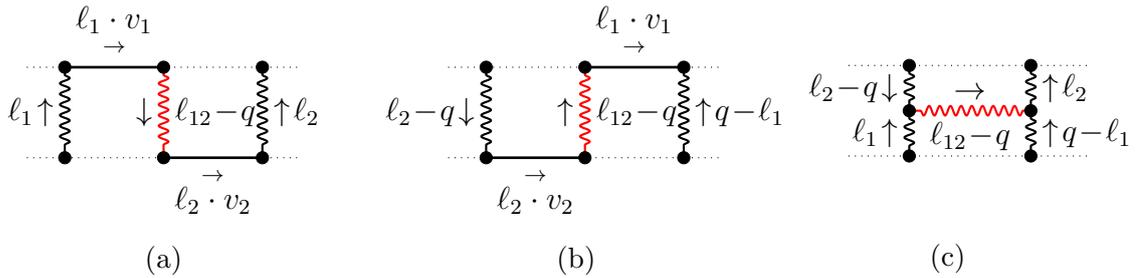
\begin{figure*}[t!]
  \centering
  \begin{subfigure}{.35\textwidth}
    \centering
    \begin{tikzpicture}[baseline={(current bounding box.center)},scale=1]
      \coordinate (inA) at (0.2,.6);
      \coordinate (outA) at (3.8,.6);
      \coordinate (inB) at (0.2,-.6);
      \coordinate (outB) at (3.8,-.6);
      \coordinate (xA) at (.7,.6);
      \coordinate (yA) at (2,.6);
      \coordinate (yzA) at (1.5,.6);
      \coordinate (zA) at (3.3,.6);
      \coordinate (xB) at (.7,-.6);
      \coordinate (yB) at (2,-.6);
      \coordinate (zB) at (3.3,-.6);
	   \draw [dotted] (inA) -- (outA);
	   \draw [dotted] (inB) -- (outB);
	   \draw [draw=none] (xA) to[out=40,in=140] (zA);
	   \draw [zParticleF] (xA) -- (yA) node [midway, above] {$\underset{\rightarrow}{\ell_1\cdot v_1}$};
	   \draw [zParticleF] (yB) -- (zB) node [midway, below] {$\overset{\rightarrow}{\ell_2\cdot v_2}$};
	   \draw [photon] (xA) -- (xB) node [midway, left] {$\ell_1\!\uparrow$};
	   \draw [photonRed] (yA) -- (yB) node [midway, right] {$\color{black}\ell_{12}\!-\!q$} node [midway, left] {$\color{black}\downarrow$};
	   \draw [photon] (zA) -- (zB) node [midway, right] {$\uparrow\!\ell_2$};
      \draw [fill] (xA) circle (.08);
      \draw [fill] (yA) circle (.08);
      \draw [fill] (zA) circle (.08);
      \draw [fill] (xB) circle (.08);
      \draw [fill] (yB) circle (.08);
      \draw [fill] (zB) circle (.08);
    \end{tikzpicture}
    \caption{}
  \end{subfigure}
  \begin{subfigure}{.35\textwidth}
    \centering
    \begin{tikzpicture}[baseline={(current bounding box.center)},scale=1]
      \coordinate (inA) at (0.2,.6);
      \coordinate (outA) at (3.8,.6);
      \coordinate (inB) at (0.2,-.6);
      \coordinate (outB) at (3.8,-.6);
      \coordinate (xA) at (.7,.6);
      \coordinate (yA) at (2,.6);
      \coordinate (yzA) at (1.5,.6);
      \coordinate (zA) at (3.3,.6);
      \coordinate (xB) at (.7,-.6);
      \coordinate (yB) at (2,-.6);
      \coordinate (zB) at (3.3,-.6);
	   \draw [dotted] (inA) -- (outA);
	   \draw [dotted] (inB) -- (outB);
	   \draw [draw=none] (xA) to[out=40,in=140] (zA);
	   \draw [zParticleF] (yA) -- (zA) node [midway, above] {$\underset{\rightarrow}{\ell_1\cdot v_1}$};
	   \draw [zParticleF] (xB) -- (yB) node [midway, below] {$\overset{\rightarrow}{\ell_2\cdot v_2}$};
	   \draw [photon] (xA) -- (xB) node [midway, left] {$\ell_2\!-\!q\!\downarrow$};
	   \draw [photonRed] (yA) -- (yB) node [midway, right] {$\color{black}\ell_{12}\!-\!q$} node [midway, left] {$\color{black}\uparrow$};
	   \draw [photon] (zA) -- (zB) node [midway, right] {$\uparrow\!q\!-\!\ell_1$};
      \draw [fill] (xA) circle (.08);
      \draw [fill] (yA) circle (.08);
      \draw [fill] (zA) circle (.08);
      \draw [fill] (xB) circle (.08);
      \draw [fill] (yB) circle (.08);
      \draw [fill] (zB) circle (.08);
    \end{tikzpicture}
    \caption{}
  \end{subfigure}
	\raisebox{-0.4cm}{
	\begin{subfigure}{0.26\textwidth}
		\centering
		\begin{tikzpicture}[baseline={(current bounding box.center)},scale=1]
		  \coordinate (inA) at (0.4,.6);
		  \coordinate (outA) at (3.6,.6);
		  \coordinate (inB) at (0.4,-.6);
		  \coordinate (outB) at (3.6,-.6);
		  \coordinate (xA) at (1.2,.6);
		  \coordinate (x0) at (1.2,0);
		  \coordinate (xyA) at (1.5,.6);
		  \coordinate (yA) at (2,.6);
		  \coordinate (yzB) at (2.5,-.6);
		  \coordinate (zA) at (2.8,.6);
		  \coordinate (xB) at (1.2,-.6);
		  \coordinate (yB) at (2,-.6);
		  \coordinate (zB) at (2.8,-.6);
		  \coordinate (z0) at (2.8,0);
		  \draw [dotted] (inA) -- (outA);
		  \draw [dotted] (inB) -- (outB) node [midway,below] {$\color{white}\overset{\rightarrow}{\ell_2\cdot v_2}$};
		  \draw [photon] (xA) -- (x0) node [midway,left] {$\ell_2\!-\!q\!\downarrow$};
		  \draw [photon] (x0) -- (xB) node [midway,left] {$\ell_1\!\uparrow$};
		  \draw [photonRed] (z0) -- (x0) node [midway,above] {$\color{black}\rightarrow$} node [midway,below] {$\color{black}\ell_{12}\!-\!q$};
		  \draw [photon] (zA) -- (z0) node [midway,right] {$\uparrow\!\ell_2$};
		  \draw [photon] (z0) -- (zB) node [midway,right] {$\uparrow\!q\!-\!\ell_1$};
		  \draw [fill] (xA) circle (.08);
		  \draw [fill] (x0) circle (.08);
		  \draw [fill] (z0) circle (.08);
		  \draw [fill] (zA) circle (.08);
		  \draw [fill] (xB) circle (.08);
		  \draw [fill] (zB) circle (.08);
		\end{tikzpicture}
		\caption{}
	  \end{subfigure}
	  }
  \caption{\small
    A selection of three diagrams involved in the integral family $I_{n_1,n_2,\ldots,n_7}^{(\sigma_1;\sigma_2;\sigma_3)}$~\eqref{eq:integralFamilies},
	with $\ell_{12}^\mu=\ell_1^\mu+\ell_2^\mu$.
	Wiggly lines propagating in the bulk represent massless propagators $(\ell^2\pm{\rm sgn}(\ell^0)i0)^{-n}$;
	solid lines propagating on the worldlines (dotted) represent linear propagators $(\omega\pm i0)^{-n}$.   	
	The propagator highlighted in red corresponds with $D_3$~\eqref{eq:integralFamilies},
	and is the only \emph{active} bulk propagator ---
	i.e.~the only one with the ability to go on shell.
  }
  \label{fig:iDiags}
\end{figure*}

Having shown that complete WQFT observables
naturally involve retarded propagators we now
study a family of Feynman integrals involving these instead of
the usual time-symmetric Feynman type.
We focus on the following class of integrals
relevant for deriving classical observables
in the gravitational scattering of two comparable-mass bodies at 3PM order:
\begin{align}\label{eq:integralFamilies}
  &I^{(\sigma_1;\sigma_2;\sigma_3)}_{n_1,n_2,\ldots,n_7}
  :=\int_{\ell_1,\ell_2}\!\frac{\dd(\ell_1\cdot v_2)\dd(\ell_2\cdot v_1)}
  {D_1^{n_1}D_2^{n_2}\cdots D_7^{n_7}},\\
  &D_1=\ell_1\!\cdot\! v_1\!+\!\sigma_1i0\,, \,\,
  D_2=\ell_2\!\cdot\! v_2\!+\!\sigma_2i0\,,\,\,
  D_3=(\ell_1+\ell_2-q)^2+\sigma_3\,{\rm sgn}(\ell_1^0+\ell_2^0-q^0)i0\,,\nn\\
  &D_4=\ell_1^2\,, \quad
  D_5=\ell_2^2\,, \quad
  D_6=(\ell_1-q)^2\,,\quad
  D_7=(\ell_2-q)^2\,,\nn
\end{align}
where $\dd(\omega):=2\pi\delta(\omega)$,
$\int_k:=\int\frac{\d^Dk}{(2\pi)^D}$ in $D=4-2\eps$ dimensions
and $\sigma_i=\pm1$ determine the $i0$ prescription.
The two velocities $v_i^\mu$ are time-like vectors satisfying $v_i^2=1$;
the exchanged momentum $q^\mu$ is a space-like vector satisfying $q\cdot v_i=0$.
In general, the integrals are functions only of $\gamma=v_1\cdot v_2$
and $|q|:=\sqrt{-q\cdot q}$;
as the latter dependence is trivially fixed on dimensional grounds
we can safely set $|q|=1$.
A selection of three WQFT Feynman graphs giving rise to these integrals is shown in \Fig{fig:iDiags}.

We begin with some general remarks.
The two linearized propagators $D_1$ and $D_2$ are associated with
deflection modes propagating on the worldlines;
the quadratic propagators $D_3$--$D_7$ with massless particles
propagating in the bulk.
Of these five propagators only the first, $D_3$, can go on-shell
over the constrained (by the $\dd$-functions) domain of integration;
for the others $D_i\neq0$ generically as we may choose a frame
that ensures these propagators become Euclidean,
i.e.~$\ell_i^2=-\Bell_i^2$.
We therefore refer to $D_3$ as an \emph{active} propagator,
and track $i0$ prescriptions only on $D_{1,2,3}$.
As we shall see, the velocity scaling of $k^\mu:=\ell_1^\mu+\ell_2^\mu-q^\mu$
in $D_3$ determines our region of integration (potential or radiation) in the static limit $v\to0$.

Feynman integrals with retarded propagators
are always purely real or imaginary (pseudoreal).
In the current context this is necessarily true,
as we will construct real physical observables
as linear combinations of these integrals with (pseudo)real coefficients.
We can also show it directly by taking the complex conjugate,
which sends $i0\to-i0$ everywhere:
\begin{align}
  &I^{(\sigma_1;\sigma_2;\sigma_3)*}_{n_1,n_2,\ldots,n_7}\nn\\
  &=\int_{\ell_1,\ell_2}\!\frac{\dd(\ell_1\cdot v_2)\dd(\ell_2\cdot v_1)}
  {(\ell_1\!\cdot\! v_1\!-\!\sigma_1i0)^{n_1}(\ell_2\!\cdot\! v_2\!-\!\sigma_2i0)^{n_2}
  ((\ell_1\!+\!\ell_2\!-\!q)^2\!-\!\sigma_3\,{\rm sgn}(\ell_1^0\!+\!\ell_2^0\!-\!q^0)i0)^{n_3}\cdots}\nn\\
  &=\int_{\ell_1,\ell_2}\!\frac{\dd(\ell_1\cdot v_2)\dd(\ell_2\cdot v_1)}
  {(-\ell_1\!\cdot\! v_1\!-\!\sigma_1i0)^{n_1}(-\ell_2\!\cdot\! v_2\!-\!\sigma_2i0)^{n_2}
  ((\ell_1\!+\!\ell_2\!-\!q)^2\!+\!\sigma_3\,{\rm sgn}(\ell_1^0\!+\!\ell_2^0\!-\!q^0)i0)^{n_3}\cdots}\nn\\
  &=(-1)^{n_1+n_2}I^{(\sigma_1;\sigma_2;\sigma_3)}_{n_1,n_2,\ldots,n_7}\,.
\end{align}
In the second step we have replaced $\ell^\mu_i\to-\ell^\mu_i$, $q^\mu\to-q^\mu$.\footnote{The
latter is possible as $I^{(\sigma_1;\sigma_2;\sigma_3)}_{n_1,n_2,\ldots,n_7}$ only depends on $q^\mu$ via $|q|$.}
We relied here on the velocities $v_i^\mu$ being real, time-like vectors:
so this argument applies strictly in the physical regime $1<\gamma<\infty$.
We see a clear separation into two sub-families of integrals:
those with even or odd values of $n_1+n_2$,
which are purely real or imaginary.

There is also a close relationship to integrals
involving the Feynman $i0$ prescription.
We remind ourselves that Feynman and retarded propagators
may be decomposed according to\footnote{One
	should be careful applying such distributional identities when products of propagators are involved;
	however, in the present context there is no such issue.
}
\begin{subequations}
\begin{align}
\frac1{k^2+i0}&=
{\cal P}\frac1{k^2}-i\pi\delta(k^2)\,,\\
\frac1{k^2+\sigma\,{\rm sgn}(k^0)i0}&=
{\cal P}\frac1{k^2}-i\pi\sigma\,{\rm sgn}(k^0)\delta(k^2)\,.
\end{align}
\end{subequations}
Here ${\cal P}$ refers to a principal value prescription.
Specializing to $n_1=n_2=0$ (no worldline propagators) and $n_3=1$:
\begin{align}\label{eq:feynret}
  I^{(\sigma_1;\sigma_2;\sigma_3)}_{0,0,1,n_4,n_5,n_6,n_7}
  &=\int_{\ell_1,\ell_2}\!\frac{\dd(\ell_1\cdot v_2)\dd(\ell_2\cdot v_1)}
  {(\ell_1^2)^{n_4}(\ell_2^2)^{n_5}((\ell_1-q)^2)^{n_6}((\ell_2-q)^2)^{n_7}}\times\\
  &\qquad\left({\cal P}\frac1{(\ell_1+\ell_2-q)^2}-
  i\pi\sigma_3{\rm sgn}(\ell_1^0+\ell_2^0-q^0)\delta((\ell_1+\ell_2-q)^2)\right)\,.\nn
\end{align}
The on-shell contribution proportional to $\sigma_3$ vanishes by symmetry.
The entire integral is therefore given by its principal value part,
and thus the real part of the same integral
where a Feynman $i0$ prescription is used instead.

To obtain explicit expressions for these integrals we take a well-studied approach
\cite{Herrmann:2021lqe,Herrmann:2021tct,DiVecchia:2021bdo,Bjerrum-Bohr:2021vuf,Brandhuber:2021eyq}:
seek linear identities between the integrals arising from symmetries and
integration-by-parts (IBP) relations, 
and reduce to a basis of masters \cite{Smirnov:2008iw,Gehrmann:1999as}.
Expressions are then recovered by solving differential equations (DEs) in $\gamma$  \cite{Henn:2013pwa}.
Finally, constants of integration are fixed by comparison with the static limit
$v\to0$, where $v$ is the relative velocity and $\gamma=(1-v^2)^{-1/2}$
using the method of regions \cite{Beneke:1997zp}.
The only aspects specifically affected by the retarded $i0$ prescription
are the symmetry relations
and the fixing of boundary conditions in the static limit.

\subsection{Linear identities}

We begin with the symmetry relations.
Naively there are eight families of integrals,
corresponding to the possible sign choices on $\sigma_i$.
However, the integrals satisfy
\begin{subequations}
\begin{align}
  I^{(\sigma_1;\sigma_2;\sigma_3)}_{n_1,n_2,n_3,n_4,n_5,n_6,n_7}&=
  I^{(\sigma_1;\sigma_2;\sigma_3)}_{n_1,n_2,n_3,n_6,n_7,n_4,n_5}
  \qquad\qquad\text{(shift $\ell^\mu_i\to\ell^\mu_i+q^\mu$)}\,,\\
  I^{(\sigma_1;\sigma_2;\sigma_3)}_{n_1,n_2,n_3,n_4,n_5,n_6,n_7}&=(-1)^{n_1+n_2}
  I^{(-\sigma_1;-\sigma_2;-\sigma_3)}_{n_1,n_2,n_3,n_4,n_5,n_6,n_7}
  \,\,\,\,\text{(flip $\ell^\mu_i\to-\ell^\mu_i$, $q^\mu\to-q^\mu$)}\,,\label{eq:relB}\\
  I^{(\sigma_1;\sigma_2;\sigma_3)}_{n_1,n_2,n_3,n_4,n_5,n_6,n_7}&=
  I^{(\sigma_2;\sigma_1;\sigma_3)}_{n_2,n_1,n_3,n_5,n_4,n_7,n_6}
  \qquad\qquad\text{(exchange $\ell^\mu_1\leftrightarrow\ell^\mu_2$, $v^\mu_1\leftrightarrow v^\mu_2$)}\,.\label{eq:relC}
\end{align}
\end{subequations}
Using the second two relations \eqref{eq:relB} and \eqref{eq:relC}
we are left with three independent families,
which we choose as
\begin{align}\label{eq:sigmas}
  I_{n_1,n_2,\ldots,n_7}=I_{n_1,n_2,\ldots,n_7}^{(+;+;+)}\,,\quad
  I'_{n_1,n_2,\ldots,n_7}=I_{n_1,n_2,\ldots,n_7}^{(-;-;+)}\,,\quad
  I''_{n_1,n_2,\ldots,n_7}=I_{n_1,n_2,\ldots,n_7}^{(+;-;+)}\,.
\end{align}
When one of the propagators $D_1$--$D_3$ is raised
to a zero or negative power,
i.e.~that propagator is now a numerator,
then there is overlap between the different families:
\begin{subequations}\label{eq:overlap}
\begin{align}
  I^{(+;\sigma_2;\sigma_3)}_{n_1,n_2,\ldots,n_7}&=
  I^{(-;\sigma_2;\sigma_3)}_{n_1,n_2,\ldots,n_7}\,,
  \quad n_1\leq0\,,\\
  I^{(\sigma_1;+;\sigma_3)}_{n_1,n_2,\ldots,n_7}&=
  I^{(\sigma_1;-;\sigma_3)}_{n_1,n_2,\ldots,n_7}\,,
  \quad n_2\leq0\,,\\
  I^{(\sigma_1;\sigma_2;+)}_{n_1,n_2,\ldots,n_7}&=
  I^{(\sigma_1;\sigma_2;-)}_{n_1,n_2,\ldots,n_7}\,,
  \quad n_3\leq0\,.
\end{align}
\end{subequations}
In these cases the $i0$ prescription is irrelevant.

The other identities are IBP relations,
conveniently generated using on-the-market tools including
\texttt{FIRE}~\cite{Smirnov:2019qkx},
\texttt{LiteRed}~\cite{Lee:2012cn,Lee:2013mka}
and \texttt{KIRA}~\cite{Maierhofer:2017gsa,Klappert:2020nbg}.
As these are fairly conventional we will not go into detail
regarding our use of them here;
the only subtlety is our handling of the $\dd(\ell_1\cdot v_2)$ and $\dd(\ell_2\cdot v_1)$
present in \eqn{eq:integralFamilies}.
Following \rcites{Herrmann:2021lqe,Herrmann:2021tct},
when generating IBPs we generalize them to include derivatives:
\begin{equation}
	\frac{\dd^{(n)}(\omega)}{(-1)^nn!}=
	\frac{i}{(\omega+i0)^{n+1}}-\frac{i}{(\omega-i0)^{n+1}}\,.
\end{equation}
Four-dimensional Lorentz covariance is maintained by
treating these $\dd$-functions as cut propagators.
From this perspective our integrals thus have nine propagators instead of seven,
but in general we always choose masters
adhering to the schematic form \eqref{eq:integralFamilies}.

\subsection{Canonical bases}

While in principle any linearly independent set of integrals
would suffice as a set of master integrals,
it is convenient to choose a so-called canonical basis \cite{Henn:2013pwa}.
A suitable basis of real master integrals
(also used in \rcite{Bjerrum-Bohr:2021vuf}, $\epsilon=2-\sfrac{D}2$) is
\begin{subequations}\label{eq:bMasters}
\begin{align}
{\cal I}_1&=
2\epsilon^2I_{0,0,0,1,1,1,1}\,,\\
{\cal I}_2&=
2\epsilon^2\sqrt{\gamma^2-1}I_{0,0,1,0,0,1,1}\,,\\
{\cal I}_3&=
2\epsilon\sqrt{\gamma^2-1}I_{0,0,2,0,0,1,1}\,,\\
{\cal I}_4&=
-4I_{-1,-1,3,0,0,1,1}+(1+2\epsilon)\gamma I_{0,0,2,0,0,1,1}\,,\\
{\cal I}_5&=
\frac{2(4\epsilon-1)(2\epsilon-1)}{\sqrt{\gamma^2-1}}I_{0,0,1,0,1,0,1}\,,\\
{\cal I}_6&=
2\epsilon^2\sqrt{\gamma^2-1}I_{0,0,1,1,1,1,1}\,,\\
{\cal I}_7&=
-8\epsilon^2I_{-1,-1,1,1,1,1,1}+4\epsilon^2\gamma I_{0,0,0,1,1,1,1}\,,\\
{\cal I}_8&=
-\epsilon^2(\gamma^2-1)I_{1,1,1,0,0,1,1}\,,
\end{align}
\end{subequations}
with ${\cal I}'_i$ and ${\cal I}''_i$ similarly defined in terms of
$I'_{n_1,n_2,\ldots,n_7}$ and $I''_{n_1,n_2,\ldots,n_7}$~\eqref{eq:sigmas}.
Introducing $x=\gamma-\sqrt{\gamma^2-1}$,
so that $0<x<1$ in the physical region,
the defining property of this canonical basis is that the
resulting system of DEs takes an $\eps$-factorized form:
\begin{equation}\label{eq:diffEqn}
\frac{\d\vec{{\cal I}}}{\d x}=
\epsilon\left(\frac{A}{x}+\frac{B_+}{1+x}-\frac{B_-}{1-x}\right)\vec{\cal I}\,,
\end{equation}
which one observes by explicitly differentiating the master integrals
\eqref{eq:bMasters} and reducing the resulting integrals via IBPs to the same basis.
The set of poles $\{x,1+x,1-x\}$ is the symbol alphabet;
the constant matrices are
\begin{align}
A=\left(
\begin{array}{cccccccc}
 0 & 0 & 0 & 0 & 0 & 0 & 0 & 0 \\
 0 & -6 & 0 & -1 & 0 & 0 & 0 & 0 \\
 0 & 0 & 2 & -2 & 0 & 0 & 0 & 0 \\
 0 & 12 & 2 & 0 & 0 & 0 & 0 & 0 \\
 0 & 0 & 0 & 0 & 2 & 0 & 0 & 0 \\
 0 & 0 & 4 & 2 & 4 & 2 & -2 & 0 \\
 0 & 12 & 8 & 0 & 8 & 2 & -2 & 0 \\
 0 & 0 & -1 & 0 & 0 & 0 & 0 & 0 \\
\end{array}
\right)\,,\quad
B_\pm=\left(
\begin{array}{cccccccc}
 0 & 0 & 0 & 0 & 0 & 0 & 0 & 0 \\
 0 & 6 & 0 & 0 & 0 & 0 & 0 & 0 \\
 0 & 0 & -2 & 0 & 0 & 0 & 0 & 0 \\
 0 & 0 & 0 & 0 & 0 & 0 & 0 & 0 \\
 0 & 0 & 0 & 0 & -2 & 0 & 0 & 0 \\
 0 & 0 & -4 & 0 & -4 & -2 & 0 & 0 \\
 \pm4 & 0 & 0 & 0 & 0 & 0 & 2 & 0 \\
 0 & 0 & 0 & 0 & 0 & 0 & 0 & 0 \\
\end{array}
\right)\,,
\end{align}
and for the $\vec{{\cal I}}'$, $\vec{{\cal I}}''$ integrals we have
$A=A'=A''$, $B_\pm=B'_\pm=B''_\pm$.
This $\eps$-factorized form of the DEs is
solved as a series expansion in $\eps$:
\begin{align}
	\vec{{\cal I}}=\vec{{\cal I}}^{(0)}+\eps\,\vec{{\cal I}}^{(1)}+\cO(\eps^2)\,,
\end{align}
with solution
\begin{align}
	\vec{{\cal I}}^{(0)}=\vec{c}^{\,(0)}\,, &&
	\vec{{\cal I}}^{(1)}=\left(A\log(x)+B_+\log(1+x)-B_-\log(1-x)\right)\vec{{\cal I}}^{(0)}+\vec{c}^{\,(1)}\,.
\end{align}
The overlap relations \eqref{eq:overlap} also tell us that
\begin{equation}
  {\cal I}_i={\cal I}'_i={\cal I}''_i\,,\quad 1\leq i\leq7\,,
\end{equation}
while ${\cal I}_8$, ${\cal I}'_8$ and ${\cal I}''_8$ each have separate expressions.
The only remaining subtlety is fixing the constants of integration $\vec{c}^{\,(i)}$,
a point to which we will return in \sec{sec:regions}.

A canonical basis for the purely imaginary integrals is
\begin{subequations}
\begin{align}
{\cal I}_9&=-2\eps\sqrt{\gamma^2-1}I_{1,0,1,0,1,0,1}\,,\\
{\cal I}_{10}&=-2\eps\sqrt{\gamma^2-1}I_{1,0,1,1,1,0,0}\,,\\
{\cal I}_{11}&=\frac{1-2\eps}3I_{1,0,1,1,0,1,0}\,,\\
{\cal I}_{12}&=I_{-1,0,2,0,0,1,1}\,,\\
{\cal I}_{13}&=-\eps I_{-1,0,1,1,1,1,1}\,.
\end{align}
\end{subequations}
These five integrals also obey DEs of the form \eqref{eq:diffEqn} but with $5\times5$ matrices
\begin{align}
A=\left(
\begin{array}{ccccc}
 0 & 0 & 6 & 0 & 0 \\
 0 & 0 & 0 & -4 & 0 \\
 0 & 0 & 2 & 0 & 0 \\
 0 & 0 & 0 & -2 & 0 \\
 0 & 0 & 0 & -2 & 0
\end{array}
\right)\,,\,\,
B_+=\left(
\begin{array}{ccccc}
 0 & 0 & 0 & 0 & 0 \\
 0 & 0 & 0 & 0 & 0 \\
 0 & 0 & -2 & 0 & 0 \\
 0 & 0 & 0 & 6 & 0 \\
 0 & 0 & -6 & 0 & 2
\end{array}
\right)\,,\,\,
B_-=\left(
\begin{array}{ccccc}
 0 & 0 & 0 & 0 & 0 \\
 0 & 0 & 0 & 0 & 0 \\
 0 & 0 & -2 & 0 & 0 \\
 0 & 0 & 0 & -2 & 0 \\
 0 & 0 & 6 & 4 & -2
\end{array}
\right)\,,
\end{align}
and $A=A'$, $B_\pm=B'_\pm$.
Similar expressions for
${\cal I}'_i$ and ${\cal I}''_i$ are again defined in terms of
$I'_{n_1,n_2,\ldots,n_7}$ and $I''_{n_1,n_2,\ldots,n_7}$; 
in the latter case we require two additional masters
\begin{subequations}
\begin{align}
  {\cal I}''_{14}&=-2\eps\sqrt{\gamma^2-1}I^{(+;-;+)}_{0,1,1,1,0,1,0}\,,\\
  {\cal I}''_{15}&=-2\eps\sqrt{\gamma^2-1}I^{(+;-;+)}_{0,1,1,1,1,0,0}\,,
\end{align}
\end{subequations}
in order to have a complete basis.
So in this case the DEs involve $7\times 7$ matrices.
Fortunately, all seven of these integrals are given by the overlap relations \eqref{eq:overlap}:
\begin{subequations}
\begin{align}
{\cal I}''_i&={\cal I}_i\,,\quad 9\leq i\leq13\,,\\
{\cal I}''_{14}&={\cal I}'_{9}\,,\quad
{\cal I}''_{15}={\cal I}'_{10}\,.
\end{align}
\end{subequations}
We therefore focus on the ${\cal I}_i$ and ${\cal I}'_i$ families.

\subsection{Method of regions}\label{sec:regions}

Our strategy for fixing boundary conditions on the integrals is to evaluate
them to leading order in the static limit where $v\to0$.
However, this expansion requires caution as the two limits $\eps\to0$ and $v\to0$ do not generally commute \cite{Henn:2014qga,Bjerrum-Bohr:2021vuf}.
In order to expand the integrand in $v\to0$ before integration
we use the \emph{method of regions} \cite{Smirnov:2012gma,Bern:2021yeh,Herrmann:2021tct,Becher:2014oda}.
Here, one identifies all singular regions in the static limit
and writes the full integral as a sum of contributions from each region \cite{Beneke:1997zp}.
Each of these contributions satisfy the same differential equations but behave differently in the static limit,
which leads to different boundary conditions.

Regions in the static limit are characterized by velocity scalings of the loop momenta,
with their time-like and spatial components scaling as
\begin{subequations}\label{eq:regG}
	\begin{align}
	  \ell^\text{pot}_i&=(\ell_i^0,\Bell_i)\sim(v,1)\,,\label{eq:potG}\\
	  \ell_i^\text{rad}&=(\ell_i^0,\Bell_i)\sim(v,v)\,.\label{eq:radG}
	\end{align}
\end{subequations}
These we refer to these as potential and radiative momenta:
only radiative momenta can go on-shell, $\ell_i^2=0$.
In the present context, we note that only the momentum $k^\mu=\ell_1^\mu+\ell_2^\mu-q^\mu$,
carried by the active $D_3$ propagator~\eqref{eq:integralFamilies},
may be radiative;
for all other internal momenta the potential scaling~\eqref{eq:potG} is mandated
by the presence of $\dd(\ell_i\cdot v_j)$.
Thus, in the static limit the behavior of $D_3$
characterizes the entire integral,
splitting it into a sum of two regions:
\begin{align}\label{eq:regions}
	I^{(\sigma_1;\sigma_2;\sigma_3)}_{n_1,n_2,\ldots,n_7}=
	I^{(\sigma_1;\sigma_2;\sigma_3){\rm pot}}_{n_1,n_2,\ldots,n_7}+I^{(\sigma_1;\sigma_2;\sigma_3){\rm rad}}_{n_1,n_2,\ldots,n_7}\,,
\end{align}
which we again refer to as potential and radiation.
In these two regions the integrand may be expanded in $v$ assuming the momentum scalings~\eqref{eq:regG},
and boundary conditions are fixed by matching to the leading-order term in this expansion.
The resulting integrals are simpler as their dependence on $\gamma=(1-v^2)^{-1/2}$ is trivialized.

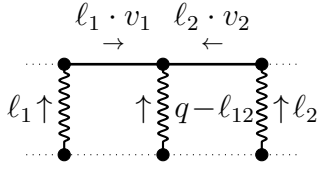
\begin{figure*}[t!]
	\centering
	\begin{subfigure}{.35\textwidth}
	  \centering
	  \begin{tikzpicture}[baseline={(current bounding box.center)},scale=1]
		\coordinate (inA) at (0.2,.6);
		\coordinate (outA) at (3.8,.6);
		\coordinate (inB) at (0.2,-.6);
		\coordinate (outB) at (3.8,-.6);
		\coordinate (xA) at (.7,.6);
		\coordinate (yA) at (2,.6);
		\coordinate (yzA) at (1.5,.6);
		\coordinate (zA) at (3.3,.6);
		\coordinate (xB) at (.7,-.6);
		\coordinate (yB) at (2,-.6);
		\coordinate (zB) at (3.3,-.6);
		 \draw [dotted] (inA) -- (outA);
		 \draw [dotted] (inB) -- (outB);
		 \draw [draw=none] (xA) to[out=40,in=140] (zA);
		 \draw [zParticleF] (xA) -- (yA) node [midway, above] {$\underset{\rightarrow}{\ell_1\cdot v_1}$};
		 \draw [zParticleF] (yA) -- (zA) node [midway, above] {$\underset{\leftarrow}{\ell_2\cdot v_2}$};
		 \draw [photon] (xA) -- (xB) node [midway, left] {$\ell_1\!\uparrow$};
		 \draw [photon] (yA) -- (yB) node [midway, right] {$q\!-\!\ell_{12}$} node [midway, left] {$\uparrow$};
		 \draw [photon] (zA) -- (zB) node [midway, right] {$\uparrow\!\ell_2$};
		\draw [fill] (xA) circle (.08);
		\draw [fill] (yA) circle (.08);
		\draw [fill] (zA) circle (.08);
		\draw [fill] (xB) circle (.08);
		\draw [fill] (yB) circle (.08);
		\draw [fill] (zB) circle (.08);
	  \end{tikzpicture}
	\end{subfigure}
	\caption{\small
	  An example of a diagram contributing to the test-body integral family $J_{n_1,n_2,\ldots,n_7}^{(\sigma_1;\sigma_2)}$
		\eqref{eq:testBodyFamily}, in which none of the massless lines can go on-shell.
	}
	\label{fig:examplePot}
\end{figure*}

Let us first analyze the potential region (also discussed in \rcite{Jakobsen:2022fcj}).
Here all internal momenta live in the potential region~\eqref{eq:potG},
so it makes no difference whether the $D_3$ propagator is advanced or retarded.
In the static limit to leading order in $v$
\begin{align}
  \label{eq:potL}
  I^{(\sigma_1;\sigma_2;\sigma_3){\rm pot}}_{n_1,n_2,\ldots,n_7}
  =
  (-1)^{n_1}
  J^{(-\sigma_1;\sigma_2)}_{n_1,n_2,\ldots,n_7}
  +
  \mathcal{O}(v^{2-n_1-n_2})
  \ ,
\end{align}
where the integrals $J^{(\sigma_1;\sigma_2)}_{n_1,n_2,\ldots,n_7}\sim\cO(v^{-n_1-n_2})$ are given by
\begin{align}\label{eq:testBodyFamily}
\begin{aligned}
  &
  J^{(\sigma_1;\sigma_2)}_{n_1,n_2,\ldots,n_7}
  :=\int_{\ell_1,\ell_2}\!\frac{\dd(\ell_1\cdot v_2)\dd(\ell_2\cdot v_2)}
  {D_1^{n_1}D_2^{n_2}\cdots D_7^{n_7}},
  \\
  &D_1=\ell_1\!\cdot\! v_1\!+\!\sigma_1i0\,, \,\,
  D_2=\ell_2\!\cdot\! v_1\!+\!\sigma_2i0\,,\,\,
  D_3=(\ell_1+\ell_2-q)^2\,,\\
  &D_4=\ell_1^2\,, \quad
  D_5=\ell_2^2\,, \quad
  D_6=(\ell_1-q)^2\,,\quad
  D_7=(\ell_2-q)^2\,.
\end{aligned}
\end{align}
This integral family is familiar: in \rcite{Jakobsen:2022fcj} integrals of this kind
were relevant for computing observables in the probe limit $m_1\ll m_2$.
With respect to IBP and symmetry relations there are three master integrals:
\begin{subequations}\label{eq:easyMasters}
	\begin{align}
		J^{(+;\pm)}_{0,0,1,1,1,0,0}&=
		-(4\pi)^{-3+2\eps}\frac{\Gamma^3(\frac12-\eps)\Gamma(2\eps)}{\Gamma(\frac32-3\eps)}\,,\\
		J^{(+;\pm)}_{1,0,1,1,1,0,0}&=
		(4\pi)^{-\frac52+2\eps}
		\frac{i}{2\sqrt{\gamma^2-1}}
		\frac{\Gamma(\frac12-2\eps)\Gamma^2(\frac12-\eps)\Gamma(-\eps)\Gamma(\frac12+2\eps)}
		{\Gamma(\frac12-3\eps)\Gamma(1-2\eps)}\,,\\
		J^{(+;+)}_{1,1,1,1,1,0,0}&=
		-2J^{(+;-)}_{1,1,1,1,1,0,0}
		=(4\pi)^{-2+2\eps}\frac{\Gamma^3(-\eps)\Gamma(1+2\eps)}{3(\gamma^2-1)\Gamma(-3\eps)}\,,
	\end{align}
\end{subequations}
and as promised the dependence on $\gamma$ and $\eps$ factorizes.
An example of a WQFT diagram giving rise to such integrals is displayed in \Fig{fig:examplePot}.
As none of the internal momenta can go on-shell this integral family consists only of rational functions
of the dimension $D=4-2\eps$.

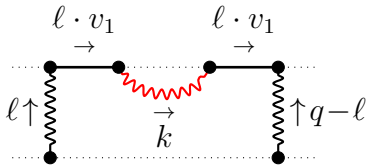
\begin{figure*}[t!]
	\centering
	\begin{subfigure}{.35\textwidth}
	  \centering
	  \begin{tikzpicture}[baseline={(current bounding box.center)},scale=1]
		\coordinate (inA) at (0,.6);
		\coordinate (outA) at (4,.6);
		\coordinate (inB) at (0,-.6);
		\coordinate (outB) at (4,-.6);
		\coordinate (xA) at (.5,.6);
		\coordinate (y1A) at (1.4,.6);
		\coordinate (y2A) at (2.6,.6);
		\coordinate (zA) at (3.5,.6);
		\coordinate (xB) at (.5,-.6);
		\coordinate (zB) at (3.5,-.6);
		 \draw [dotted] (inA) -- (outA);
		 \draw [dotted] (inB) -- (outB) node [midway, above] {$\overset{\rightarrow}{k}$};
		 \draw [draw=none] (xA) to[out=40,in=140] (zA);
		 \draw [zParticleF] (xA) -- (y1A) node [midway, above] {$\underset{\rightarrow}{\ell\cdot v_1}$};
		 \draw [zParticleF] (y2A) -- (zA) node [midway, above] {$\underset{\rightarrow}{\ell\cdot v_1}$};
		 \draw [photon] (xA) -- (xB) node [midway, left] {$\ell\!\uparrow$};
		 \draw [photon] (zA) -- (zB) node [midway, right] {$\uparrow\!q\!-\!\ell$};
		 \draw [photonRed] (y1A)to[out=-70,in=-110] (y2A);
		\draw [fill] (xA) circle (.08);
		\draw [fill] (y1A) circle (.08);
		\draw [fill] (y2A) circle (.08);
		\draw [fill] (zA) circle (.08);
		\draw [fill] (xB) circle (.08);
		\draw [fill] (zB) circle (.08);
	  \end{tikzpicture}
	\end{subfigure}
	\caption{\small
		A schematic representation of the ``mushroom'' integral family $K^{(\sigma_1;\sigma_2;\sigma_3)}_{n_1,n_2,\ldots,n_5}$,
		in which the propagator carrying momentum $k^\mu$ is radiative (red line).
	}
	\label{fig:exampleMushroom}
\end{figure*}

Next we turn to the region where $k^\mu=\ell_1^\mu+\ell_2^\mu-q^\mu$
is radiative and $\ell^\mu=\ell_1^\mu$ lives in the potential region. We find that
\begin{equation}
    I^{(\sigma_1;\sigma_2;\sigma_3){\rm rad}}_{n_1,n_2,\ldots,n_7}
  =
  K^{(\sigma_1;\sigma_2;\sigma_3)}_{n_1,n_2,n_3,n_4+n_7,n_5+n_6}
  +\mathcal{O}(v^{D+1-n_1-n_2-2n_3})\,,
\end{equation}
where the new integral family $K^{(\sigma_1;\sigma_2;\sigma_3)}_{n_1,n_2,\ldots,n_7}\sim\cO(v^{D-1-n_1-n_2-2n_3})$ is defined as
\begin{align}\label{eq:kIntegral}
  &K^{(\sigma_1;\sigma_2;\sigma_3)}_{n_1,n_2\ldots,n_5}:=
  \int_{\ell,k}\!\frac{\dd((k-\ell)\cdot v_1)\dd(\ell\cdot v_2)}
	{(\ell\cdot v_1\!+\!\sigma_1i0)^{n_1}(\ell\cdot v_1\!+\!\sigma_2i0)^{n_2}
	(k^2\!+\!\sigma_3{\rm sgn}(k^0)i0)^{n_3}(\ell^2)^{n_4}((\ell\!-\!q)^2)^{n_5}}\,.
\end{align}
These integrals (sometimes called ``mushrooms'') describe self-interaction where a worldline emits and absorbs the same graviton.
A schematic representation of these integrals is given in \Fig{fig:exampleMushroom}.
It is tempting to assume that these integrals form a subset of the $I^{(\sigma_1;\sigma_2;\sigma_3)}_{n_1,n_2,\ldots,n_7}$
family~\eqref{eq:integralFamilies}; however, this is only true when $\sigma_1=\sigma_2$.

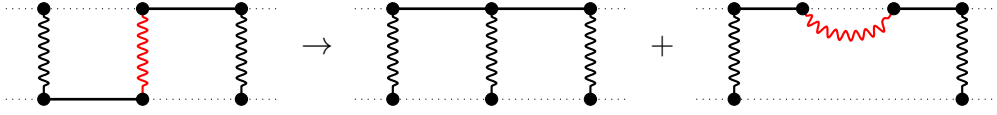
\begin{figure*}[t]
	\centering
		\begin{tikzpicture}[baseline={0},scale=1]
		  \coordinate (inA) at (0.2,.6);
		  \coordinate (outA) at (3.8,.6);
		  \coordinate (inB) at (0.2,-.6);
		  \coordinate (outB) at (3.8,-.6);
		  \coordinate (xA) at (.7,.6);
		  \coordinate (yA) at (2,.6);
		  \coordinate (yzA) at (1.5,.6);
		  \coordinate (zA) at (3.3,.6);
		  \coordinate (xB) at (.7,-.6);
		  \coordinate (yB) at (2,-.6);
		  \coordinate (zB) at (3.3,-.6);
		   \draw [dotted] (inA) -- (outA);
		   \draw [dotted] (inB) -- (outB);
		   \draw [draw=none] (xA) to[out=40,in=140] (zA);
		   \draw [zParticleF] (yA) -- (zA);
		   \draw [zParticleF] (xB) -- (yB);
		   \draw [photon] (xA) -- (xB);
		   \draw [photonRed] (yA) -- (yB);
		   \draw [photon] (zA) -- (zB);
		  \draw [fill] (xA) circle (.08);
		  \draw [fill] (yA) circle (.08);
		  \draw [fill] (zA) circle (.08);
		  \draw [fill] (xB) circle (.08);
		  \draw [fill] (yB) circle (.08);
		  \draw [fill] (zB) circle (.08);
		\end{tikzpicture}
	  $\,\,\rightarrow\,\,$
	  \begin{tikzpicture}[baseline={0},scale=1]
		\coordinate (inA) at (0.2,.6);
		\coordinate (outA) at (3.8,.6);
		\coordinate (inB) at (0.2,-.6);
		\coordinate (outB) at (3.8,-.6);
		\coordinate (xA) at (.7,.6);
		\coordinate (yA) at (2,.6);
		\coordinate (yzA) at (1.5,.6);
		\coordinate (zA) at (3.3,.6);
		\coordinate (xB) at (.7,-.6);
		\coordinate (yB) at (2,-.6);
		\coordinate (zB) at (3.3,-.6);
		 \draw [dotted] (inA) -- (outA);
		 \draw [dotted] (inB) -- (outB);
		 \draw [draw=none] (xA) to[out=40,in=140] (zA);
		 \draw [zParticleF] (xA) -- (yA);
		 \draw [zParticleF] (yA) -- (zA);
		 \draw [photon] (xA) -- (xB);
		 \draw [photon] (yA) -- (yB);
		 \draw [photon] (zA) -- (zB);
		\draw [fill] (xA) circle (.08);
		\draw [fill] (yA) circle (.08);
		\draw [fill] (zA) circle (.08);
		\draw [fill] (xB) circle (.08);
		\draw [fill] (yB) circle (.08);
		\draw [fill] (zB) circle (.08);
	  \end{tikzpicture}
	  $\,\,+\,\,$
	  \begin{tikzpicture}[baseline={0},scale=1]
		\coordinate (inA) at (0,.6);
		\coordinate (outA) at (4,.6);
		\coordinate (inB) at (0,-.6);
		\coordinate (outB) at (4,-.6);
		\coordinate (xA) at (.5,.6);
		\coordinate (y1A) at (1.4,.6);
		\coordinate (y2A) at (2.6,.6);
		\coordinate (zA) at (3.5,.6);
		\coordinate (xB) at (.5,-.6);
		\coordinate (zB) at (3.5,-.6);
		 \draw [dotted] (inA) -- (outA);
		 \draw [dotted] (inB) -- (outB);
		 \draw [draw=none] (xA) to[out=40,in=140] (zA);
		 \draw [zParticleF] (xA) -- (y1A);
		 \draw [zParticleF] (y2A) -- (zA);
		 \draw [photon] (xA) -- (xB);
		 \draw [photon] (zA) -- (zB);
		 \draw [photonRed] (y1A)to[out=-70,in=-110] (y2A);
		\draw [fill] (xA) circle (.08);
		\draw [fill] (y1A) circle (.08);
		\draw [fill] (y2A) circle (.08);
		\draw [fill] (zA) circle (.08);
		\draw [fill] (xB) circle (.08);
		\draw [fill] (zB) circle (.08);
	  \end{tikzpicture}
	  \caption{The split into potential and radiative regions~\eqref{eq:regions}.}
	  \label{fig:diagrammaticUnderstanding}
  \end{figure*} 

The split into regions~\eqref{eq:regions} can also be understood diagrammatically.
In \Fig{fig:diagrammaticUnderstanding}
we draw the two diagrams corresponding with the $J^{(\sigma_1;\sigma_2)}_{n_1,n_2,\ldots,n_7}$
and $K^{(\sigma_1;\sigma_2;\sigma_3)}_{n_1,n_2,\ldots,n_5}$ integrals
identified in the static limit.
The procedure can be interpreted as moving one of the worldline propagators between the two worldlines,
using the fact that $v_1^\mu=v_2^\mu$ to leading order in the static limit.
The only  difficulty is handling the ``active'' propagator which gives rise to two different choices:
it can either connect the two worldlines, or the same worldline twice.
These are precisely the potential and radiation regions.

\subsection{Mushroom integrals}

We analyze the mushroom integrals $K^{(\sigma_1;\sigma_2;\sigma_3)}_{n_1,n_2,\ldots,n_5}$
\eqref{eq:kIntegral} in more detail,
as these are sensitive to the $i0$ prescription of the active bulk propagator.
They are nested single-scale one-loop integrals.
Without loss of generality we choose $\sigma_1=1$, $\sigma_2=-1$ and $\sigma_3=1$,
so that $n_1$ and $n_2$ count powers of retarded and advanced worldline propagators respectively,
while $n_3$ counts powers of the retarded bulk propagator.
There is a nested tadpole integral:
\begin{align}
	\int_k\frac{\dd((k-\ell)\cdot v_1)}{(k^2+{\rm sgn}(k^0)i0)^{n_3}}&=
	\int_\mathbf{k}\frac{1}{((\ell\cdot v_1+i0)^2-\mathbf{k}^2)^{n_3}}\\
	&=
        \bigg(
        \frac{
          e^{-i\pi}
        }{
          4\pi
        }\bigg)^{\!\! \frac{D-1}{2}}
        \frac{\Gamma(n_3-\frac{D-1}2)}{\Gamma(n_3)}
	(\ell\cdot v_1+i0)^{D-1-2n_3}\,.\nn
\end{align}
The retarded massless propagator with power $n_3$ is turned into a retarded worldline propagator with power $2n_3-D+1$.
An important realization here is that we can obtain ${\rm sgn}(k^0)$ by projecting $k^\mu$
with any time-like four-vector: so in this case
${\rm sgn}(k^0)i0=(k\cdot v_1)i0=(\ell\cdot v_1)i0$ (on support of the $\dd$-function constraints).
Otherwise one may use the standard tadpole integral,
identifying $(0-i \ell\cdot v_1)=-i(\ell\cdot v_1+i 0)$ as an imaginary mass.

Integration on the active momentum $k^\mu$ thus results in additional powers of the retarded worldline propagator.
The integral now reads:
\begin{align}
	&K^{(+;-;+)}_{n_1,n_2,n_3,n_4,n_5}\\
	&=
  \left(
  \frac{e^{-i\pi}}{4\pi}
  \right)^{\!\! \frac{D-1}2}
  \frac{\Gamma(n_3-\frac{D-1}2)}{\Gamma(n_3)}
\int_\ell\frac{\dd(\ell\cdot v_2)}
	{(\ell\cdot v_1\!+\!i0)^{n_1+2n_3-D+1}(\ell\cdot v_1\!-\!i0)^{n_2}
	  (\ell^2)^{n_4}((\ell\!-\!q)^2)^{n_5}}\,.\nn
\end{align}
This integral is similar to the ones encountered at 2PM~\cite{Jakobsen:2021zvh},
the only difference being the presence of both retarded and advanced worldline propagators.
It can be integrated directly using Schwinger parametrization, and one finds
\begin{align}
  &K^{(+;-;+)}_{n_1,n_2,n_3,n_4,n_5}=
  (4\pi)^{1-D}
  \frac{\Gamma(n_3-\frac{D-1}2)}{\Gamma(n_3)}
  \frac{
    \cos(\sfrac\pi2(n_1-n_2-D+1))
  }{
    \cos(\sfrac\pi2(n_1+n_2-D+1))
  }\times\label{eq:kResult}\\
  &\qquad
  (-i)^{n_1-n_2+2n_3+2n_4+2n_5}
  \left(\frac{4}{\gamma^2-1}\right)^{\frac{n_1+n_2+2n_3-D+1}2}
  \Gamma_{n_1+n_2+2n_3-D+1,n_4,n_5}(D-1)\,.\nn
\end{align}
The factor $\Gamma_{n_1,n_2,n_3}(D)$ is given by
\begin{align}
	&\Gamma_{n_1,n_2,n_3}(D):=
	\frac{\Gamma(n_2+n_3+\sfrac{n_1}2-\sfrac{D}2)\Gamma(\sfrac{n_1}2)}
	{2\Gamma(n_1)\Gamma(n_2)\Gamma(n_3)}
	\frac{\Gamma(\sfrac{D}2-n_2-\sfrac{n_1}2)\Gamma(\sfrac{D}2-n_3-\sfrac{n_1}2)}
	{\Gamma(D-n_1-n_2-n_3)}\,,
\end{align}
and arises from the 2PM 1-loop integral ---
see e.g.~\rcite{Jakobsen:2021zvh}.

\subsection{Integral expressions}
\label{sec:integralExp}

Here we present expressions for the master integrals of the
$I^{(\sigma_1;\sigma_2;\sigma_3)}_{n_1,n_2,\ldots,n_7}$ family.
In each case we give results only up to the order in $\eps$ that we find is necessary
for producing observables.
The b-type master integrals \eqref{eq:bMasters} are given by\footnote{We
	thank Gregor K\"alin for checking the ${\cal I}_5$ integral.
}
\begin{subequations}\label{eq:bResults}
\begin{align}
	\tilde{\cal I}_1=\tilde{\cal I}'_1=\tilde{\cal I}''_1&=
	\cO(\eps^2)\,,\\
	\tilde{\cal I}_2=\tilde{\cal I}'_2=\tilde{\cal I}''_2&=
	\sfrac{\eps}2\log(x)+\cO(\eps^2)\,,\\
	\tilde{\cal I}_3=\tilde{\cal I}'_3=\tilde{\cal I}''_3&=
	-\sfrac12+\cO(\eps)\,,\\
	\tilde{\cal I}_4=\tilde{\cal I}'_4=\tilde{\cal I}''_4&=
	-\sfrac12+\cO(\eps)\,,\\
	\tilde{\cal I}_5=\tilde{\cal I}'_5=\tilde{\cal I}''_5&=
	\sfrac12+\cO(\eps)\,,\\
	\tilde{\cal I}_6=\tilde{\cal I}'_6=\tilde{\cal I}''_6&=
	-\eps\log(x)+\cO(\eps^2)\,,\\
	\tilde{\cal I}_7=\tilde{\cal I}'_7=\tilde{\cal I}''_7&=
	\cO(\eps^2)\,,\\
	\tilde{\cal I}_8&=
	-\sfrac12+\sfrac{\eps}2\log(x)+\cO(\eps^2)\,,\\
	\tilde{\cal I}'_8&=
	-\sfrac12+\sfrac{\eps}2\log(x)+\cO(\eps^2)\,,\\
	\tilde{\cal I}''_8&=1+\sfrac{\eps}2\log(x)+\cO(\eps^2)\,,
\end{align}
\end{subequations}
with the normalization
$\tilde{\cal I}_i:=(4\pi)^{2-2\epsilon}e^{2\gamma_{\rm E}\epsilon}{\cal I}_i$.
As expected, all of these are real in the physical region $0<x<1$:
the first seven are given by the real part of the corresponding integrals
with Feynman $i0$ prescription~\cite{Bjerrum-Bohr:2021vuf},
which carry both real and imaginary parts
when evaluated in the physical region.
The v-type integrals are
\begin{subequations}\label{eq:vResults}
	\begin{align}
	  \tilde{\cal I}_9=\tilde{\cal I}''_9&=
	  \cO(\eps^0)\,,\\
	  \tilde{\cal I}_{10}=\tilde{\cal I}''_{10}&=
	  +\sfrac{i\pi}{2}
          -i\pi\log(2x)\eps+\cO(\eps^2)
          \,,\\
	  \tilde{\cal I}_{11}=\tilde{\cal I}'_{11}=\tilde{\cal I}''_{11}&=
	  -\sfrac{i\pi}{6}
          -\sfrac{i\pi}3(\log(8x)-\log(1-x^2))\eps
          +\cO(\eps^2)\,,\\
	  \tilde{\cal I}_{12}=\tilde{\cal I}'_{12}=\tilde{\cal I}''_{12}&=
	  +\sfrac{i\pi}{4}
          -\sfrac{i\pi}2(\log(2x)+\log(1-x)-3\log(1+x))\eps
          +\cO(\eps^2)\,,\\
	  \tilde{\cal I}_{13}=\tilde{\cal I}'_{13}=\tilde{\cal I}''_{13}&=
	  -\sfrac{i\pi}2(\log(4x)-2\log(1+x))\eps+\cO(\eps^2)\,,\\
	  \tilde{\cal I}'_9
          =\tilde{\cal I}''_{14}
          &=\cO(\eps^0)\,,\\
	  \tilde{\cal I}'_{10}=\tilde{\cal I}''_{15}
          &=
	  -\sfrac{i\pi}{2}-i\pi\log(\sfrac{x}{2})\eps+\cO(\eps^2)\,,
	\end{align}
\end{subequations}
and as expected are purely imaginary.
We see that ${\cal I}_i={\cal I}'_i$ for $11\leq i\leq13$,
which is due to all three of these integrals vanishing in the potential region: 
this gives rise to an additional ``hidden'' symmetry.
When assembling physical observables,
$\log(1-x)$ generically cancels between the integrals ${\cal I}_{11}$ and ${\cal I}_{12}$.
This gives rise to the two commonly occurring combinations:
\begin{subequations}
\begin{align}
	{\rm arccosh}\gamma&=-\log(x)\,,\\
	\log\bigg(\frac{\gamma+1}2\bigg)&=2\log(1+x)-\log(4x)\,,
\end{align}
\end{subequations}
which we will use to present our results in \sec{sec:tidal} in a compact manner.

\section{Radiation-reacted tidal effects}
\label{sec:tidal}

With the technology for performing retarded integrals at 3PM order now established,
let us use it to derive gravitational observables from a scattering event.
We focus on the inclusion of tidal effects in the point-particle action:
\begin{align}\label{eq:action}
	S^{(i)}_{\rm pp}+S^{(i)}_{\rm tidal}=m_i \int\!\d\tau \left[-\frac{1}2g_{\mu\nu}\dot x_i^{\mu}\dot x_i^{\nu}
	+c_{E^2}^{(i)} E_{\mu \nu}^{(i)} E^{(i) \mu \nu}+c_{B^2}^{(i)} B_{\mu \nu}^{(i)} B^{(i) \mu \nu}\right]\,,
\end{align}
where $c^{(i)}_{E^2}$ and $c^{(i)}_{B^2}$ are the quadrupole Love numbers and
$E_{\mu \nu}^{(i)}:= R_{\mu \alpha \nu \beta} \dot{x}_i^{ \alpha} \dot{x}_i^{ \beta}$,
$B_{\mu \nu}^{(i)}:= R^{*}_{\mu \alpha \nu \beta} \dot{x}_i^{ \alpha} \dot{x}_i^{ \beta}$
are the electromagnetic curvature tensors with
$R^*_{\mu \alpha \nu \beta}:= \sfrac{1}{2} \epsilon_{\nu \beta \rho \sigma} {R_{\mu \alpha}}^{\rho \sigma}$.
As explained in \rcites{Mogull:2020sak,Jakobsen:2021zvh},
we read off Feynman rules by expressing the constituent fields in Fourier space:
$h_{\mu \nu}(x)=\int_k e^{-ik \cdot x} h_{\mu \nu}(k)$ and
$z_i^\mu (\tau) = \int_{\omega} e^{-i \omega \tau} z_i^{\mu}(\omega)$.
Interaction vertices stemming from the point-particle action involve only a single graviton:
\begin{align}
	&\begin{tikzpicture}[baseline={(current bounding box.center)}]
	  \coordinate (in) at (-1,0);
	  \coordinate (out1) at (1,0);
	  \coordinate (out2) at (1,0.5);
	  \coordinate (out3) at (1,0.9);
	  \coordinate (x) at (0,0);
	  \node (k) at (0,-1.3) {$h_{\mu\nu}(k)$};
	  \draw (out1) node [right] {$z^{\rho_1}(\omega_1)$};
	  \draw (out2) node [right] {$\!\!\!\vdots$};
	  \draw (out3) node [right] {$z^{\rho_n}(\omega_n)$};
	  \draw [dotted] (in) -- (x);
	  \draw [zUndirected] (x) -- (out1);
	  \draw [zUndirected] (x) to[out=30,in=180] (out3);
	  \draw [graviton] (x) -- (k);
	  \draw [fill] (x) circle (.08);
	  \end{tikzpicture}=
	i^{n-1}m\kappa\,
	e^{ik\cdot b}\dd\bigg(k\cdot v+\sum_{i=1}^n\omega_i\bigg)\times\\[-20pt]
	&\qquad\qquad\,\,
	\left(\frac12\left(\prod_{i=1}^nk_{\rho_i}\!\right)\!v^\mu v^\nu+
	\sum_{i=1}^n\omega_i\left(\prod_{j\neq i}^nk_{\rho_j}\right)\!
	v^{(\mu}\delta^{\nu)}_{\rho_i}+
	\sum_{i<j}^n\omega_i\omega_j\left(\prod_{l\neq i,j}^nk_{\rho_l}\right)
	\delta^{(\mu}_{\rho_i}\delta^{\nu)}_{\rho_j}\right)\!,\nn
\end{align}
while those arising from the tidal corrections involve a minimum of two gravitons:
\begin{subequations}\label{eq:tidalVertices}
\begin{align}
	\begin{tikzpicture}[baseline={(current bounding box.center)}]
	  \coordinate (in) at (-1,0);
	  \coordinate (out1) at (1,0);
	  \coordinate (out2) at (1,0.5);
	  \coordinate (out3) at (1,0.9);
	  \coordinate (x) at (0,0);
	  \node (k1) at (-.9,-1.3) {$h_{\mu_1\nu_1}(k_1)$};
	  \node (k2) at (.9,-1.3) {$h_{\mu_2\nu_2}(k_2)$};
	  \draw (out1) node [right] {$z^{\rho_1}(\omega_1)$};
	  \draw (out2) node [right] {$\!\!\!\vdots$};
	  \draw (out3) node [right] {$z^{\rho_n}(\omega_n)$};
	  \draw [dotted] (in) -- (x);
	  \draw [zUndirected] (x) -- (out1);
	  \draw [zUndirected] (x) to[out=30,in=180] (out3);
	  \draw [graviton] (x) -- (k1);
	  \draw [graviton] (x) -- (k2);
	  \draw [fill] (x) +(-.12,-.12) rectangle ++(.12,.12);
	  \end{tikzpicture}&=2 i^{n+1}m\kappa^2 e^{iq\cdot b} \dd\bigg(q \cdot v+\sum_{i=1}^n\omega_i\bigg) \times \\[-30pt]
	  &\left[c_{E^2}^{(i)}{R^{(1)}}_{\epsilon \alpha \delta \beta,\mu_1 \nu_1}{R^{(1)}}_\gamma {{}^\alpha}_\xi {{}^{\beta}}_{,\mu_2 \nu_2}\right. \nn \\
	  &\qquad\left.\quad+c_{B^2}^{(i)}{{R^*}^{(1)}}_{\epsilon \alpha \delta \beta,\mu_1 \nu_1}{{R^*}}^{(1)}_\gamma {{}^\alpha}_\xi {{}^{\beta}}_{,\mu_2 \nu_2} \right] {T^{\epsilon \delta \gamma \xi}}_{\rho_1...\rho_n} \,,\nn \\
	  \begin{tikzpicture}[baseline={(current bounding box.center)}]
		\coordinate (in) at (-1,0);
		\coordinate (out1) at (1,0);
		\coordinate (out2) at (1,0.5);
		\coordinate (out3) at (1,0.9);
		\coordinate (x) at (0,0);
		\node (k1) at (-1.7,-1.3) {$h_{\mu_1\nu_1}(k_1)$};
		\node (k2) at (0,-1.7) {$h_{\mu_2\nu_2}(k_2)$};
		\node (k3) at (1.7,-1.3) {$h_{\mu_3\nu_3}(k_3)$};
		\draw (out1) node [right] {$z^{\rho_1}(\omega_1)$};
		\draw (out2) node [right] {$\!\!\!\vdots$};
		\draw (out3) node [right] {$z^{\rho_n}(\omega_n)$};
		\draw [dotted] (in) -- (x);
		\draw [zUndirected] (x) -- (out1);
		\draw [zUndirected] (x) to[out=30,in=180] (out3);
		\draw [graviton] (x) -- (k1);
		\draw [graviton] (x) -- (k2);
		\draw [graviton] (x) -- (k3);
		\draw [fill] (x) +(-.12,-.12) rectangle ++(.12,.12);
		\end{tikzpicture}=&i^{n+1}m\kappa^3 e^{i q\cdot b} \dd\bigg(q\cdot v+\sum_{i=1}^n\omega_i\bigg) \times \\[-30pt]
		&\sum_{\sigma \in S_3}\left[c_{E^2}^{(i)}\left({R^{(2)}}_{\epsilon \alpha \delta \beta,\mu_{\sigma_1} \nu_{\sigma_1}\mu_{\sigma_2} \nu_{\sigma_2}}{R^{(1)}}_\gamma {{}^\alpha}_\xi {{}^{\beta}}_{,\mu_{\sigma_3} \nu_{\sigma_3}}\right. \right. \nn \\
		&\left. \left.  +{R^{(1)}}_{\epsilon \alpha \delta \beta,\mu_{\sigma_1} \nu_{\sigma_1}}{R^{(2)}}_\gamma {{}^\alpha}_\xi {{}^{\beta}}_{,\mu_{\sigma_2} \nu_{\sigma_2}\mu_{\sigma_3} \nu_{\sigma_3}}\right)\right. \nn \\
		& \left. +c_{B^2}^{(i)}\left({{R^*}^{(2)}}_{\epsilon \alpha \delta \beta,\mu_{\sigma_1} \nu_{\sigma_1}\mu_{\sigma_2} \nu_{\sigma_2}}{{R^*}}^{(1)}_\gamma {{}^\alpha}_\xi {{}^{\beta}}_{,\mu_{\sigma_3} \nu_{\sigma_3}}\right.\right.\nn \\
		&  \left.\left.+{{R^*}^{(1)}}_{\epsilon \alpha \delta \beta,\mu_{\sigma_1} \nu_{\sigma_1}}{{R^*}}^{(2)}_\gamma {{}^\alpha}_\xi {{}^{\beta}}_{,\mu_{\sigma_2} \nu_{\sigma_2}\mu_{\sigma_3} \nu_{\sigma_3}}\right) \right]{T^ {\epsilon \delta \gamma \xi}}_{\rho_1...\rho_n}\,.\nn
\end{align}
\end{subequations}
Both expressions include
\begin{align}
	&{T^{\epsilon \sigma \gamma \xi}}_{\rho_1...\rho_n}=\left(\prod_{i=1}^n q_{\rho_i}\!\right)\!v^\epsilon v^\sigma v^\gamma v^\xi+
	4\sum_{i=1}^n\omega_i\left(\prod_{j\neq i}^nq_{\rho_j}\right)\!
	v^{(\epsilon}v^\sigma v^\gamma \delta^{\xi)}_{\rho_i} \nn \\
	&\qquad+ 24\sum_{i<j}^n\omega_i\omega_j\left(\prod_{k\neq i,j}^n q_{\rho_k}\right)
	v^{(\epsilon}v^\sigma \delta^{\gamma}_{\rho_i}\delta^{\xi)}_{\rho_j}+ 24\sum_{i<j<k}^n\omega_i\omega_j \omega_k\left(\prod_{l\neq i,j,k}^n q_{\rho_l} \right)
	v^{(\epsilon}\delta^\sigma_{\rho_i} \delta^{\gamma}_{\rho_j}\delta^{\xi)}_{\rho_k} \nn \\
	&\qquad+24\sum_{i<j<k<l}^n \omega_i\omega_j \omega_k \omega_l \left(\prod_{m\neq i,j,k,l}^n q_{\rho_m} \right)
	\delta^{(\epsilon}_{\rho_i}\delta^\sigma_{\rho_j} \delta^{\gamma}_{\rho_k}\delta^{\xi)}_{\rho_l}\,,
\end{align}
where $q^\mu =\sum_{i}k_i^\mu$ is the total momentum of all emitted gravitons and
\begin{align}
{R^{(n)}}_{\alpha \beta \rho \sigma, \mu_1 \nu_1...\mu_n \nu_n}:=
\dfrac{\delta^n {R^{(n)}}_{\alpha \beta \rho \sigma}}{\delta h^{\mu_1 \nu_1}\cdots\delta h^{\mu_n \nu_n}}.
\end{align}
${R^{(n)}}_{\alpha \beta \rho \sigma}$ is given by the $n$'th order of $\kappa=\sqrt{32 \pi G}$
in a PM expansion of the curvature tensor where we replace the graviton field
by its Fourier transform $h_{\mu \nu}(x)\rightarrow h_{\mu \nu}(-k)$,
and similarly for the dual of the curvature tensor.
The complete set of Feynman rules also includes bulk interactions arising from the
$D$-dimensional Einstein-Hilbert action and gauge-fixing term:
\begin{align}\label{eq:SEH}
	S_{\rm EH}=-\frac2{\kappa^2}\int\!\d^Dx\sqrt{-g}\,R\,, &&
	S_{\rm gf}=
	\int\!\d^Dx\big(\partial_\nu h^{\mu\nu}-\sfrac12\partial^\mu{h^\nu}_\nu\big)^2\,,
\end{align}
where the gauge-fixing constraint is
$\partial_\nu h^{\mu\nu}=\sfrac12\partial^\mu{h^\nu}_\nu$.
Expressions for the retarded graviton and worldline propagators were provided in
\eqns{eq:gravProp}{eq:Propagators}.

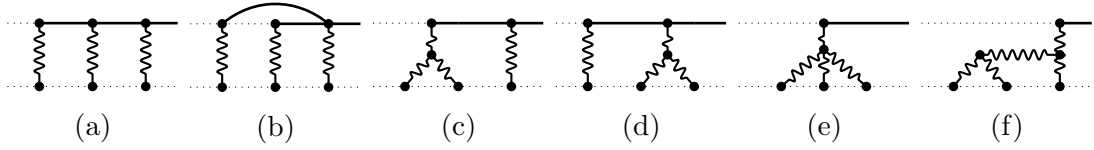
\begin{figure*}[t]
	\centering
	\begin{subfigure}{.15\textwidth}
	  \centering
	  \begin{tikzpicture}[baseline={(current bounding box.center)},scale=.7]
			\coordinate (inA) at (0.4,.6);
			\coordinate (outA) at (3.6,.6);
			\coordinate (inB) at (0.4,-.6);
			\coordinate (outB) at (3.6,-.6);
			\coordinate (xA) at (1,.6);
			\coordinate (xyA) at (1.5,.6);
			\coordinate (yA) at (2,.6);
			\coordinate (yzA) at (1.5,.6);
			\coordinate (zA) at (3,.6);
			\coordinate (xB) at (1,-.6);
			\coordinate (yB) at (2,-.6);
			\coordinate (zB) at (3,-.6);
			\draw [fill] (xA) circle (.08);
			\draw [fill] (yA) circle (.08);
			\draw [fill] (zA) circle (.08);
			\draw [fill] (xB) circle (.08);
			\draw [fill] (yB) circle (.08);
			\draw [fill] (zB) circle (.08);
			\draw [dotted] (inA) -- (outA);
			\draw [dotted] (inB) -- (outB);
			\draw [zParticleF] (zA) -- (outA);
			\draw [draw=none] (xA) to[out=40,in=140] (zA);
			\draw [zParticleF] (xA) -- (yA);
			\draw [zParticleF] (yA) -- (zA);
			\draw [photon] (xA) -- (xB);
			\draw [photon] (yA) -- (yB);
			\draw [photon] (zA) -- (zB);
		\end{tikzpicture}
	  \caption{}
	\end{subfigure}
	\begin{subfigure}{.15\textwidth}
	  \centering
	  \begin{tikzpicture}[baseline={(current bounding box.center)},scale=.7]
			\coordinate (inA) at (0.4,.6);
			\coordinate (outA) at (3.6,.6);
			\coordinate (inB) at (0.4,-.6);
			\coordinate (outB) at (3.6,-.6);
			\coordinate (xA) at (1,.6);
			\coordinate (xyA) at (1.5,.6);
			\coordinate (yA) at (2,.6);
			\coordinate (yzA) at (1.5,.6);
			\coordinate (zA) at (3,.6);
			\coordinate (xB) at (1,-.6);
			\coordinate (yB) at (2,-.6);
			\coordinate (zB) at (3,-.6);
			\draw [fill] (xA) circle (.08);
			\draw [fill] (yA) circle (.08);
			\draw [fill] (zA) circle (.08);
			\draw [fill] (xB) circle (.08);
			\draw [fill] (yB) circle (.08);
			\draw [fill] (zB) circle (.08);
			\draw [dotted] (inA) -- (outA);
			\draw [dotted] (inB) -- (outB);
			\draw [zParticleF] (zA) -- (outA);
			\draw [zParticleF] (xA) to[out=40,in=140] (zA);
			\draw [zParticleF] (yA) -- (zA);
			\draw [photon] (xA) -- (xB);
			\draw [photon] (yA) -- (yB);
			\draw [photon] (zA) -- (zB);
		\end{tikzpicture}
	  \caption{}
	\end{subfigure}
	\begin{subfigure}{.15\textwidth}
	  \centering
	  \begin{tikzpicture}[baseline={(current bounding box.center)},scale=.7]
			\coordinate (inA) at (0.4,.6);
			\coordinate (outA) at (3.6,.6);
			\coordinate (inB) at (0.4,-.6);
			\coordinate (outB) at (3.6,-.6);
			\coordinate (xA) at (1,.6);
			\coordinate (xyA) at (1.5,.6);
			\coordinate (xy0) at (1.5,0);
			\coordinate (yA) at (2,.6);
			\coordinate (yzA) at (1.5,.6);
			\coordinate (zA) at (3,.6);
			\coordinate (xB) at (1,-.6);
			\coordinate (yB) at (2,-.6);
			\coordinate (zB) at (3,-.6);
			\draw [fill] (xyA) circle (.08);
			\draw [fill] (xy0) circle (.08);
			\draw [fill] (zA) circle (.08);
			\draw [fill] (xB) circle (.08);
			\draw [fill] (yB) circle (.08);
			\draw [fill] (zB) circle (.08);
			\draw [dotted] (inA) -- (outA);
			\draw [dotted] (inB) -- (outB);
			\draw [zParticleF] (zA) -- (outA);
			\draw [draw=none] (xA) to[out=40,in=140] (zA);
			\draw [zParticleF] (xyA) -- (yA);
			\draw [zParticleF] (yA) -- (zA);
			\draw [photon] (xy0) -- (xyA);
			\draw [photon] (xy0) -- (xB);
			\draw [photon] (xy0) -- (yB);
			\draw [photon] (zA) -- (zB);
		\end{tikzpicture}
	  \caption{}
	\end{subfigure}
	\begin{subfigure}{.15\textwidth}
	  \centering
	  \begin{tikzpicture}[baseline={(current bounding box.center)},scale=.7]
			\coordinate (inA) at (0.4,.6);
			\coordinate (outA) at (3.6,.6);
			\coordinate (inB) at (0.4,-.6);
			\coordinate (outB) at (3.6,-.6);
			\coordinate (xA) at (1,.6);
			\coordinate (xyA) at (1.5,.6);
			\coordinate (yA) at (2,.6);
			\coordinate (yzA) at (2.5,.6);
			\coordinate (yz0) at (2.5,0);
			\coordinate (zA) at (3,.6);
			\coordinate (xB) at (1,-.6);
			\coordinate (yB) at (2,-.6);
			\coordinate (zB) at (3,-.6);
			\draw [fill] (xA) circle (.08);
			\draw [fill] (yzA) circle (.08);
			\draw [fill] (yz0) circle (.08);
			\draw [fill] (xB) circle (.08);
			\draw [fill] (yB) circle (.08);
			\draw [fill] (zB) circle (.08);
			\draw [dotted] (inA) -- (outA);
			\draw [dotted] (inB) -- (outB);
			\draw [zParticleF] (zA) -- (outA);
			\draw [draw=none] (xA) to[out=40,in=140] (zA);
			\draw [zParticleF] (xA) -- (yA);
			\draw [zParticleF] (yA) -- (zA);
			\draw [photon] (xA) -- (xB);
			\draw [photon] (yz0) -- (yzA);
			\draw [photon] (yz0) -- (yB);
			\draw [photon] (yz0) -- (zB);
		\end{tikzpicture}
	  \caption{}
	\end{subfigure}
	\begin{subfigure}{.15\textwidth}
	  \centering
	  \begin{tikzpicture}[baseline={(current bounding box.center)},scale=.7]
			\coordinate (inA) at (0.4,.6);
			\coordinate (outA) at (3.6,.6);
			\coordinate (inB) at (0.4,-.6);
			\coordinate (outB) at (3.6,-.6);
			\coordinate (xA) at (1,.6);
			\coordinate (xyA) at (1.5,.6);
			\coordinate (yA) at (2,.6);
			\coordinate (y0) at (2,0.1);
			\coordinate (yzA) at (1.5,.6);
			\coordinate (zA) at (3,.6);
			\coordinate (xB) at (1.2,-.6);
			\coordinate (yB) at (2,-.6);
			\coordinate (zB) at (2.8,-.6);
			\draw [fill] (yA) circle (.08);
			\draw [fill] (xB) circle (.08);
			\draw [fill] (yB) circle (.08);
			\draw [fill] (zB) circle (.08);
			\draw [fill] (y0) circle (.08);
			\draw [dotted] (inA) -- (outA);
			\draw [dotted] (inB) -- (outB);
			\draw [zParticleF] (zA) -- (outA);
			\draw [draw=none] (xA) to[out=40,in=140] (zA);
			\draw [zParticleF] (yA) -- (zA);
			\draw [photon] (y0) -- (yA);
			\draw [photon] (y0) -- (xB);
			\draw [photon] (y0) -- (yB);
			\draw [photon] (y0) -- (zB);
		\end{tikzpicture}
	  \caption{}
	\end{subfigure}
	\begin{subfigure}{.15\textwidth}
	  \centering
	  \begin{tikzpicture}[baseline={(current bounding box.center)},scale=.7]
			\coordinate (inA) at (0.4,.6);
			\coordinate (outA) at (3.6,.6);
			\coordinate (inB) at (0.4,-.6);
			\coordinate (outB) at (3.6,-.6);
			\coordinate (xA) at (1,.6);
			\coordinate (xyA) at (1.5,.6);
			\coordinate (xy0) at (1.5,0);
			\coordinate (z0) at (3,0);
			\coordinate (yA) at (2,.6);
			\coordinate (yzA) at (1.5,.6);
			\coordinate (zA) at (3,.6);
			\coordinate (xB) at (1,-.6);
			\coordinate (yB) at (2,-.6);
			\coordinate (zB) at (3,-.6);
			\draw [fill] (xy0) circle (.08);
			\draw [fill] (zA) circle (.08);
			\draw [fill] (xB) circle (.08);
			\draw [fill] (yB) circle (.08);
			\draw [fill] (zB) circle (.08);
			\draw [fill] (z0) circle (.08);
			\draw [dotted] (inA) -- (outA);
			\draw [dotted] (inB) -- (outB);
			\draw [zParticleF] (zA) -- (outA);
			\draw [photon] (xy0) -- (z0);
			\draw [draw=none] (xA) to[out=40,in=140] (zA);
			\draw [photon] (xy0) -- (xB);
			\draw [photon] (xy0) -- (yB);
			\draw [photon] (zA) -- (zB);
		\end{tikzpicture}
	  \caption{}
	\end{subfigure}
	\caption{\small The six diagrams contributing to the $m_1m_2^3$ component of $\Delta p_1^{(3)\mu}$
	in the absence of tidal corrections.
	The upper worldline is one continuous fluctuation and hence we have test-body motion.
	}
	\label{fig:testBodyDiags}
  \end{figure*}

\begin{figure*}[t]
  \centering
  \begin{subfigure}{.2\textwidth}
    \centering
    \begin{tikzpicture}[baseline={(current bounding box.center)},scale=.7]
  		\coordinate (inA) at (0.4,.6);
  		\coordinate (outA) at (3.6,.6);
  		\coordinate (inB) at (0.4,-.6);
  		\coordinate (outB) at (3.6,-.6);
  		\coordinate (xA) at (1,.6);
  		\coordinate (xyA) at (1.5,.6);
  		\coordinate (yA) at (2,.6);
  		\coordinate (yzA) at (1.5,.6);
  		\coordinate (zA) at (3,.6);
  		\coordinate (xB) at (1,-.6);
  		\coordinate (yB) at (2,-.6);
  		\coordinate (zB) at (3,-.6);
  		\draw [fill] (xyA) +(-.12,-.12) rectangle ++(.12,.12);
  		\draw [fill] (zA) circle (.08);
  		\draw [fill] (xB) circle (.08);
  		\draw [fill] (yB) circle (.08);
  		\draw [fill] (zB) circle (.08);
  		\draw [dotted] (inA) -- (outA);
  		\draw [dotted] (inB) -- (outB);
  		\draw [zParticleF] (zA) -- (outA);
  		\draw [draw=none] (xA) to[out=40,in=140] (zA);
  		\draw [zParticleF] (xyA) -- (yA);
  		\draw [zParticleF] (yA) -- (zA);
  		\draw [photon] (xyA) -- (xB);
  		\draw [photon] (xyA) -- (yB);
  		\draw [photon] (zA) -- (zB);
  	\end{tikzpicture}
	\caption{}
  \end{subfigure}
  \begin{subfigure}{.2\textwidth}
    \centering
    \begin{tikzpicture}[baseline={(current bounding box.center)},scale=.7]
  		\coordinate (inA) at (0.4,.6);
  		\coordinate (outA) at (3.6,.6);
  		\coordinate (inB) at (0.4,-.6);
  		\coordinate (outB) at (3.6,-.6);
  		\coordinate (xA) at (1,.6);
  		\coordinate (xyA) at (1.5,.6);
  		\coordinate (yA) at (2,.6);
  		\coordinate (yzA) at (2.5,.6);
  		\coordinate (zA) at (3,.6);
  		\coordinate (xB) at (1,-.6);
  		\coordinate (yB) at (2,-.6);
  		\coordinate (zB) at (3,-.6);
  		\draw [fill] (xA) circle (.08);
  		\draw [fill] (yzA) +(-.12,-.12) rectangle ++(.12,.12);
  		\draw [fill] (xB) circle (.08);
  		\draw [fill] (yB) circle (.08);
  		\draw [fill] (zB) circle (.08);
  		\draw [dotted] (inA) -- (outA);
  		\draw [dotted] (inB) -- (outB);
  		\draw [zParticleF] (zA) -- (outA);
  		\draw [draw=none] (xA) to[out=40,in=140] (zA);
  		\draw [zParticleF] (xA) -- (yA);
  		\draw [zParticleF] (yA) -- (zA);
  		\draw [photon] (xA) -- (xB);
  		\draw [photon] (yzA) -- (yB);
  		\draw [photon] (yzA) -- (zB);
  	\end{tikzpicture}
	\caption{}
  \end{subfigure}
  \begin{subfigure}{.2\textwidth}
    \centering
    \begin{tikzpicture}[baseline={(current bounding box.center)},scale=.7]
  		\coordinate (inA) at (0.4,.6);
  		\coordinate (outA) at (3.6,.6);
  		\coordinate (inB) at (0.4,-.6);
  		\coordinate (outB) at (3.6,-.6);
  		\coordinate (xA) at (1,.6);
  		\coordinate (xyA) at (1.5,.6);
  		\coordinate (yA) at (2,.6);
  		\coordinate (yzA) at (1.5,.6);
  		\coordinate (zA) at (3,.6);
  		\coordinate (xB) at (1,-.6);
  		\coordinate (yB) at (2,-.6);
  		\coordinate (zB) at (3,-.6);
  		\draw [fill] (yA)+(-.12,-.12) rectangle ++(.12,.12);
  		\draw [fill] (xB) circle (.08);
  		\draw [fill] (yB) circle (.08);
  		\draw [fill] (zB) circle (.08);
  		\draw [dotted] (inA) -- (outA);
  		\draw [dotted] (inB) -- (outB);
  		\draw [zParticleF] (zA) -- (outA);
  		\draw [draw=none] (xA) to[out=40,in=140] (zA);
  		\draw [zParticleF] (yA) -- (zA);
  		\draw [photon] (yA) -- (xB);
  		\draw [photon] (yA) -- (yB);
  		\draw [photon] (yA) -- (zB);
  	\end{tikzpicture}
	\caption{}
  \end{subfigure}
  \begin{subfigure}{.2\textwidth}
    \centering
    \begin{tikzpicture}[baseline={(current bounding box.center)},scale=.7]
  		\coordinate (inA) at (0.4,.6);
  		\coordinate (outA) at (3.6,.6);
  		\coordinate (inB) at (0.4,-.6);
  		\coordinate (outB) at (3.6,-.6);
  		\coordinate (xA) at (1,.6);
  		\coordinate (xyA) at (1.5,.6);
  		\coordinate (xy0) at (1.5,0);
  		\coordinate (z0) at (3,0);
  		\coordinate (yA) at (2,.6);
  		\coordinate (yzA) at (1.5,.6);
  		\coordinate (zA) at (3,.6);
  		\coordinate (xB) at (1,-.6);
  		\coordinate (yB) at (2,-.6);
  		\coordinate (zB) at (3,-.6);
  		\draw [fill] (xy0) circle (.08);
  		\draw [fill] (zA) +(-.12,-.12) rectangle ++(.12,.12);
  		\draw [fill] (xB) circle (.08);
  		\draw [fill] (yB) circle (.08);
  		\draw [fill] (zB) circle (.08);
  		\draw [dotted] (inA) -- (outA);
  		\draw [dotted] (inB) -- (outB);
  		\draw [zParticleF] (zA) -- (outA);
  		\draw [photon] (xy0)to[out=0,in=-140] (zA);
  		\draw [draw=none] (xA) to[out=40,in=140] (zA);
  		\draw [photon] (xy0) -- (xB);
  		\draw [photon] (xy0) -- (yB);
  		\draw [photon] (zA) -- (zB);
  	\end{tikzpicture}
	\caption{}
  \end{subfigure}
    \caption{\small The four types of diagrams contributing to the test-body $m_1m_2^3$ components
  	of $\Delta p_1^{(3)\mu}$ linear in tidal coefficients.}
  \label{fig:testBodyDiagsTid}
\end{figure*}
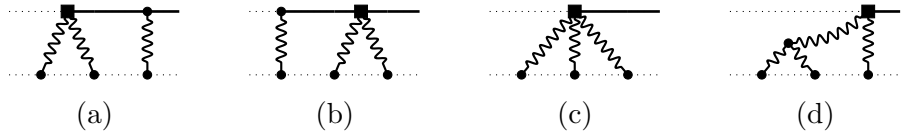

\begin{figure*}[t!]
	\centering
	\begin{subfigure}{0.13\textwidth}
	  \centering
	  \begin{tikzpicture}[baseline={(current bounding box.center)},scale=.6]
		\coordinate (inA) at (0.4,.6);
		\coordinate (outA) at (3.6,.6);
		\coordinate (inB) at (0.4,-.6);
		\coordinate (outB) at (3.6,-.6);
		\coordinate (xA) at (1,.6);
		\coordinate (xyA) at (1.5,.6);
		\coordinate (yA) at (2,.6);
		\coordinate (yzA) at (1.5,.6);
		\coordinate (zA) at (3,.6);
		\coordinate (xB) at (1,-.6);
		\coordinate (yB) at (2,-.6);
		\coordinate (zB) at (3,-.6);
		\draw [dotted] (inA) -- (outA);
		\draw [dotted] (inB) -- (outB);
		\draw [zParticleF] (zA) -- (outA);
		\draw [draw=none] (xA) to[out=40,in=140] (zA);
		\draw [zParticleF] (xB) -- (yB);
		\draw [zParticleF] (yA) -- (zA);
		\draw [photon] (xA) -- (xB);
		\draw [photonRed] (yA) -- (yB);
		\draw [photon] (zA) -- (zB);
		\draw [fill] (xA) circle (.08);
		\draw [fill] (yA) circle (.08);
		\draw [fill] (zA) circle (.08);
		\draw [fill] (xB) circle (.08);
		\draw [fill] (yB) circle (.08);
		\draw [fill] (zB) circle (.08);
	  \end{tikzpicture}
	  \caption{}
	\end{subfigure}
	\begin{subfigure}{0.13\textwidth}
	  \centering
	  \begin{tikzpicture}[baseline={(current bounding box.center)},scale=.6]
		\coordinate (inA) at (0.4,.6);
		\coordinate (outA) at (3.6,.6);
		\coordinate (inB) at (0.4,-.6);
		\coordinate (outB) at (3.6,-.6);
		\coordinate (xA) at (1,.6);
		\coordinate (xyA) at (1.5,.6);
		\coordinate (yA) at (2,.6);
		\coordinate (yzA) at (1.5,.6);
		\coordinate (zA) at (3,.6);
		\coordinate (xB) at (1,-.6);
		\coordinate (yB) at (2,-.6);
		\coordinate (zB) at (3,-.6);
		 \draw [dotted] (inA) -- (outA);
		 \draw [dotted] (inB) -- (outB);
		 \draw [zParticleF] (zA) -- (outA);
		 \draw [draw=none] (xA) to[out=40,in=140] (zA);
		 \draw [zParticleF] (xA) -- (yA);
		 \draw [zParticleF] (yB) -- (zB);
		 \draw [photon] (xA) -- (xB);
		 \draw [photonRed] (yA) -- (yB);
		 \draw [photon] (zA) -- (zB);
		\draw [fill] (xA) circle (.08);
		\draw [fill] (yA) circle (.08);
		\draw [fill] (zA) circle (.08);
		\draw [fill] (xB) circle (.08);
		\draw [fill] (yB) circle (.08);
		\draw [fill] (zB) circle (.08);
	  \end{tikzpicture}
	  \caption{}
	\end{subfigure}
	\begin{subfigure}{0.13\textwidth}
	  \centering
	  \begin{tikzpicture}[baseline={(current bounding box.center)},scale=.6]
		\coordinate (inA) at (0.4,.6);
		\coordinate (outA) at (3.6,.6);
		\coordinate (inB) at (0.4,-.6);
		\coordinate (outB) at (3.6,-.6);
		\coordinate (xA) at (1,.6);
		\coordinate (xyA) at (1.5,.6);
		\coordinate (yA) at (2,.6);
		\coordinate (yzA) at (1.5,.6);
		\coordinate (zA) at (3,.6);
		\coordinate (xB) at (1,-.6);
		\coordinate (yB) at (2,-.6);
		\coordinate (zB) at (3,-.6);
		\draw [dotted] (inA) -- (outA);
		\draw [dotted] (inB) -- (outB);
		\draw [zParticleF] (zA) -- (outA);
		\draw [draw=none] (xA) to[out=40,in=140] (zA);
		\draw [zParticleF] (xA) to[out=40,in=140] (zA);
		\draw [zParticleF] (yB) -- (zB);
		\draw [photon] (xA) -- (xB);
		\draw [photon] (yA) -- (yB);
		\draw [photonRed] (zA) -- (zB);
		\draw [fill] (xA) circle (.08);
		\draw [fill] (yA) circle (.08);
		\draw [fill] (zA) circle (.08);
		\draw [fill] (xB) circle (.08);
		\draw [fill] (yB) circle (.08);
		\draw [fill] (zB) circle (.08);
	  \end{tikzpicture}
	  \caption{}
	\end{subfigure}
	\begin{subfigure}{0.13\textwidth}
	  \centering
	  \begin{tikzpicture}[baseline={(current bounding box.center)},scale=.6]
		\coordinate (inA) at (0.4,.6);
		\coordinate (outA) at (3.6,.6);
		\coordinate (inB) at (0.4,-.6);
		\coordinate (outB) at (3.6,-.6);
		\coordinate (xA) at (1,.6);
		\coordinate (xyA) at (1.5,.6);
		\coordinate (yA) at (2,.6);
		\coordinate (yzB) at (2.5,-.6);
		\coordinate (yz0) at (2.5,0);
		\coordinate (zA) at (3,.6);
		\coordinate (xB) at (1,-.6);
		\coordinate (yB) at (2,-.6);
		\coordinate (zB) at (3,-.6);
		\draw [dotted] (inA) -- (outA);
		\draw [dotted] (inB) -- (outB);
		\draw [zParticleF] (zA) -- (outA);
		\draw [draw=none] (xA) to[out=40,in=140] (zA);
		\draw [zParticleF] (xA) -- (yA);
		\draw [photon] (xA) -- (xB);
		\draw [photon] (yzB) -- (yz0);
		\draw [photonRed] (yA) -- (yz0);
		\draw [photon] (zA) -- (yz0);
		\draw [fill] (xA) circle (.08);
		\draw [fill] (yA) circle (.08);
		\draw [fill] (zA) circle (.08);
		\draw [fill] (xB) circle (.08);
		\draw [fill] (yzB) circle (.08);
		\draw [fill] (yz0) circle (.08);
	  \end{tikzpicture}
	  \caption{}
	\end{subfigure}
	\begin{subfigure}{0.13\textwidth}
	  \centering
	  \begin{tikzpicture}[baseline={(current bounding box.center)},scale=.6]
		\coordinate (inA) at (0.4,.6);
		\coordinate (outA) at (3.6,.6);
		\coordinate (inB) at (0.4,-.6);
		\coordinate (outB) at (3.6,-.6);
		\coordinate (xA) at (1,.6);
		\coordinate (xyA) at (1.5,.6);
		\coordinate (yA) at (2,.6);
		\coordinate (yzB) at (2.5,-.6);
		\coordinate (yz0) at (2.5,0);
		\coordinate (zA) at (3,.6);
		\coordinate (xB) at (1,-.6);
		\coordinate (yB) at (2,-.6);
		\coordinate (zB) at (3,-.6);
		\draw [dotted] (inA) -- (outA);
		\draw [dotted] (inB) -- (outB);
		\draw [zParticleF] (zA) -- (outA);
		\draw [zParticleF] (xA) to[out=40,in=140] (zA);
		\draw [photon] (xA) -- (xB);
		\draw [photon] (yzB) -- (yz0);
		\draw [photon] (yA) -- (yz0);
		\draw [photonRed] (zA) -- (yz0);
		\draw [fill] (xA) circle (.08);
		\draw [fill] (yA) circle (.08);
		\draw [fill] (zA) circle (.08);
		\draw [fill] (xB) circle (.08);
		\draw [fill] (yzB) circle (.08);
		\draw [fill] (yz0) circle (.08);
	  \end{tikzpicture}
	  \caption{}
	\end{subfigure}
	\raisebox{-0.15cm}{
	\begin{subfigure}{0.13\textwidth}
	  \centering
	  \begin{tikzpicture}[baseline={(current bounding box.center)},scale=.6]
		\coordinate (inA) at (0.4,.6);
		\coordinate (outA) at (3.6,.6);
		\coordinate (inB) at (0.4,-.6);
		\coordinate (outB) at (3.6,-.6);
		\coordinate (xA) at (1,.6);
		\coordinate (xyB) at (1.5,-.6);
		\coordinate (xy0) at (1.5,0);
		\coordinate (yA) at (2,.6);
		\coordinate (yzA) at (1.5,.6);
		\coordinate (zA) at (3,.6);
		\coordinate (xB) at (1,-.6);
		\coordinate (yB) at (2,-.6);
		\coordinate (zB) at (3,-.6);
		\draw [dotted] (inA) -- (outA);
		\draw [dotted] (inB) -- (outB);
		\draw [zParticleF] (zA) -- (outA);
		\draw [zParticleF] (yA) -- (zA);
		\draw [photon] (xA) -- (xy0);
		\draw [photonRed] (yA) -- (xy0);
		\draw [photon] (xyB) -- (xy0);
		\draw [photon] (zA) -- (zB);
		\draw [fill] (xA) circle (.08);
		\draw [fill] (yA) circle (.08);
		\draw [fill] (zA) circle (.08);
		\draw [fill] (xyB) circle (.08);
		\draw [fill] (xy0) circle (.08);
		\draw [fill] (zB) circle (.08);
	  \end{tikzpicture}
	  \caption{}
	\end{subfigure}}
	\raisebox{-0.15cm}{
	\begin{subfigure}{0.13\textwidth}
	  \centering
	  \begin{tikzpicture}[baseline={(current bounding box.center)},scale=.6]
		\coordinate (inA) at (0.4,.6);
		\coordinate (outA) at (3.6,.6);
		\coordinate (inB) at (0.4,-.6);
		\coordinate (outB) at (3.6,-.6);
		\coordinate (xA) at (1,.6);
		\coordinate (xyA) at (1.5,.6);
		\coordinate (xy0) at (1.5,0);
		\coordinate (yA) at (2,.6);
		\coordinate (yzA) at (1.5,.6);
		\coordinate (zA) at (3,.6);
		\coordinate (xB) at (1,-.6);
		\coordinate (yB) at (2,-.6);
		\coordinate (zB) at (3,-.6);
		 \draw [dotted] (inA) -- (outA);
		 \draw [dotted] (inB) -- (outB);
		 \draw [zParticleF] (zA) -- (outA);
		 \draw [zParticleF] (yB) -- (zB);
		 \draw [photon] (xy0) -- (xyA);
		 \draw [photon] (xy0) -- (xB);
		 \draw [photonRed] (xy0) -- (yB);
		 \draw [photon] (zA) -- (zB);
		\draw [fill] (xyA) circle (.08);
		\draw [fill] (xy0) circle (.08);
		\draw [fill] (zA) circle (.08);
		\draw [fill] (xB) circle (.08);
		\draw [fill] (yB) circle (.08);
		\draw [fill] (zB) circle (.08);
	  \end{tikzpicture}
	  \caption{}
	\end{subfigure}}
	\begin{subfigure}{0.13\textwidth}
	  \centering
	  \begin{tikzpicture}[baseline={(current bounding box.center)},scale=.6]
		\coordinate (inA) at (0.4,.6);
		\coordinate (outA) at (3.6,.6);
		\coordinate (inB) at (0.4,-.6);
		\coordinate (outB) at (3.6,-.6);
		\coordinate (xA) at (1,.6);
		\coordinate (xyA) at (1.5,.6);
		\coordinate (yA) at (2,.6);
		\coordinate (yzA) at (2.5,.6);
		\coordinate (yz0) at (2.5,0);
		\coordinate (zA) at (3,.6);
		\coordinate (xB) at (1,-.6);
		\coordinate (yB) at (2,-.6);
		\coordinate (zB) at (3,-.6);
		\draw [dotted] (inA) -- (outA);
		\draw [dotted] (inB) -- (outB);
		\draw [zParticleF] (zA) -- (outA);
		\draw [zParticleF] (xB) -- (yB);
		\draw [zParticleF] (yzA) -- (zA);
		\draw [photon] (xA) -- (xB);
		\draw [photon] (yz0) -- (yzA);
		\draw [photonRed] (yz0) -- (yB);
		\draw [photon] (yz0) -- (zB);
		\draw [fill] (xA) circle (.08);
		\draw [fill] (yzA) circle (.08);
		\draw [fill] (yz0) circle (.08);
		\draw [fill] (xB) circle (.08);
		\draw [fill] (yB) circle (.08);
		\draw [fill] (zB) circle (.08);
	  \end{tikzpicture}
	  \caption{}
	\end{subfigure}
	\begin{subfigure}{0.13\textwidth}
	  \centering
	  \begin{tikzpicture}[baseline={(current bounding box.center)},scale=.6]
		\coordinate (inA) at (0.4,.6);
		\coordinate (outA) at (3.6,.6);
		\coordinate (inB) at (0.4,-.6);
		\coordinate (outB) at (3.6,-.6);
		\coordinate (xA) at (1.2,.6);
		\coordinate (x0) at (1.2,0);
		\coordinate (xyA) at (1.5,.6);
		\coordinate (yA) at (2,.6);
		\coordinate (yzB) at (2.5,-.6);
		\coordinate (zA) at (2.8,.6);
		\coordinate (xB) at (1.2,-.6);
		\coordinate (yB) at (2,-.6);
		\coordinate (zB) at (2.8,-.6);
		\coordinate (z0) at (2.8,0);
		\draw [dotted] (inA) -- (outA);
		\draw [dotted] (inB) -- (outB);
		\draw [zParticleF] (zA) -- (outA);
		\draw [photon] (xA) -- (xB);
		\draw [photonRed] (z0) -- (x0);
		\draw [photon] (zA) -- (zB);
		\draw [fill] (xA) circle (.08);
		\draw [fill] (x0) circle (.08);
		\draw [fill] (z0) circle (.08);
		\draw [fill] (zA) circle (.08);
		\draw [fill] (xB) circle (.08);
		\draw [fill] (zB) circle (.08);
	  \end{tikzpicture}
	  \caption{}
	\end{subfigure}
	\begin{subfigure}{0.13\textwidth}
	  \centering
	  \begin{tikzpicture}[baseline={(current bounding box.center)},scale=.6]
		\coordinate (inA) at (0.4,.6);
		\coordinate (outA) at (3.6,.6);
		\coordinate (inB) at (0.4,-.6);
		\coordinate (outB) at (3.6,-.6);
		\coordinate (xA) at (1.2,.6);
		\coordinate (x0) at (1.2,0);
		\coordinate (xyA) at (1.5,.6);
		\coordinate (yA) at (2,.6);
		\coordinate (yzB) at (2.5,-.6);
		\coordinate (zA) at (2.8,.6);
		\coordinate (xB) at (1.2,-.6);
		\coordinate (yB) at (2,-.6);
		\coordinate (zB) at (2.8,-.6);
		\coordinate (y0) at (2,0);
		\draw [dotted] (inA) -- (outA);
		\draw [dotted] (inB) -- (outB);
		\draw [zParticleF] (zA) -- (outA);
		\draw [photon] (xA) -- (y0);
		\draw [photon] (zA) -- (y0);
		\draw [photon] (xB) -- (y0);
		\draw [photon] (zB) -- (y0);
		\draw [fill] (xA) circle (.08);
		\draw [fill] (zA) circle (.08);
		\draw [fill] (xB) circle (.08);
		\draw [fill] (zB) circle (.08);
		\draw [fill] (y0) circle (.08);
	  \end{tikzpicture}
	  \caption{}
	\end{subfigure}
	\begin{subfigure}{0.13\textwidth}
	  \centering
	  \begin{tikzpicture}[baseline={(current bounding box.center)},scale=.6]
		\coordinate (inA) at (0.4,.6);
		\coordinate (outA) at (3.6,.6);
		\coordinate (inB) at (0.4,-.6);
		\coordinate (outB) at (3.6,-.6);
		\coordinate (xA) at (1.2,.6);
		\coordinate (x0) at (1.2,0);
		\coordinate (xyA) at (1.5,.6);
		\coordinate (yA) at (2,.6);
		\coordinate (yzB) at (2.5,-.6);
		\coordinate (zA) at (2.8,.6);
		\coordinate (xB) at (1.2,-.6);
		\coordinate (yB) at (2,-.6);
		\coordinate (zB) at (2.8,-.6);
		\coordinate (y0) at (2,.25);
		\coordinate (y1) at (2,-.25);
		\draw [dotted] (inA) -- (outA);
		\draw [dotted] (inB) -- (outB);
		\draw [zParticleF] (zA) -- (outA);
		\draw [photon] (xA) -- (y0);
		\draw [photon] (zA) -- (y0);
		\draw [photon] (xB) -- (y1);
		\draw [photon] (zB) -- (y1);
		\draw [photon] (y0) -- (y1);
		\draw [fill] (xA) circle (.08);
		\draw [fill] (zA) circle (.08);
		\draw [fill] (xB) circle (.08);
		\draw [fill] (zB) circle (.08);
		\draw [fill] (y0) circle (.08);
		\draw [fill] (y1) circle (.08);
	  \end{tikzpicture}
	  \caption{}
	\end{subfigure}
	\raisebox{0.12cm}{
	\begin{subfigure}{0.13\textwidth}
	  \centering
	  \begin{tikzpicture}[baseline={(current bounding box.center)},scale=.6]
		\coordinate (inA) at (0.4,.6);
		\coordinate (outA) at (3.6,.6);
		\coordinate (inB) at (0.4,-.6);
		\coordinate (outB) at (3.6,-.6);
		\coordinate (xA) at (1,.6);
		\coordinate (xyA) at (1.5,.6);
		\coordinate (yA) at (2,.6);
		\coordinate (yzA) at (2.5,.6);
		\coordinate (zA) at (3,.6);
		\coordinate (xB) at (1,-.6);
		\coordinate (yB) at (2,-.6);
		\coordinate (zB) at (3,-.6);
		 \draw [dotted] (inA) -- (outA);
		 \draw [dotted] (inB) -- (outB);
		 \draw [zParticleF] (zA) -- (outA);
		 \draw [draw=none] (xA) to[out=40,in=140] (zA);
		 \draw [zParticleF] (xA) -- (xyA);
		 \draw [zParticleF] (yzA) -- (zA);
		 \draw [photon] (xA) -- (xB);
		 \draw [photonRed] (xyA)  to[out=40,in=140] (yzA);
		 \draw [photon] (zA) -- (zB);
		\draw [fill] (xA) circle (.08);
		\draw [fill] (zA) circle (.08);
		\draw [fill] (zB) circle (.08);
		\draw [fill] (xB) circle (.08);
		\draw [fill] (xyA) circle (.08);
		\draw [fill] (yzA) circle (.08);
	  \end{tikzpicture}
	  \caption{}
	\end{subfigure}
	\begin{subfigure}{0.13\textwidth}
	  \centering
	  \begin{tikzpicture}[baseline={(current bounding box.center)},scale=.6]
		\coordinate (inA) at (0.4,.6);
		\coordinate (outA) at (3.6,.6);
		\coordinate (inB) at (0.4,-.6);
		\coordinate (outB) at (3.6,-.6);
		\coordinate (xA) at (1,.6);
		\coordinate (xyA) at (1.5,.6);
		\coordinate (yA) at (2,.6);
		\coordinate (yzA) at (2.5,.6);
		\coordinate (zA) at (3,.6);
		\coordinate (xB) at (1,-.6);
		\coordinate (yB) at (2,-.6);
		\coordinate (zB) at (3,-.6);
		\draw [dotted] (inA) -- (outA);
		\draw [dotted] (inB) -- (outB);
		\draw [zParticleF] (zA) -- (outA);
		\draw [draw=none] (xA) to[out=40,in=140] (zA);
		\draw [zParticleF] (xB) -- (xyB);
		\draw [zParticleF] (yzB) -- (zB);
		\draw [photon] (xA) -- (xB);
		\draw [photonRed] (xyB)  to[out=40,in=140] (yzB);
		\draw [photon] (zA) -- (zB);
		\draw [fill] (xA) circle (.08);
		\draw [fill] (zA) circle (.08);
		\draw [fill] (zB) circle (.08);
		\draw [fill] (xB) circle (.08);
		\draw [fill] (xyB) circle (.08);
		\draw [fill] (yzB) circle (.08);
	  \end{tikzpicture}
	  \caption{}
	\end{subfigure}
	\begin{subfigure}{0.13\textwidth}
	  \centering
	  \begin{tikzpicture}[baseline={(current bounding box.center)},scale=.6]
		\coordinate (inA) at (0.4,.6);
		\coordinate (outA) at (3.6,.6);
		\coordinate (inB) at (0.4,-.6);
		\coordinate (outB) at (3.6,-.6);
		\coordinate (xA) at (1,.6);
		\coordinate (xyA) at (1.5,.6);
		\coordinate (yA) at (2,.6);
		\coordinate (yzA) at (2.5,.6);
		\coordinate (zA) at (3,.6);
		\coordinate (xB) at (1,-.6);
		\coordinate (yB) at (2,-.6);
		\coordinate (zB) at (3,-.6);
		\draw [dotted] (inA) -- (outA);
		\draw [dotted] (inB) -- (outB);
		\draw [zParticleF] (zA) -- (outA);
		\draw [draw=none] (xA) to[out=40,in=140] (zA);
		\draw [zParticleF] (xA) -- (xyA);
		\draw [zParticleF] (yzA) -- (zA);
		\draw [photon] (xA) -- (xB);
		\draw [photonRed] (xyA)  to[out=40,in=140] (zA);
		\draw [photon] (yzA) -- (yzB);
		\draw [fill] (xA) circle (.08);
		\draw [fill] (zA) circle (.08);
		\draw [fill] (yzB) circle (.08);
		\draw [fill] (xB) circle (.08);
		\draw [fill] (xyA) circle (.08);
		\draw [fill] (yzA) circle (.08);
	  \end{tikzpicture}
	  \caption{}
	\end{subfigure}}
	\caption{\small
	  The 14 types of diagrams contributing to the $m_1^2 m_2^2$ components of the 3PM gravitational
	  impulse $\Delta p_1^{(3)\mu}$ without tidal corrections.
	  All diagrams except the last, (n), are associated with the comparable-mass family
	  $I^{(\sigma_1;\sigma_2;\sigma_3)}_{n_1,n_2,\ldots,n_7}$
	  \eqref{eq:integralFamilies};
	  diagrams (l)--(n) are associated with $K^{(\sigma_1;\sigma_2;\sigma_3)}_{n_1,n_2,\ldots,n_5}$ family~\eqref{eq:kIntegral}.
	}
	\label{fig:splitDiags}
  \end{figure*}
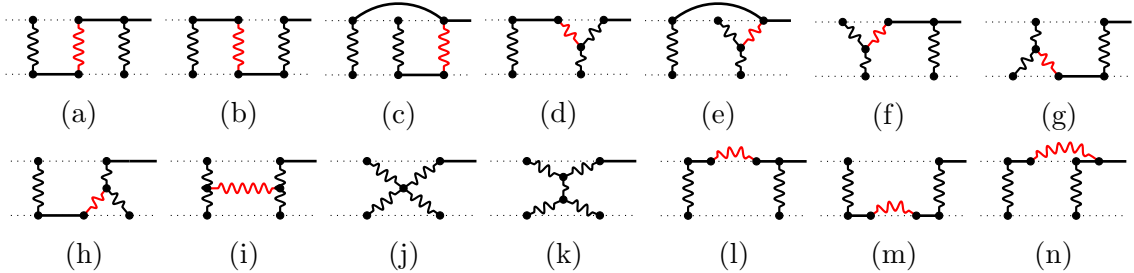

\begin{figure*}[t!]
  \centering
    \raisebox{+0.15cm}{
  \begin{subfigure}{0.13\textwidth}
    \centering
    \begin{tikzpicture}[baseline={(current bounding box.center)},scale=.6]
  		\coordinate (inA) at (0.4,.6);
  		\coordinate (outA) at (3.6,.6);
  		\coordinate (inB) at (0.4,-.6);
  		\coordinate (outB) at (3.6,-.6);
  		\coordinate (xA) at (1,.6);
  		\coordinate (xyB) at (1.5,-.6);
  		\coordinate (yA) at (2,.6);
  		\coordinate (yzA) at (1.5,.6);
  		\coordinate (zA) at (3,.6);
  		\coordinate (xB) at (1,-.6);
  		\coordinate (yB) at (2,-.6);
  		\coordinate (zB) at (3,-.6);
  		\draw [dotted] (inA) -- (outA);
  		\draw [dotted] (inB) -- (outB);
  		\draw [zParticleF] (zA) -- (outA);
  		\draw [draw=none] (xA) to[out=40,in=140] (zA);
  		\draw [zParticleF] (yA) -- (zA);
  		\draw [photon] (xA) -- (xyB);
  		\draw [photonRed] (yA) -- (xyB);
  		\draw [photon] (zA) -- (zB);
  		\draw [fill] (xA) circle (.08);
  		\draw [fill] (yA) circle (.08);
  		\draw [fill] (zA) circle (.08);
  		\draw [fill] (xyB)+(-0.08,-0.08) rectangle ++(.12,.12);
  		\draw [fill] (zB) circle (.08);
  	\end{tikzpicture}
	\caption{}
  \end{subfigure}
  \begin{subfigure}{0.13\textwidth}
    \centering
    \begin{tikzpicture}[baseline={(current bounding box.center)},scale=.6]
  		\coordinate (inA) at (0.4,.6);
  		\coordinate (outA) at (3.6,.6);
  		\coordinate (inB) at (0.4,-.6);
  		\coordinate (outB) at (3.6,-.6);
  		\coordinate (xA) at (1,.6);
  		\coordinate (xyA) at (1.5,.6);
  		\coordinate (yA) at (2,.6);
  		\coordinate (yzB) at (2.5,-.6);
  		\coordinate (zA) at (3,.6);
  		\coordinate (xB) at (1,-.6);
  		\coordinate (yB) at (2,-.6);
  		\coordinate (zB) at (3,-.6);
  		\draw [dotted] (inA) -- (outA);
  		\draw [dotted] (inB) -- (outB);
  		\draw [zParticleF] (zA) -- (outA);
  		\draw [draw=none] (xA) to[out=40,in=140] (zA);
  		\draw [zParticleF] (xA) -- (yA);
  		\draw [photon] (xA) -- (xB);
  		\draw [photonRed] (yA) -- (yzB);
  		\draw [photon] (zA) -- (yzB);
  		\draw [fill] (xA) circle (.08);
  		\draw [fill] (yA) circle (.08);
  		\draw [fill] (zA) circle (.08);
  		\draw [fill] (xB) circle (.08);
  		\draw [fill] (yzB)+(-0.08,-0.08) rectangle ++(.12,.12);
  	\end{tikzpicture}
	\caption{}
  \end{subfigure}
  \begin{subfigure}{0.13\textwidth}
    \centering
    \begin{tikzpicture}[baseline={(current bounding box.center)},scale=.6]
  		\coordinate (inA) at (0.4,.6);
  		\coordinate (outA) at (3.6,.6);
  		\coordinate (inB) at (0.4,-.6);
  		\coordinate (outB) at (3.6,-.6);
  		\coordinate (xA) at (1,.6);
  		\coordinate (xyA) at (1.5,.6);
  		\coordinate (yA) at (2,.6);
  		\coordinate (yzA) at (1.5,.6);
  		\coordinate (yzB) at (2.5,-.6);
  		\coordinate (zA) at (3,.6);
  		\coordinate (xB) at (1,-.6);
  		\coordinate (yB) at (2,-.6);
  		\coordinate (zB) at (3,-.6);
		\draw [dotted] (inA) -- (outA);
		\draw [dotted] (inB) -- (outB);
		\draw [zParticleF] (zA) -- (outA);
		\draw [zParticleF] (xA) to[out=40,in=140] (zA);
		\draw [photon] (xA) -- (xB);
		\draw [photon] (yA) -- (yzB);
		\draw [photonRed] (zA) -- (yzB);
  		\draw [fill] (xA) circle (.08);
  		\draw [fill] (yA) circle (.08);
  		\draw [fill] (zA) circle (.08);
  		\draw [fill] (xB) circle (.08);
  		\draw [fill] (yzB)+(-0.08,-0.08) rectangle ++(.12,.12);
  	\end{tikzpicture}
	\caption{}
  \end{subfigure}}
  \begin{subfigure}{0.13\textwidth}
    \centering
    \begin{tikzpicture}[baseline={(current bounding box.center)},scale=.6]
  		\coordinate (inA) at (0.4,.6);
  		\coordinate (outA) at (3.6,.6);
  		\coordinate (inB) at (0.4,-.6);
  		\coordinate (outB) at (3.6,-.6);
  		\coordinate (xA) at (1,.6);
  		\coordinate (xyA) at (1.5,.6);
  		\coordinate (yA) at (2,.6);
  		\coordinate (yzA) at (1.5,.6);
  		\coordinate (zA) at (3,.6);
  		\coordinate (xB) at (1,-.6);
  		\coordinate (yB) at (2,-.6);
  		\coordinate (zB) at (3,-.6);
  		\draw [dotted] (inA) -- (outA);
  		\draw [dotted] (inB) -- (outB);
  		\draw [zParticleF] (zA) -- (outA);
  		\draw [zParticleF] (yB) -- (zB);
  		\draw [photon] (xyA) -- (xB);
  		\draw [photonRed] (xyA) -- (yB);
  		\draw [photon] (zA) -- (zB);
  		\draw [fill] (xyA)+(-0.08,-0.08) rectangle ++(.12,.12);
  		\draw [fill] (zA) circle (.08);
  		\draw [fill] (xB) circle (.08);
  		\draw [fill] (yB) circle (.08);
  		\draw [fill] (zB) circle (.08);
  	\end{tikzpicture}
	\caption{}
  \end{subfigure}
  \begin{subfigure}{0.13\textwidth}
    \centering
    \begin{tikzpicture}[baseline={(current bounding box.center)},scale=.6]
  		\coordinate (inA) at (0.4,.6);
  		\coordinate (outA) at (3.6,.6);
  		\coordinate (inB) at (0.4,-.6);
  		\coordinate (outB) at (3.6,-.6);
  		\coordinate (xA) at (1,.6);
  		\coordinate (xyA) at (1.5,.6);
  		\coordinate (yA) at (2,.6);
  		\coordinate (yzA) at (2.5,.6);
  		\coordinate (zA) at (3,.6);
  		\coordinate (xB) at (1,-.6);
  		\coordinate (yB) at (2,-.6);
  		\coordinate (zB) at (3,-.6);
  		\draw [dotted] (inA) -- (outA);
  		\draw [dotted] (inB) -- (outB);
  		\draw [zParticleF] (zA) -- (outA);
  		\draw [zParticleF] (xB) -- (yB);
  		\draw [zParticleF] (yzA) -- (zA);
  		\draw [photon] (xA) -- (xB);
  		\draw [photonRed] (yzA) -- (yB);
  		\draw [photon] (yzA) -- (zB);
  		\draw [fill] (xA) circle (.08);
  		\draw [fill] (yzA)+(-0.08,-0.08) rectangle ++(.12,.12);
  		\draw [fill] (xB) circle (.08);
  		\draw [fill] (yB) circle (.08);
  		\draw [fill] (zB) circle (.08);
  	\end{tikzpicture}
	\caption{}
  \end{subfigure}
  \begin{subfigure}{0.13\textwidth}
    \centering
    \begin{tikzpicture}[baseline={(current bounding box.center)},scale=.6]
  		\coordinate (inA) at (0.4,.6);
  		\coordinate (outA) at (3.6,.6);
  		\coordinate (inB) at (0.4,-.6);
  		\coordinate (outB) at (3.6,-.6);
  		\coordinate (xA) at (1.2,.6);
  		\coordinate (x0) at (1.2,0);
  		\coordinate (xyA) at (1.5,.6);
  		\coordinate (yA) at (2,.6);
  		\coordinate (yzB) at (2.5,-.6);
  		\coordinate (zA) at (2.8,.6);
  		\coordinate (xB) at (1.2,-.6);
  		\coordinate (yB) at (2,-.6);
  		\coordinate (zB) at (2.8,-.6);
  		\coordinate (z0) at (2.8,0);
		\draw [dotted] (inA) -- (outA);
		\draw [dotted] (inB) -- (outB);
		\draw [zParticleF] (zA) -- (outA);
		\draw [photon] (xA) -- (xB);
		\draw [photonRed] (x0) to[out=0,in=140] (zB);
		\draw [photon] (zA) -- (zB);
  		\draw [fill] (xA) circle (.08);
  		\draw [fill] (x0) circle (.08);
  		\draw [fill] (zA) circle (.08);
  		\draw [fill] (xB) circle (.08);
  		\draw [fill] (zB)+(-0.08,-0.08) rectangle ++(.12,.12);
  	\end{tikzpicture}
	\caption{}
  \end{subfigure}
  \begin{subfigure}{0.13\textwidth}
    \centering
    \begin{tikzpicture}[baseline={(current bounding box.center)},scale=.6]
  		\coordinate (inA) at (0.4,.6);
  		\coordinate (outA) at (3.6,.6);
  		\coordinate (inB) at (0.4,-.6);
  		\coordinate (outB) at (3.6,-.6);
  		\coordinate (xA) at (1.2,.6);
  		\coordinate (x0) at (1.2,0);
  		\coordinate (xyA) at (1.5,.6);
  		\coordinate (yA) at (2,.6);
  		\coordinate (yzB) at (2.5,-.6);
  		\coordinate (zA) at (2.8,.6);
  		\coordinate (xB) at (1.2,-.6);
  		\coordinate (yB) at (2,-.6);
  		\coordinate (zB) at (2.8,-.6);
  		\coordinate (z0) at (2.8,0);
		\draw [dotted] (inA) -- (outA);
		\draw [dotted] (inB) -- (outB);
		\draw [zParticleF] (zA) -- (outA);
		\draw [photon] (xA) -- (xB);
		\draw [photonRed] (x0) to[out=0,in=-140] (zA);
		\draw [photon] (zA) -- (zB);
  		\draw [fill] (xA) circle (.08);
  		\draw [fill] (x0) circle (.08);
  		\draw [fill] (zA)+(-0.08,-0.08) rectangle ++(.12,.12);
  		\draw [fill] (xB) circle (.08);
  		\draw [fill] (zB) circle (.08);
  	\end{tikzpicture}
	\caption{}
  \end{subfigure}
  \raisebox{-0.15cm}{
  \begin{subfigure}{0.13\textwidth}
    \centering
    \begin{tikzpicture}[baseline={(current bounding box.center)},scale=.6]
  		\coordinate (inA) at (0.4,.6);
  		\coordinate (outA) at (3.6,.6);
  		\coordinate (inB) at (0.4,-.6);
  		\coordinate (outB) at (3.6,-.6);
  		\coordinate (xA) at (1.2,.6);
  		\coordinate (x0) at (1.2,0);
  		\coordinate (xyA) at (1.5,.6);
  		\coordinate (yA) at (2,.6);
  		\coordinate (yzB) at (2.5,-.6);
  		\coordinate (zA) at (2.8,.6);
  		\coordinate (xB) at (1.2,-.6);
  		\coordinate (yB) at (2,-.6);
  		\coordinate (zB) at (2.8,-.6);
  		\coordinate (z0) at (2.8,0);
		\draw [dotted] (inA) -- (outA);
		\draw [dotted] (inB) -- (outB);
		\draw [zParticleF] (zA) -- (outA);
		\draw [photon] (xA) -- (xB);
		\draw [photonRed] (xA) to[out=-40,in=180] (z0);
		\draw [photon] (zA) -- (zB);
  		\draw [fill] (xA)+(-0.08,-0.08) rectangle ++(.12,.12);
  		\draw [fill] (z0) circle (.08);
  		\draw [fill] (zA) circle (.08);
  		\draw [fill] (xB) circle (.08);
  		\draw [fill] (zB) circle (.08);
  	\end{tikzpicture}
	\caption{}
  \end{subfigure}
  \begin{subfigure}{0.13\textwidth}
    \centering
    \begin{tikzpicture}[baseline={(current bounding box.center)},scale=.6]
  		\coordinate (inA) at (0.4,.6);
  		\coordinate (outA) at (3.6,.6);
  		\coordinate (inB) at (0.4,-.6);
  		\coordinate (outB) at (3.6,-.6);
  		\coordinate (xA) at (1.2,.6);
  		\coordinate (x0) at (1.2,0);
  		\coordinate (xyA) at (1.5,.6);
  		\coordinate (yA) at (2,.6);
  		\coordinate (yzB) at (2.5,-.6);
  		\coordinate (zA) at (2.8,.6);
  		\coordinate (xB) at (1.2,-.6);
  		\coordinate (yB) at (2,-.6);
  		\coordinate (zB) at (2.8,-.6);
  		\coordinate (z0) at (2.8,0);
		\draw [dotted] (inA) -- (outA);
		\draw [dotted] (inB) -- (outB);
		\draw [zParticleF] (zA) -- (outA);
		\draw [photon] (xA) -- (xB);
		\draw [photonRed] (xB) to[out=40,in=180] (z0);
		\draw [photon] (zA) -- (zB);
  		\draw [fill] (xA) circle (.08);
  		\draw [fill] (z0) circle (.08);
  		\draw [fill] (zA) circle (.08);
  		\draw [fill] (xB)+(-0.08,-0.08) rectangle ++(.12,.12);
  		\draw [fill] (zB) circle (.08);
  	\end{tikzpicture}
	\caption{}
  \end{subfigure}}
  \begin{subfigure}{0.13\textwidth}
    \centering
    \begin{tikzpicture}[baseline={(current bounding box.center)},scale=.6]
  		\coordinate (inA) at (0.4,.6);
  		\coordinate (outA) at (3.6,.6);
  		\coordinate (inB) at (0.4,-.6);
  		\coordinate (outB) at (3.6,-.6);
  		\coordinate (xA) at (1,.6);
  		\coordinate (xyA) at (1.5,.6);
  		\coordinate (yA) at (2,.6);
  		\coordinate (yzA) at (1.5,.6);
  		\coordinate (zA) at (3,.6);
  		\coordinate (xB) at (1,-.6);
  		\coordinate (yB) at (2,-.6);
  		\coordinate (zB) at (3,-.6);
		\draw [dotted] (inA) -- (outA);
		\draw [dotted] (inB) -- (outB);
		\draw [zParticleF] (zA) -- (outA);
		\draw [draw=none] (xA) to[out=40,in=140] (zA);
		\draw [zParticleF] (yA) -- (zA);
		\draw [photon] (xA) -- (xB);
		\draw [photonRed] (xA) to[out=40,in=140] (yA);
		\draw [photon] (zA) -- (zB);
  		\draw [fill] (xA) +(-0.08,-0.08) rectangle ++(.12,.12);
  		\draw [fill] (yA) circle (.08);
  		\draw [fill] (zA) circle (.08);
  		\draw [fill] (xB) circle (.08);
  		\draw [fill] (zB) circle (.08);
  	\end{tikzpicture}
	\caption{}
  \end{subfigure}
  \begin{subfigure}{0.13\textwidth}
    \centering
    \begin{tikzpicture}[baseline={(current bounding box.center)},scale=.6]
  		\coordinate (inA) at (0.4,.6);
  		\coordinate (outA) at (3.6,.6);
  		\coordinate (inB) at (0.4,-.6);
  		\coordinate (outB) at (3.6,-.6);
  		\coordinate (xA) at (1,.6);
  		\coordinate (xyA) at (1.5,.6);
  		\coordinate (yA) at (2,.6);
  		\coordinate (yzA) at (1.5,.6);
  		\coordinate (zA) at (3,.6);
  		\coordinate (xB) at (1,-.6);
  		\coordinate (yB) at (2,-.6);
  		\coordinate (zB) at (3,-.6);
		\draw [dotted] (inA) -- (outA);
		\draw [dotted] (inB) -- (outB);
		\draw [zParticleF] (zA) -- (outA);
		\draw [draw=none] (xA) to[out=40,in=140] (zA);
		\draw [zParticleF] (xA) -- (yA);
		\draw [photon] (xA) -- (xB);
		\draw [photonRed] (yA) to[out=40,in=140] (zA);
		\draw [photon] (zA) -- (zB);
  		\draw [fill] (zA) +(-0.08,-0.08) rectangle ++(.12,.12);
  		\draw [fill] (yA) circle (.08);
  		\draw [fill] (xA) circle (.08);
  		\draw [fill] (xB) circle (.08);
  		\draw [fill] (zB) circle (.08);
  	\end{tikzpicture}
	\caption{}
  \end{subfigure}
  \begin{subfigure}{0.13\textwidth}
    \centering
    \begin{tikzpicture}[baseline={(current bounding box.center)},scale=.6]
  		\coordinate (inA) at (0.4,.6);
  		\coordinate (outA) at (3.6,.6);
  		\coordinate (inB) at (0.4,-.6);
  		\coordinate (outB) at (3.6,-.6);
  		\coordinate (xA) at (1,.6);
  		\coordinate (xyA) at (1.5,.6);
  		\coordinate (yA) at (2.3,.6);
  		\coordinate (yzA) at (1.5,.6);
  		\coordinate (zA) at (3,.6);
  		\coordinate (xB) at (1,-.6);
  		\coordinate (yB) at (2.3,-.6);
  		\coordinate (zB) at (3,-.6);
		\draw [dotted] (inA) -- (outA);
		\draw [dotted] (inB) -- (outB);
		\draw [zParticleF] (zA) -- (outA);
		\draw [draw=none] (xA) to[out=40,in=140] (zA);
		\draw [zParticleF] (yA) -- (zA);
		\draw [photon] (xA) -- (xB);
		\draw [photonRed] (xA) to[out=40,in=140] (zA);
		\draw [photon] (yA) -- (yB);
  		\draw [fill] (xA) +(-0.08,-0.08) rectangle ++(.12,.12);
  		\draw [fill] (yA) circle (.08);
  		\draw [fill] (zA) circle (.08);
  		\draw [fill] (xB) circle (.08);
  		\draw [fill] (yB) circle (.08);
  	\end{tikzpicture}
	\caption{}
  \end{subfigure}
  \begin{subfigure}{0.13\textwidth}
    \centering
    \begin{tikzpicture}[baseline={(current bounding box.center)},scale=.6]
  		\coordinate (inA) at (0.4,.6);
  		\coordinate (outA) at (3.6,.6);
  		\coordinate (inB) at (0.4,-.6);
  		\coordinate (outB) at (3.6,-.6);
  		\coordinate (xA) at (1,.6);
  		\coordinate (xyA) at (1.5,.6);
  		\coordinate (yA) at (2,.6);
  		\coordinate (yzA) at (1.5,.6);
  		\coordinate (zA) at (3,.6);
  		\coordinate (xB) at (1,-.6);
  		\coordinate (yB) at (2,-.6);
  		\coordinate (zB) at (3,-.6);
		\draw [dotted] (inA) -- (outA);
		\draw [dotted] (inB) -- (outB);
		\draw [zParticleF] (zA) -- (outA);
		\draw [draw=none] (xA) to[out=40,in=140] (zA);
		\draw [zParticleF] (yB) -- (zB);
		\draw [photon] (xA) -- (xB);
		\draw [photonRed] (xB)  to[out=40,in=140] (yB);
		\draw [photon] (zA) -- (zB);
  		\draw [fill] (xA) circle (.08);
  		\draw [fill] (zA) circle (.08);
  		\draw [fill] (xB) +(-0.08,-0.08) rectangle ++(.12,.12);
  		\draw [fill] (yB) circle (.08);
  		\draw [fill] (zB) circle (.08);
  	\end{tikzpicture}
  	\caption{}
  \end{subfigure}
  \begin{subfigure}{0.13\textwidth}
    \centering
    \begin{tikzpicture}[baseline={(current bounding box.center)},scale=.6]
  		\coordinate (inA) at (0.4,.6);
  		\coordinate (outA) at (3.6,.6);
  		\coordinate (inB) at (0.4,-.6);
  		\coordinate (outB) at (3.6,-.6);
  		\coordinate (xA) at (1,.6);
  		\coordinate (xyA) at (1.5,.6);
  		\coordinate (yA) at (2,.6);
  		\coordinate (yzA) at (1.5,.6);
  		\coordinate (zA) at (3,.6);
  		\coordinate (xB) at (1,-.6);
  		\coordinate (yB) at (2,-.6);
  		\coordinate (zB) at (3,-.6);
  		\draw [dotted] (inA) -- (outA);
  		\draw [dotted] (inB) -- (outB);
  		\draw [zParticleF] (zA) -- (outA);
  		\draw [draw=none] (xA) to[out=40,in=140] (zA);
  		\draw [zParticleF] (xB) -- (yB);
  		\draw [photon] (xA) -- (xB);
  		\draw [photonRed] (yB)  to[out=40,in=140] (zB);
  		\draw [photon] (zA) -- (zB);
  		\draw [fill] (xA) circle (.08);
  		\draw [fill] (zA) circle (.08);
  		\draw [fill] (xB) circle (.08);
  		\draw [fill] (yB) circle (.08);
  		\draw [fill] (zB) +(-0.08,-0.08) rectangle ++(.12,.12);
  	\end{tikzpicture}
	\caption{}
  \end{subfigure}
    \caption{\small The 14 types of diagrams contributing to the $m_1^2m_2^2$ 
	comparable-mass components of $\Delta p_1^{(3)\mu}$ linear in tidal coefficients.}
  \label{fig:splitDiagsTid}
\end{figure*}
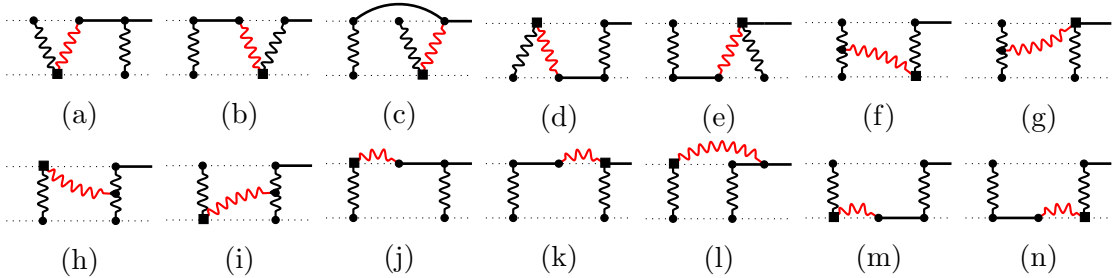

\subsection{Impulse}

Our main goal is to calculate the impulse (deflection) on the first body,
including radiation-reaction effects.
This is recovered from the WQFT using:
\begin{align}
	\Delta p^\mu_1 =-m_1 \omega^2 \braket{z_1^\mu(\omega)}|_{\omega =0}\,,
\end{align}
where the expectation value was discussed in \eqn{eq:ininZ1}
and $\Delta p_i^\mu=\sum_nG^n\Delta p_i^{(n)\mu}$ in the PM expansion.
As the results for $\Delta p_i^{(1)\mu}$ and $\Delta p_i^{(2)\mu}$ are well-established
--- tidal effects beginning at 2PM order~\cite{Bini:2020flp,Haddad:2020que,Cheung:2020sdj,Kalin:2020lmz} ---
we focus here on the 3PM component $\Delta p_i^{(3)\mu}$.
This will allow us to use the retarded integrals derived in \sec{sec:retInt}.
Results for $\Delta p_i^{(1)\mu}$ and $\Delta p_i^{(2)\mu}$ are
included in the ancillary file attached to the \texttt{arXiv} submission of this paper.

We proceed by drawing all diagrams with a single (cut) outgoing $z_1^\mu$ line.
There are four categories of diagrams,
displayed in Figs.~\ref{fig:testBodyDiags}--\ref{fig:splitDiagsTid}.
All retarded propagators, both in the bulk and on the worldlines,
point towards the outgoing line: from cause to effect.
In particular: Figs.~\ref{fig:testBodyDiags} and \ref{fig:testBodyDiagsTid}
contain the contributions to $\Delta p_i^{(3)\mu}$ in the test-body limit $m_1\ll m_2$,
with and without the tidal corrections respectively.
These involve the integral family $J^{(\sigma_1;\sigma_2)}_{n_1,n_2,\ldots,n_7}$,
consisting only of potential modes.
Figs.~\ref{fig:splitDiags} and \ref{fig:splitDiagsTid} contain the comparable-mass diagrams.
These latter contributions are modified by the inclusion of radiative effects,
and involve the integral families $I^{(\sigma_1;\sigma_2;\sigma_3)}_{n_1,n_2,\ldots,n_7}$
and $K^{(\sigma_1;\sigma_2;\sigma_3)}_{n_1,n_2,\ldots,n_5}$ discussed in \sec{sec:retInt}.
There is also a third category of diagrams (not drawn) relevant in the other
test-body limit $m_1\gg m_2$; however, as these are related by symmetry to those in
Figs.~\ref{fig:testBodyDiags} and \ref{fig:testBodyDiagsTid} we do not need to calculate
them explicitly.

All integrands can be expressed as a Fourier transform over integrals of the kind
discussed in \sec{sec:retInt}:
\begin{align}
  \int_q e^{i q\cdot b} \dd(q \cdot v_1) \dd(q \cdot v_2) |q|^n
  \{I^{(\sigma_1;\sigma_2;\sigma_3)}_{n_1,...n_7},J^{(\sigma_1;\sigma_2)}_{n_1,...n_7},
  K^{(\sigma_1;\sigma_2;\sigma_3)}_{n_1,...n_5}\}\,,
\end{align}
where $q^\mu$ is the total momentum exchanged via gravitons between the two worldlines.
To bring the diagrams into this form,
we need to resolve four-dimensional delta functions in the bulk
and integrate over the energies $\omega_i$ of worldline propagators.
As explained in \rcite{Jakobsen:2022fcj},
any leftover components of $\ell_i^\mu$ may be conveniently resolved on a basis of
$w_i^\mu$ and $q^\mu$, where $w_i^\mu$ are the dual velocities satisfying $v_i\cdot w_j=\delta_{ij}$:
\begin{align}
	w_1^\mu=\frac{\gamma v_2^\mu-v_1^\mu}{\gamma^2-1}\,, &&
	w_2^\mu=\frac{\gamma v_1^\mu-v_2^\mu}{\gamma^2-1}\,.
\end{align}
After reducing to the master integrals given in \sec{sec:integralExp},
our last step is to perform the $q$-Fourier transform.

Our final result for $\Delta p_{1}^{\mu}$ up to 3PM order is given in the ancillary file.
It takes the generic form~\cite{Saketh:2021sri}:
\begin{align}
	\Delta p_1^{\mu}&=p_\infty\sin\theta\frac{b^\mu}{|b|}\!+\!
	(\cos\theta\!-\!1)\frac{m_1m_2}{E^2}[(\gamma m_1+m_2)v_1^\mu\!-\!(\gamma m_2+m_1)v_2^\mu]
	-v_2\cdot P_{\rm rad}\,w_2^\mu\,,
\end{align}
where the center-of-mass momentum is $p_\infty= \mu\sqrt{\gamma^2-1}/\Gamma$,
$\mu=M \nu = m_1 m_2 /M$, $M=m_1+m_2$ and $\Gamma = E/M = \sqrt{1+2\nu(\gamma-1)}$.
All terms proportional to the impact parameter $b^\mu$, both conservative and radiative,
arise from the real integrals~\eqref{eq:bResults};
terms proportional to $v_i^\mu$, including $P_{\rm rad}^\mu$, come from the imaginary integrals~\eqref{eq:vResults}.
Here $\theta$ is the scattering angle in the center-of-mass frame,
and $P_{\rm rad}^\mu$ is the radiated four-momentum:
\begin{align}
    \sin{\left( \dfrac{\theta}{2} \right)}= \dfrac{\sqrt{- \Delta p_1^2}}{2 p_\infty}\,,&&
	P_{\rm rad}^\mu=-\Delta p_1^\mu-\Delta p_2^\mu\,.
\end{align}
The entire dynamics is therefore encoded by $\theta$ and $P_{\rm rad}^\mu$, which we present below.

\subsection{Scattering angle}

In the absence of tidal effects, the complete scattering angle $\theta$ up to $\cO(G^3)$ is
\begin{subequations}
\begin{align}
	\frac{\theta_{\rm cons}}{\Gamma}&=\frac{GM}{|b|}\frac{2(2\gamma^2-1)}{\gamma^2-1}
	+\left(\frac{GM}{|b|}\right)^2\frac{3\pi(5\gamma^2-1)}{4(\gamma^2-1)}\\
	&\quad+\left(\frac{GM}{|b|}\right)^3
	\left(2
	\frac{
	  64\gamma^6-120\gamma^4+60\gamma^2-5
	}{
	  3(\gamma^2-1)^3
	}
	\Gamma^2\right.\nn\\
	&\quad\qquad\qquad\qquad-\left.
	\frac{8\nu
	\gamma
	(14\gamma^2+25)}{3(\gamma^2-1)}
	-
	8\nu
	\frac{
	  (4\gamma^4-12\gamma^2-3)
	}{
	  (\gamma^2-1)
	}
	\frac{\text{arccosh}\gamma}{\sqrt{\gamma^2-1}}\right)+\cO(G^4)
	\,,\nn\\
	\frac{\theta_{\rm rad}}{\Gamma}
&=\left(\frac{GM}{|b|}\right)^3\frac{4\nu(2\gamma^2-1)^2}{(\gamma^2-1)^{3/2}}
\Big(-\frac83+\frac1{v^2}+\frac{(3v^2-1)}{v^3}{\rm arccosh}\gamma\Big)+\cO(G^4)\,,
\end{align}
\end{subequations}
where $\theta=\theta_{\rm cons}+\theta_{\rm rad}$ has a finite high-energy $\gamma\to\infty$ limit:
\begin{align}
	\theta\xrightarrow{\gamma\to\infty}4\frac{GE}{|b|}+\frac{32}3\left(\frac{GE}{|b|}\right)^3\,.
\end{align}
This is the  well-known result of Amati, Ciafaloni and Veneziano~\cite{Amati:1990xe},
the radiative correction $\theta_{\rm rad}$ being required in order to cancel a logarithmic divergence
that otherwise appears in this limit~\cite{Damour:2020tta}.
With the inclusion of tidal effects, only the conservative part of the angle is modified up to 3PM order,
\ie~$\theta_{\rm tidal}=\theta_{\rm tidal,cons}+\cO(G^4)$:
\begin{align}
	\dfrac{\theta_{\rm tidal}}{\Gamma}=&\dfrac{45 \pi G^2M^2}{128|b|^6  (\gamma^2-1)}\left[(11-30\gamma^2+35\gamma^4)\left(c_{E^2}^{+}-\delta c_{E^2}^{-}\right)\right.\notag \\
	&\qquad \qquad \qquad  \left. +(-5-30\gamma^2+35\gamma^4)\left(c_{B^2}^{+}-\delta c_{B^2}^{-}\right)\right] \notag \\
		&+ \dfrac{48 G^3 M^3}{|b|^7 (\gamma^2-1)^2} \bigg[\dfrac{1}{35}\left(-5+72 \gamma^2-192 \gamma^4 +160 \gamma^6\right) \left( (1+\delta^2) c_{E^2}^{+}-2\delta c_{E^2}^{-}\right) \notag \\
		&\qquad \qquad \qquad \,+\dfrac{1}{35}\left(2+30 \gamma^2-192 \gamma^4 +160 \gamma^6\right) \left((1+\delta^2)c_{B^2}^{+}-2\delta c_{B^2}^{-}\right) \notag \\
		&\qquad \qquad\qquad\,+\dfrac{2 \nu \gamma}{5 (\gamma^2-1)}\left(4047+9426 \gamma^2 +804 \gamma^4 -104 \gamma^6 +32 \gamma ^8 \right)  c_{E^2}^{+} \notag \\
		&\qquad \qquad\qquad\, +\dfrac{4 \nu \gamma }{5  (\gamma^2-1)}\left(2006 + 4733 \gamma^2 +392 \gamma^4 -52 \gamma^6 +16 \gamma^8 \right)  c_{B^2}^{+} \notag \\
		&\qquad\qquad\qquad\, -\dfrac{6  \nu}{ \gamma^2-1} \frac{{\rm arccosh}\gamma}{\sqrt{\gamma^2-1}}\left[(33+474\gamma^2+440 \gamma^4)c_{E^2}^{+} \right. \notag \\
		&\qquad \qquad \qquad \qquad \qquad \qquad \qquad \left. +(32+474 \gamma^2+440 \gamma^4)c_{B^2}^{+} \right] \bigg]+\mathcal{O}(G^4)\,,\nn
\end{align}
with $\delta=(m_2-m_1)/M$ and $c_{{E^2/B^2}}^{\pm}=c_{E^2/B^2}^{(2)}\pm c_{E^2/B^2}^{(1)}$ ---
see also \rcites{Cheung:2020sdj,Kalin:2020lmz}.
It has a finite high-energy limit:
\begin{align}
    \theta_{\rm tidal}\xrightarrow{\gamma\to\infty}
	\left(\frac{GE}{|b|}\right)^7\dfrac{384(\kappa_{E^2}+\kappa_{B^2}) }{5 \nu^2}\,, &&
	\kappa_{E^2/B^2}=\dfrac{c_{E^2/B^2}^{+}}{G^4M^4} \,,
\end{align}
where the dimensionless parameters $\kappa_{E^2/B^2}$ account for the mass dependence of the Love numbers $c_{E^2/B^2}^{(a)}$.

The absence of a radiative part of the tidal correction to the scattering angle at 3PM
order is explained using the linear response relation~\cite{Bini:2012ji,Damour:2020tta,Bini:2021gat}:
\begin{align}\label{eq:biniDamour}
    \theta_{\rm rad}=
	- \dfrac{1}{2} \dfrac{\partial \theta_{\rm cons}}{\partial E} E_{\rm rad}
	- \dfrac{1}{2} \dfrac{\partial \theta_{\rm cons}}{\partial J} J_{\rm rad}\,.
\end{align}
This predicts the radiative part of the scattering angle $\theta_{\rm rad}$ given knowledge of the
radiated energy $E_{\rm rad}$ and angular momentum $J_{\rm rad}$.
As $E_{\rm rad}=P_{\rm rad}^0$ (in the center-of-mass frame) begins at 3PM order,
to deduce the 3PM contribution to $\theta_{\rm rad}$ we need only $J_{\rm rad}$ at 2PM order.
As we shall see in \sec{sec:waveform}, the absence of a \emph{wave memory} in the tidal correction 
to the 2PM waveform guarantees that $J_{\rm tidal, rad}=\cO(G^3)$,
hence $\theta_{\rm tidal, rad}=\cO(G^4)$.

\subsection{Radiated momentum}

The radiated four-momentum without finite-size effects up to $\mathcal{O}(G^3)$ is
(see also refs.~\cite{Herrmann:2021lqe,Herrmann:2021tct,Riva:2021vnj}):
\begin{align}\label{eq:radMom}
	P_{\rm rad}^\mu=\dfrac{G^3m_1^2m_2^2 \pi}{|b|^3} \dfrac{v_1^\mu+v_2^\mu}{\gamma+1} \mathcal{E}(\gamma)+\cO(G^4)\,,
\end{align}
with the scalar function $\mathcal{E}(\gamma)$ giving the
radiated energy in the center-of-mass (com) frame
$E_{\rm rad}=P_{\rm rad}\cdot v_{\rm com}=P_{\rm rad}\cdot(m_1v_1^\mu +m_2 v_2^\mu)/E$:
\begin{align}
	E_{\rm rad}=\dfrac{G^3\pi M^4\nu^2}{|b|^3 \Gamma} \mathcal{E}(\gamma) \,, &&
	\mathcal{E}(\gamma)=e_1+e_2 \log{\left(\frac{\gamma+1}{2}\right)} +e_{3}\frac{{\rm arccosh}\gamma}{\sqrt{\gamma^2-1}}\,.
\end{align}
The coefficients are
\begin{subequations}
	\begin{align}
		e_1&= \dfrac{210 \gamma^6-552\gamma^5+339 \gamma^4 -912\gamma^3+3148\gamma^2-3336\gamma+1151}{48 (\gamma^2-1)^{3/2}}\,,\\
		e_2&= -\dfrac{35 \gamma^4+60\gamma^3-150 \gamma^2+76\gamma-5}{8 \sqrt{\gamma^2-1}}\,,\\
		e_3&= \dfrac{\gamma(2\gamma^2-3)\left( 35 \gamma^4-30\gamma^2+11\right)}{16 (\gamma^2-1)^{3/2}}\,.
	\end{align}
\end{subequations}
Incorporating leading-order tidal effects the correction to the radiated four-momentum
$P_{\rm rad,tidal}^\mu$ at this order is given by
\begin{align}\label{eq:radMomTid}
	P_{\rm rad,tidal}^{\mu} =& \dfrac{ G^3 \pi m_1^2 m_2^2}{ |b|^7 }
	\left[\left(c_{E^2}^{(1)} \mathcal{A}_{E^2}+c_{E^2}^{(2)} \mathcal{B}_{E^2}+c_{B^2}^{(1)} \mathcal{A}_{B^2}+c_{B^2}^{(2)} \mathcal{B}_{B^2} \right) \dfrac{\gamma v_2^\mu -v_1^\mu}{\sqrt{\gamma^2-1}} + \left( 1\leftrightarrow 2 \right) \right]\,,
\end{align}
and the scalar factors $\mathcal{A}_x$ depend only on $\gamma$:
\begin{subequations}
	\begin{align}
	\mathcal{A}_x=&a_{1,x}+a_{2,x} \log{\left(\frac{\gamma+1}{2}\right)} +a_{3,x}\dfrac{{\rm arccosh}{ \gamma}}{\sqrt{\gamma^2-1}}\,,\\
	a_{1,E^2}=&\frac{15 }{128 (\gamma-1) (\gamma +1)^4}\left(937 \gamma ^9+1551 \gamma ^8-2463 \gamma ^7-5645 \gamma ^6 +20415 \gamma ^5\right. \notag \\
	&\left. +65965 \gamma ^4-349541 \gamma ^3+535057 \gamma ^2-360356 \gamma +92160\right)\,,  \\
	a_{1,B^2}=&\frac{15 }{256 (\gamma +1)^4}\left(1559 \gamma ^{8}+3716 \gamma ^7-1630 \gamma ^6-11660 \gamma ^5+28288 \gamma ^4\right. \notag \\
	& \left. +155292 \gamma ^3 -543442 \gamma ^2+535212 \gamma -180775 \right)\,, \\
	a_{2,E^2}=&\sfrac{225}{32}\left(21 \gamma ^4-14 \gamma ^2+9\right)\,,\\
	a_{2,B^2}=&\sfrac{1575 }{32} \left(3 \gamma ^4-2 \gamma ^2-1\right)\,,\\
	a_{3,x}=&-\frac{\gamma  \left(2 \gamma ^2-3\right)}{4 \left(\gamma ^2-1\right)} a_{2,x}\,.
	\end{align}
\end{subequations}
Finally, ${\cal B}_x$ are rational functions of $\gamma$:
\begin{subequations}
	\begin{align}
	\mathcal{B}_{E^2}=&\frac{45(\gamma -1) }{64{(\gamma +1)^4}} \left( 42 \gamma ^8+210 \gamma ^7+315 \gamma ^6-105 \gamma ^5-944 \gamma ^4-1528 \gamma ^3 \right. \nn \\[-8pt]
	&\qquad \qquad \qquad \qquad \quad \left.+22011 \gamma ^2  -33201 \gamma +16272\right)\,,\\
	\mathcal{B}_{B^2}=&\,-\frac{45(\gamma -1)^2 \left(105 \gamma ^5+630 \gamma ^4+1840 \gamma ^3+3690 \gamma ^2-17769 \gamma +15984\right)}{64(\gamma +1)^4}\,.
	\end{align}
\end{subequations}
The tidal correction to the radiated energy is given by
\begin{align}
	E_{\rm rad, tidal}=&  \dfrac{ G^3 \pi M^3 \nu^2 \sqrt{\gamma^2-1}}{ |b|^7  \Gamma}  \left[\mathcal{A}_{E^2} \left(m_1 c_{E^2}^{(1)}+m_2 c_{E^2}^{(2)} \right)+\mathcal{B}_{E^2} \left(m_1 c_{E^2}^{(2)}+m_2 c_{E^2}^{(1)} \right) \right. \nn\\
	&\left. \qquad \qquad \qquad \quad \; \; \, +\mathcal{A}_{B^2} \left(m_1 c_{B^2}^{(1)}+m_2 c_{B^2}^{(2)} \right)+\mathcal{B}_{B^2} \left(m_1 c_{B^2}^{(2)}+m_2 c_{B^2}^{(1)} \right) \right]\,.
\end{align}
These results all agree with \rcite{Mougiakakos:2022sic}.
\subsection{Waveform}\label{sec:waveform}

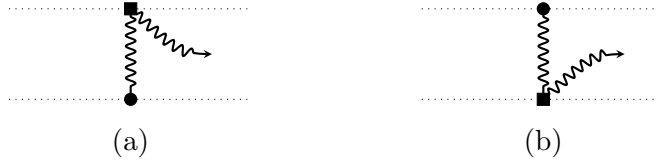
\begin{figure*}[t]
  \centering
  \begin{subfigure}{.35\textwidth}
    \centering
    \begin{tikzpicture}[baseline={(current bounding box.center)},scale=1]
  		\coordinate (inA) at (0.4,.6);
  		\coordinate (outA) at (3.6,.6);
  		\coordinate (inB) at (0.4,-.6);
  		\coordinate (outB) at (3.6,-.6);
  		\coordinate (xA) at (1,.6);
  		\coordinate (xyA) at (1.5,.6);
  		\coordinate (yA) at (2,.6);
  		\coordinate (yzA) at (1.5,.6);
  		\coordinate (zA) at (3,.6);
		\coordinate (z0) at (3,0);
  		\coordinate (xB) at (1,-.6);
  		\coordinate (yB) at (2,-.6);
  		\coordinate (zB) at (3,-.6);
  		\draw [fill] (yA) +(-0.08,-0.08) rectangle ++(.08,.08);
  		\draw [fill] (yB) circle (.08);
  		\draw [-stealth] (z0)--(3.07,0);
  		\draw [dotted] (inA) -- (outA);
  		\draw [dotted] (inB) -- (outB);
  		\draw [draw=none] (xA) to[out=40,in=140] (zA);
  		\draw [photon] (yA) -- (yB);
  		\draw [photon] (yA) to[out=-40,in=180] (z0);
  	\end{tikzpicture}
	\caption{}
  \end{subfigure}
  \begin{subfigure}{.35\textwidth}
    \centering
    \begin{tikzpicture}[baseline={(current bounding box.center)},scale=1]
  		\coordinate (inA) at (0.4,.6);
  		\coordinate (outA) at (3.6,.6);
  		\coordinate (inB) at (0.4,-.6);
  		\coordinate (outB) at (3.6,-.6);
  		\coordinate (xA) at (1,.6);
  		\coordinate (xyA) at (1.5,.6);
  		\coordinate (yA) at (2,.6);
  		\coordinate (yzA) at (1.5,.6);
  		\coordinate (zA) at (3,.6);
		\coordinate (z0) at (3,0);
  		\coordinate (xB) at (1,-.6);
  		\coordinate (yB) at (2,-.6);
  		\coordinate (zB) at (3,-.6);
  		\draw [fill] (yB) +(-0.08,-0.08) rectangle ++(.08,.08);
  		\draw [fill] (yA) circle (.08);
		\draw [-stealth] (z0)--(3.07,0);

  		\draw [dotted] (inA) -- (outA);
  		\draw [dotted] (inB) -- (outB);
  		\draw [draw=none] (xA) to[out=40,in=140] (zA);
  		\draw [photon] (yA) -- (yB);
  		\draw [photon] (yB) to[out=40,in=180] (z0);
  	\end{tikzpicture}
	\caption{}
  \end{subfigure}
  \caption{\small The two diagrams contributing to tidal effects at $\mathcal{O}(G^2)$ to the waveform}
  \label{fig:WavTid}
\end{figure*}

As a final application of the new WQFT Feynman rules~\eqref{eq:tidalVertices}
we compute the leading-order $G^2$ gravitational waveform produced by
a scattering of two compact objects, corrected by tidal effects.
This result was also recently obtained in \rcite{Mougiakakos:2022sic} ---
besides its inherent usefulness,
it provides us with a means to check the radiated linear momentum $P_{\rm rad}^\mu$~\eqref{eq:radMom}
and confirm that the correction to the radiated angular momentum $J_{\rm rad}$
from tidal effects vanishes at leading order $\cO(G^2)$.
Following \rcites{Jakobsen:2021smu,Jakobsen:2021lvp}
the gauge-invariant frequency-domain waveform $4G\,\eps^{\mu\nu}{S}_{\mu\nu}(k^{\mu}=\Omega\rho^\mu)$
is extracted from the WQFT via
\begin{align}
	{S}_{\mu\nu}(k)= \frac{2}{\kappa} k^{2} \vev{h_{\mu\nu}(k)}\, ,
\end{align}
where $\Omega$ is the gravitational wave frequency and
$\rho^{\mu}=(1,\hat{\bf x})$, $\mathbf{\hat x}:=\mathbf{x}/r$ being
a unit vector pointing towards the observer.
We work in the \emph{wave zone}, where $r=|\mathbf{x}|$ is much larger
than all other length scales in the problem.

We find it advantageous to study the time-domain waveform
$f(u,\mathbf{\hat{x}})$, which is given by a Fourier transform: 
\begin{equation}\label{eq:startingpoint}
\kappa \epsilon^{\mu\nu} h_{\mu\nu}=\frac{f(u,\mathbf{\hat{x}})}{r} = \frac{4G}{r}
\int_\Omega e^{-i\Omega u} \,\epsilon^{\mu\nu}\,
S_{\mu\nu}(k) \Bigr |_{k^{\mu}=\Omega\rho^{\mu}}\, ,
\end{equation}
and $u=t-r$ is the retarded time.
The polarization tensor factorizes as
$\epsilon^{\mu\nu}=\sfrac12\epsilon^{\mu}\epsilon^{\nu}$,
where $\epsilon\cdot\epsilon=\epsilon\cdot\rho=0$;
in a PM decomposition $f=\sum_nG^nf^{(n)}$ we seek the 2PM component $f^{(2)}$.
As the time-domain waveform in the absence of tidal effects has
already been computed in~\rcites{Jakobsen:2021smu,Jakobsen:2021lvp,Mougiakakos:2021ckm},
we focus here only on the tidal corrections.
The two contributing diagrams are drawn in \Fig{fig:WavTid}.

The integrals required are simpler than those used in~\rcites{Jakobsen:2021smu,Jakobsen:2021lvp}:
we require only derivatives with respect to $\tilde{b}^\mu$ of
\begin{align}
	\int_q \dd(q \cdot v_1) \dfrac{e^{-i q \cdot \tilde{b}}}{q^2}=
	-\dfrac{1}{4 \pi|\tilde{\mathbf{b}}|_1}\,,
\end{align}
where $\tilde{b}^\mu$ is the shifted impact parameter:
\begin{align}
	\tilde{b}^\mu=\tilde{b}_2^\mu-\tilde{b}_1^\mu\,, &&
	\tilde{b}_i^\mu=b_i^\mu+u_i v_i^\mu\,, &&
	u_i=\frac{\rho\cdot(x-b_i)}{\rho\cdot v_i}\,,
\end{align}
and $u_i$ is the retarded time in the rest frame of the $i$'th body.
We have also introduced
\begin{align}\label{eq:modBDef}
	|\tilde{\mathbf{b}}|_{1,2}&:=\sqrt{-\tilde{b}_\mu P_{1,2}^{\mu\nu}\tilde{b}_\nu}
	=\sqrt{|b|^2+(\gamma^2-1)u_{2,1}^2}\,,
\end{align}
which are the lengths of the shifted impact parameter $\tilde{b}^\mu$ in the two rest frames,
and $P_i^{\mu \nu}=\eta^{\mu \nu}-v_i^\mu v_i^\nu$.
The tidal correction to the waveform $f=\sum_nG^nf^{(n)}$ is given by:
\begin{align}\label{eq:radresult}
        f^{(2)}_{\rm tidal}=
		m_1 m_2 \frac{c_{E^2}^{(1)}W_{1,E^2}+ c_{B^2}^{(1)} W_{1,B^2}}{(|\tilde{\mathbf{b}}|_2)^9}+(1\leftrightarrow 2)\,,
\end{align}
where the coefficients are given by
\begin{subequations}
\begin{align}
	W_{1,E^2}=&- 12  (\gamma ^2-1 ) |b|^2 (-10 u_1  ( b \cdot \epsilon  \, v_1 \cdot \rho +u_2  \, v_1 \cdot \epsilon \,  v_2 \cdot \rho  )   \nn \\
	&\times [-3 \gamma  \, v_1 \cdot \rho \,  v_2 \cdot \epsilon \,  (2 \gamma ^2-1 )  v_1 \cdot \rho \,  v_1 \cdot \epsilon +3 \gamma   \, v_1 \cdot \epsilon \,  v_2 \cdot \rho ]\nn \\
	&+5  (2 \gamma ^2-1 ) [ b \cdot \epsilon \,  v_1 \cdot \rho   +u_2  \, v_1 \cdot \epsilon \, v_2 \cdot \rho ]^2\nn \\
	&-[u_1^2  (3  (10 \gamma ^2-17 )  (v_1 \cdot \rho )^2  (v_2 \cdot \epsilon )^2  \nn \\
	&+ (v_1 \cdot \epsilon )^2  ( (5-10 \gamma ^2 )  (v_1 \cdot \rho )^2-30 \gamma   \, v_1 \cdot \rho \,  v_2 \cdot \rho +3  (10 \gamma ^2-17 )  (v_2 \cdot \rho )^2 )\nn \\
	& +  6  \, v_1 \cdot \rho \, v_1 \cdot \epsilon \, v_2 \cdot \epsilon   (5 \gamma  \, v_1 \cdot \rho + (17-10 \gamma ^2 )  v_2 \cdot \rho  ) )]) \nn \\
	&+24  (\gamma ^2-1 )^2 u_1^2 (-10 u_1 [ b \cdot \epsilon \, v_1 \cdot \rho  +u_2  \, v_1 \cdot \epsilon \, v_2 \cdot \rho ]  \nn \\
	&\times  [-2 \gamma   \, v_1 \cdot \rho \, v_2 \cdot \epsilon  + (6 \gamma ^2-3 )  v_1 \cdot \rho \,  v_1 \cdot \epsilon +2 \gamma   \, v_1 \cdot \epsilon \, v_2 \cdot \rho ]\nn\\
	&+15  (2 \gamma ^2-1 ) [ b \cdot \epsilon \, v_1 \cdot \rho  + u_2  \, v_1 \cdot \epsilon \, v_2 \cdot \rho ]^2\nn \\
	&+u_1^2 [6   (v_1 \cdot \rho )^2  (v_2 \cdot \epsilon )^2+ (v_1 \cdot \epsilon )^2  (15  (2 \gamma ^2-1 )  (v_1 \cdot \rho )^2   \nn \\
	&+ 20 \gamma   \, v_1 \cdot \rho   \, v_2 \cdot \rho +6  (v_2 \cdot \rho )^2 )-4  \, v_1 \cdot \rho \, v_1 \cdot \epsilon \,  v_2 \cdot \epsilon   (5 \gamma   \, v_1 \cdot \rho  \nn \\
	&+      3  \, v_2 \cdot \rho  )])-12 (5 \gamma ^2-7 ) |b|^4 [ v_1 \cdot \rho \, v_2 \cdot \epsilon - v_1 \cdot \epsilon \,   (v_2 \cdot \rho )]^2\!  ,\\
	W_{1,B^2}=&   120  (\gamma ^2-1 )  (  b \cdot \epsilon  [  v_2 \cdot \rho -\gamma   v_1 \cdot \rho ]   \nn \\
	& +u_1  \, v_1 \cdot \rho  [ \gamma   \, v_1 \cdot \epsilon - v_2 \cdot \epsilon ]+u_2  \, v_2 \cdot \rho  [ v_2 \cdot \epsilon  -\gamma   \, v_1 \cdot \epsilon ]) \nn \\
	&\times ( |b|^2 [ \gamma  \, b \cdot \epsilon \, v_1 \cdot \rho +3 u_1  \, v_1 \cdot \rho \, v_2 \cdot \epsilon   \nn \\
	&-  \gamma  u_1  \, v_1 \cdot \rho \, v_1 \cdot \epsilon -3 u_1  \, v_1 \cdot \epsilon \, v_2 \cdot \rho
	+\gamma  u_2  \, v_1 \cdot \epsilon \, v_2 \cdot \rho ] \nn \\
	&+2  (\gamma ^2-1 ) u_1^2 [u_1  (-2  \, v_1 \cdot \rho \, v_2 \cdot \epsilon +3 \gamma   \, v_1 \cdot \rho \, v_1 \cdot \epsilon +2  v_1 \cdot \epsilon \, v_2 \cdot \rho  ) \nn \\
	&-  3 \gamma   ( b \cdot \epsilon \, v_1 \cdot \rho +u_2  \, v_1 \cdot \epsilon \, v_2 \cdot \rho  )])\,.
\end{align}
\end{subequations}
We have confirmed this result agrees with the recent \rcite{Mougiakakos:2022sic}.

From the complete waveform $f^{(2)}=f^{(2)}_{\rm pp}+f^{(2)}_{\rm tidal}$
we may deduce both the radiated linear and angular momentum~\cite{Damour:2020tta,Bonga:2018gzr}:
\begin{align}
	P^{\mu}_{\text{rad}}&=
	\frac1{32\pi G}\int\!\d u \d\sigma [\dot{f}_{ij}]^{2}\rho^{\mu}\label{eq:momdef}\,,
	\quad \text{where } f=f_{ij}\epsilon^{ij}\,,\\
	J^{\text{rad}}_{ij}&=\frac1{8\pi G}\int\!\d u \d\sigma \left (f_{k[i}\dot{f}_{j]k} -\frac{1}{2} x_{[i}\partial_{j]}f_{kl}\dot{f}_{kl}\right )\, ,
\end{align}
with $\dot{f}_{ij}:=\partial_{u}f_{ij}$ and
$\d\sigma=\sin\theta\d\theta\d\phi$ the unit sphere measure.
These formulae are most naturally interpreted in the center-of-mass frame ---
see \rcite{Jakobsen:2021lvp} for details.
From \eqref{eq:momdef} we confirm our expression \eqref{eq:radMom} for $P_{\rm rad}^\mu$,
working up to $\cO(v^5)$ in a PN expansion in order to handle the difficult multi-scale integrals.
As for the radiated angular momentum, at leading order in $G$ this
involves an integral over the gravitational wave memory,
which vanishes at the current $\cO(G^2)$:
\begin{align}
	\Delta f_{\rm tidal}(\mathbf{\hat{x}}):=f_{\rm tidal}(+\infty,\mathbf{\hat{x}})-f_{\rm tidal}(-\infty,\mathbf{\hat{x}})
	=\mathcal{O}(G^3)\,.
\end{align}
As explained in \rcite{Jakobsen:2021smu},
only diagrams with a deflection mode $z_i^\mu$ propagating on a worldline
contribute to the memory: this is not the case for the two diagrams in \Fig{fig:WavTid}.
Hence the correction to $J_{\rm rad}$ from tidal effects begins at $\cO(G^3)$,
and using the linear response relation \eqref{eq:biniDamour} this accounts for the fact
that the radiative part of the scattering angle $\theta_{\rm rad}$ is unmodified by tidal effects
until $\cO(G^4)$.

\section{Conclusions}

In this paper we have introduced the Schwinger-Keldysh in-in formalism to the WQFT,
defining complete (including radiation-reaction) scattering observables.
This involved doubling the degrees of freedom in our action,
with two evolution operators evolving states forwards and backwards in time
between the observation point and past infinity.
Our main observation is that calculating complete observables using WQFT
with the in-in formalism boils down to a simple prescription:
draw the sum of tree-level Feynman diagrams,
with exactly the same Feynman rules as in the in-out prescription,
but with \emph{retarded propagators} all pointing towards the outgoing line
representing the operator being measured.
This ensures a flow of causality,
so that when performing calculations there is no practical need
to double degrees of freedom in the action.

This simplification is in no way surprising ---
indeed, it has been taken somewhat implicitly in previous work on the WQFT
\cite{Mogull:2020sak,Jakobsen:2021smu,Jakobsen:2021lvp,Jakobsen:2021zvh,Jakobsen:2022fcj,Shi:2021qsb}.
From a purely classical perspective,
one may view the sum over tree-level WQFT diagrams as simply a mechanism
for solving the classical equations of motion of the theory ---
the approach taken in \rcite{Saketh:2021sri} for solving classical electromagnetism 
or by Westpfahl~\cite{Westpfahl:1985tsl} and more
recently Damour~\cite{Damour:2017zjx,Damour:2019lcq} in classical gravity ---
but working in configuration rather than momentum space.
Fixing boundary conditions at past infinity then requires
the use of retarded propagators.
This illustrates a key advantage of the WQFT over PM EFT-style methods
\cite{Kalin:2020mvi,Kalin:2020fhe,Dlapa:2021npj,Dlapa:2021vgp}:
in the latter, integrating out the gravitons to leave an effective worldline action ---
separating bulk from worldline degrees of freedom ---
obscures the causal nature of the observables.
By contrast, the WQFT profits by focusing on physical observables of the theory.
While the need to double degrees of freedom in the action when using PM EFT can still be
cleverly avoided by focusing on specific radiative observables,
such as the radiated four-momentum $P_{\rm rad}^\mu$
\cite{Riva:2021vnj,Mougiakakos:2022sic,Riva:2022fru},
we prefer a more flexible approach that produces complete observables.

We are left though with a practical question,
which is the feasibility of performing loop integrals with retarded propagators.
At first glance this seems a difficult task,
given the apparent need to track signs on $i0$ for all of our graviton propagators.
However, this prescription is only relevant for \emph{active propagators}:
ones that can go on-shell over the domain of integration,
and so for which the integration contour plays a meaningful role.
In fact, at 3PM order there is no more than one active graviton propagator in any given integral
(and none at all at 2PM order, which explains these integrals being rational functions).
This simplifying property is the result of energy conservation on the worldlines,
which restricts most graviton momenta to the potential region where
$\ell_i^{\rm pot}=(\ell_i^0,\Bell_i)\sim(v,1)$.

Here there is an advantage of starting from the worldline rather than
calculating observables using scattering amplitudes in the KMOC formalism
\cite{Kosower:2018adc,Maybee:2019jus}.
While it is tempting to assume that ---
upon taking the classical $\hbar\to0$ limit in the KMOC approach ---
the loop integrands obtained match those in the WQFT,
the requirement of energy conservation is obscured from the amplitudes perspective.
Using KMOC it is therefore more natural to adopt a Feynman-type basis,
in which case the formalism also produces cut integrals.
These cuts are unnecessary when using the WQFT because,
essentially, the retarded propagators already contain them:
\begin{align}
	\frac1{k^2+{\rm sgn}(k^0)i0}&=
	\frac1{k^2+i0}+
	2i\pi\theta(-k^0)\delta(k^2)\,.
\end{align}
Furthermore, integrals with retarded propagators are purely \emph{real}
in the physical domain $1<\gamma<\infty$,
as compared with Feynman integrals that are (pseudo-)real only in the unphysical domain $-1<\gamma<1$
\cite{Herrmann:2021tct,DiVecchia:2021bdo}.
This makes sense,
given that we seek real physical observables as linear combinations of these integrals.

To illustrate the Schwinger-Keldysh WQFT formalism we calculated complete gravitational observables
in a scattering event involving two massive bodies including tidal corrections,
including the momentum impulse $\Delta p_1^\mu$, scattering angle $\theta$,
and radiated four-momentum $P_{\rm rad}^\mu$. The latter two
were also obtained recently using the worldline EFT formalism~\cite{Mougiakakos:2022sic}.
We observed that the tidal correction to the scattering angle
is unmodified by the inclusion of radiation-reaction effects,
which follows via the Bini-Damour linear response relation
\cite{Bini:2012ji,Damour:2020tta,Bini:2021gat}
from the absence of radiated angular momentum $J_{\rm rad,tid}$ at 2PM order.
As confirmation we calculated the complete time-domain gravitational waveform at 2PM
order from this scattering event, again reproducing the results of \rcite{Mougiakakos:2022sic},
and confirming both the radiated linear and angular momentum at 3PM and 2PM orders respectively.

Our use of the Schwinger-Keldysh in-in formalism,
and in particular our understanding of how to handle loop integrals with retarded propagators,
paves the way to more advanced calculations in the future.
A first step will be upgrading 3PM scattering observables
involving quadratic-in-spin (quadrupole) effects,
produced earlier this year by two of the present authors~\cite{Jakobsen:2022fcj},
to include radiation-reaction effects.
Here the radiated four-momentum has been produced~\cite{Riva:2022fru},
but the full set of observables remains unknown.
In the longer term, we see the WQFT as a powerful tool for performing 4PM calculations.
Our aim will be to produce complete observables,
thus avoiding any ambiguity between conservative and radiative contributions in the future.

\begin{acknowledgments}
	We would like to thank Gregor K\"alin, Chia-Hsien Shen and Jan Steinhoff for very insightful discussions
	and comments on the manuscript.
	We also thank the organizers of the ``High-Precision Gravitational Waves'' program
	at the Kavli Institute for Theoretical Physics (KITP) for their hospitality.
	GUJ's and GM's research is funded by the Deutsche Forschungsgemeinschaft (DFG, German Research Foundation),
	Projektnummer 417533893/GRK2575 ``Rethinking Quantum Field Theory''.
	This research was also supported in part by the National Science Foundation under Grant No.~NSF PHY-1748958.
\end{acknowledgments}

\begin{appendix}

\section{In-in propagator matrix}
\label{appA}

In order to derive the in-in propagator matrix we begin with the definition
of the generating functional using the time-evolution operators in the free theory~\eqref{WJJ}:
\be
\label{genfun}
e^{\frac{i}{\hbar} W[J_{1},J_{2}]}=
\langle 0| \hat U^{(0)}_{J_{2}}(-\infty,\infty)\, \hat U^{(0)}_{J_{1}}(\infty,-\infty) 
| 0\rangle \, ,
\ee
with
\begin{equation}
	\hat U^{(0)}_{J}(T',T) =
	\mathcal{T} \exp \Bigr [ \frac{i}{\hbar} \int_{T}^{T'}\!\!\d t\! \int \!\d^{3}x
	\left (J(x) \hat \phi_{I}(x) \right )\Bigr ] \,.
\end{equation}
Taking a variational derivative of $\hat U^{(0)}_{J}(T',T)$
with respect to $J(x)$  inserts the field operator $\hat \phi_{I}(x) $
into the time-ordered expression:
\be
\frac{\hbar}{i}\, \frac{\delta \hat U^{(0)}_{J}(T',T)}{\delta J(t,\mathbf{x})}  = \mathcal{T}
\Bigl (\hat \phi_{I}(t,\mathbf{x}) \exp \Bigr [ \frac{i}{\hbar} \int_{T}^{T'}\!\!\d t\! \int \!\d^{3}x \left (J(x) \hat \phi_{I}(x) \right )\Bigr ] \Bigr )\, ,
\ee
and similarly for the anti-time ordered case $(T>T')$. Similar relations hold for 
two functional derivatives. It is then straightforward to compute the propagator matrix
via functional derivatives of the generating functional \eqref{genfun}:
\be
\vev{\phi_{A}(x) \phi_{B}(y)}=
\frac{\delta^{2}W[J_{1},J_{2}]}{\delta J_{B}(y) \, \delta J_{A}(x)}\, \Bigr |_{J_{a=0}}=
\frac{\delta^{2}
\langle 0| \hat U^{(0)}_{J_{2}}(-\infty,\infty)\, \hat U^{(0)}_{J_{1}}(\infty,-\infty)  | 0\rangle
}{\delta J_{B}(y) \, \delta J_{A}(x) }  \,\Bigr |_{J_{a=0}}\,,
\ee
where $A,B=1,2$.
Using $\vev{0|0}=1$ one finds that
\begin{align}
\begin{aligned}
\vev{\phi_{1}(x) \phi_{1}(y)}&= \langle 0| \mathcal{T} \phi(x)\phi(y)|0\rangle\,, \qquad
\vev{\phi_{1}(x) \phi_{2}(y)}= \langle 0| \phi(y)\phi(x)|0\rangle\,, \\ 
\vev{\phi_{2}(x) \phi_{1}(y)}&= \langle 0| \phi(x)\phi(y)|0\rangle\,, \qquad\,\,\,\,\,
\vev{\phi_{2}(x) \phi_{2}(y)}= \langle 0| \mathcal{T}^{\ast}\phi(x)\phi(y)|0\rangle \, ,
\end{aligned}
\end{align}
as quoted in \eqn{12matrix}.
From here it is an easy exercise to transform this into
the Keldysh basis using $\phi_{+}= \sfrac{1}{2} (\phi_{1}+\phi_{2})$ and $\phi_{-}= \phi_{1}
-\phi_{2}$,  To do so, one uses
\begin{align}
\begin{aligned}
D_{\text{ret}}(x,y)&= \theta(x^{0}-y^{0})\vev{0|[\phi(x),\phi(y)]|0}\\
&= D_{F}(x,y) - D_{-}(x,y)
=D_{+}(x,y) - D_{D}(x,y)\,,\\ 
D_{\text{adv}}(x,y)&= \theta(y^{0}-x^{0})\vev{0|[\phi(x),\phi(y)]|0}\\
&= D_{F}(x,y) - D_{+}(x,y)
=D_{-}(x,y) - D_{D}(x,y)\,,
\end{aligned}
\end{align}
and $D_{H}(x,y) = \vev{0|\{\phi(x),\phi(y)\}|0}$ to find that
\begin{align}
\begin{aligned}
\vev{\phi_{+}(x) \phi_{-}(y)}&= D_{\text{ret}}(x,y)\,,&
\vev{\phi_{-}(x) \phi_{+}(y)}&= -D_{\text{adv}}(x,y)\,, \\ 
\vev{\phi_{+}(x) \phi_{+}(y)}&= \sfrac{1}{2} D_{H}(x,y)\,, &
\vev{\phi_{-}(x) \phi_{-}(y)}&= 0 \, .
\end{aligned}
\end{align}
This precisely reproduces \eqn{Keldyshprop}.

\end{appendix}

\bibliographystyle{JHEP}
\bibliography{../bib/wqft_spin}

\end{document}